\documentclass[fleqn,usenatbib]{mnras}

\usepackage{newtxtext,newtxmath}

\usepackage[T1]{fontenc}
\usepackage{ae,aecompl}

\usepackage{graphicx}	
\usepackage{tikz}
\usepackage{natbib}
\usepackage[nolist]{acronym}

\newcommand{\Ha}{H$\alpha$}
\newcommand{\Hb}{H$\beta$}
\newcommand{\Hy}{H$\gamma$}
\newcommand{\Hd}{H$\delta$}

\newcommand{\HI}{H\,{\sc i}}
\def\HII{\hbox{H\,{\sc ii}}}
\newcommand{\SII}{[S\,{\sc ii}]}
\newcommand{\NII}{[N\,{\sc ii}]}
\newcommand{\OI}{[O\,{\sc i}]}

\newcommand{\OIII}{[O\,{\sc iii}]}
\newcommand{\D}{$^\circ$}
\newcommand{\iaus}{IAU Symposium}
\def\arcmin{\hbox{$^\prime$}}
\def\arcsec{\hbox{$^{\prime\prime}$}}
\def\kms{km\,s$^{-1}$}

\newcommand{\FeII}{[Fe\,{\sc ii}]}

\begin{acronym}
\acro{ATCA}{Australia Telescope Compact Array}
\acro{ATNF}{Australia Telescope National Facility}
\acro{ASKAP}{Australian Square Kilometre Array Pathfinder}
\acro{CABB}{Compact Array Broadband Backend}
\acro{CC}{Core Collapse}
\acro{CSIRO}{Australian Commonwealth Scientific and Industrial Research Organisation}
\acro{IR}{Infrared}
\acro{IRAF}{Image Reduction and Analysis Facility}
\acro{ISM}{interstellar medium}
\acro{LMC}{Large Magellanic Cloud}
\acro{MCs}{Small \& Large Magellanic Clouds}
\acro{MCELS}{Magellanic Cloud Emission Line Survey}
\acro{MW}{Milky Way}
\acro{MWA}{Murchison Widefield Array}
\acro{NSF}{National Science Foundation}
\acro{NOAO}{National Optical Astronomy Observatory}
\acro{NRAO}{National Radio Astronomy Observatory}
\acro{PMN}{Parkes-MIT-NRAO}
\acro{PN}{Planetary Nebula}
\acro{PNe}{Planetary Nebulae}
\acro{PSN}{CXOU J054138.2--665817}
\acro{PSS}{CXOU J054140.4--665912}
\acro{PWN}{Pulsar Wind Nebulae}
\acro{SMC}{Small Magellanic Cloud}
\acro{SN}{Supernova}
\acro{SNe}{Supernovae}
\acro{SNRs}{Supernova Remnants}
\acro{SNR}{Supernova Remnant}
\acro{WR}{Wolf Rayet}

\end{acronym}

\title[New Optically Identified SNRs in the LMC]{New Optically Identified Supernova Remnants in the Large Magellanic Cloud}

\author[M. Yew et al.]{
Miranda Yew$^{1}$, Miroslav D.~Filipovi\'c$^{1}$,\thanks{E-mail: m.filipovic@westernsydney.edu.au} Milorad Stupar$^{1}$, Sean D.~Points$^{2}$, 
\newauthor Manami Sasaki$^{3}$, Pierre Maggi$^{4}$, Frank Haberl$^{5}$, Patrick J.~Kavanagh$^{6}$, 
\newauthor Quentin A. Parker$^{7,8}$, Evan J.~Crawford$^{1}$, Branislav Vukoti\'c$^{9}$, Dejan Uro\v sevi\'c$^{10,11}$, 
\newauthor Hidetoshi Sano$^{12,13}$, Ivo R.~Seitenzahl$^{14}$, Gavin Rowell$^{15}$, Denis Leahy$^{16}$, 
\newauthor Luke~M. Bozzetto$^{1}$, Chandreyee~Maitra$^{5}$, Howard Leverenz$^{1}$, Jeffrey L.~Payne$^{1}$, 
\newauthor Laurence A. F. Park$^{1}$, Rami Z. E.~Alsaberi$^{1}$ and Thomas G.~Pannuti$^{17}$
\\
\\
$^{1}$Western Sydney University, Locked Bag 1797, Penrith South DC, NSW 2751, Australia\\
$^{2}$Cerro Tololo Inter-American Observatory/NSF's NOIRLab, Casilla 603, La Serena, Chile\\
$^{3}$Remeis Observatory and ECAP, Universit\"{a}t Erlangen-N\"{u}rnberg, Sternwartstra{\ss}e 7, D-96049 Bamberg, Germany \\
$^{4}$Observatoire Astronomique de Strasbourg, Universit\'e de Strasbourg, CNRS, 11 rue de l'Universit\'e, F-67000 Strasbourg, France \\
$^{5}$Max-Planck-Institut f\"{u}r extraterrestrische Physik, Gie{\ss}enbachstra{\ss}e 1, D-85748 Garching, Germany \\
$^{6}$School of Cosmic Physics, Dublin Institute for Advanced Studies, 31 Fitzwilliam Place, Dublin 2, Ireland \\ 
$^{7}$Department of Physics, University of Hong Kong, Hong Kong\\
$^{8}$Laboratory for Space Research, The University of Hong Kong, Hong Kong\\
$^{9}$Astronomical Observatory, Volgina 7, PO Box 74 11060 Belgrade, Serbia\\
$^{10}$Department of Astronomy, Faculty of Mathematics, University of Belgrade, Studentski trg 16, 11000 Belgrade, Serbia\\
$^{11}$Isaac Newton Institute of Chile, Yugoslavia Branch\\
$^{12}$Institute for Advanced Research, Nagoya University, Chikusa-ku, Nagoya 464-8601, Japan\\
$^{13}$National Astronomical Observatory of Japan, Mitaka, Tokyo 181-8588, Japan\\
$^{14}$School of Science, University of New South Wales, Australian Defence Force Academy, Canberra, ACT 2600, Australia\\
$^{15}$School of Physical Sciences, University of Adelaide, North Terrace, Adelaide, SA 5005, Australia\\
$^{16}$Department of Physics and Astronomy, University of Calgary, University of Calgary, Calgary, Alberta, T2N 1N4, Canada\\
$^{17}$Space Science Center, Department of Physics, Earth Science and Space Systems Engineering, Morehead State University, Morehead, KY 40351, USA\\
}
\date{Accepted XXX. Received YYY; in original form ZZZ}

\pubyear{2020}

\begin{document}
\label{firstpage}
\pagerange{\pageref{firstpage}--\pageref{lastpage}}
\maketitle

\begin{abstract}
We present a new optical sample of three \ac{SNRs} and 16 \ac{SNR} candidates in the \ac{LMC}. These objects were originally selected using deep \Ha, \SII\ and \OIII\ narrow-band imaging. Most of the newly found objects are located in less dense regions, near or around the edges of the \ac{LMC}'s main body. Together with previously suggested MCSNR~J0541--6659, we confirm the \ac{SNR} nature for two additional new objects: MCSNR~J0522--6740 and MCSNR~J0542--7104. Spectroscopic follow-up observations for 12 of the \ac{LMC} objects confirm high \SII/\Ha\ emission-line ratios ranging from 0.5 to 1.1. We consider the candidate J0509--6402 to be a special example of the remnant of a possible type~Ia \ac{SN} which is situated some 2\D\ ($\sim$1.75~kpc) north from the main body of the \ac{LMC}. We also find that the \ac{SNR} candidates in our sample are significantly larger in size than the currently known \ac{LMC} \ac{SNRs} by a factor of $\sim$2. This could potentially imply that we are discovering a previously unknown but predicted, older class of large \ac{LMC} \ac{SNRs} that are only visible optically. Finally, we suggest that most of these \ac{LMC} \ac{SNRs} are residing in a very rarefied environment towards the end of their evolutionary span where they become less visible to radio and X-ray telescopes. 
\end{abstract}




\section{Introduction}
 \label{sec:intro}

\ac{SNRs} play a major role in our understanding of \ac{SNe}, the \ac{ISM}, and the evolution of galaxies as a whole. They are generally divided into two categories according to their progenitors: core-collapse and thermonuclear (type~Ia) \ac{SNRs}. Core-collapse \ac{SNRs}, as their name suggests, result from massive stars undergoing core collapse. Thermonuclear \ac{SNRs} are remnants of type~Ia \ac{SNe} occurring when a massive white dwarf in a binary system experiences a runaway thermonuclear fusion reaction resulting in the explosive release of nuclear energy \citep[for a review see e.g.][]{2013FrPhy...8..116H}. These two types of \ac{SNe} eject heavy elements into the \ac{ISM} and heat it. \ac{SNe} create shock waves that compress magnetic fields and efficiently accelerate particles such as energetic cosmic rays observed throughout the Galaxy. Therefore, a complete study of the properties of these two types of \ac{SNRs} in galaxies can provide an opportunity to understand the origin of cosmic rays, the star formation history and chemical evolution of galaxies.

In general, \ac{SNRs} are characterised by their simultaneous exhibition of diffuse X-ray emission, their non-thermal radio spectral index, and their high \SII/\Ha\ ratios as these signatures are produced by high-velocity shocks. We utilise the criteria given in \citet[][Sect.~4.1]{1998A&AS..130..421F} to establish bona-fide \ac{SNRs} from \ac{SNR} candidates. In the simplest form, source X-ray and/or radio confirmation in addition to the high optical \SII/\Ha\ ratio is necessary for an object to be established as a bona-fide \ac{SNR}. Otherwise, if a source satisfies only one of these three multi-frequency criteria, we consider it to be an \ac{SNR} candidate.

Various surveys of \ac{SNRs} in our Galaxy and nearby galaxies have been carried out at radio, X-ray, \ac{IR} and optical wavelengths. The first extragalactic \ac{SNR} candidates were identified in the \ac{LMC} by \citet{mathewson_healey_1964} and later confirmed with a combination of radio and optical techniques by \citet{1966MNRAS.131..371W}. To date, a total of 60 \ac{SNRs} have been confirmed in the \ac{LMC} with an additional 14 suggested candidates \citep{2016A&A...585A.162M,2017ApJS..230....2B,2019MNRAS.490.5494M}. However, sensitivity and resolution limitations severely reduce the effectiveness of the past \& present generations of radio and X-ray searches for \ac{SNRs} in galaxies beyond the \ac{MCs} \citep{1980MNRAS.193..901G, 1981ApJ...246L..61L, 1985ApJ...293..400C,1997ApJS..112...49M,1997ApJS..113..333M,2012SerAJ.184...19M,2014SerAJ.189...15G,2018A&A...620A..28S,Lin_2020,2020AN....341..156S}. As a result, optical studies have produced the largest number ($\sim$1200) of new extra-Magellanic \ac{SNR} candidates. Optical extragalactic searches for \ac{SNRs} are mainly done by using an emission line ratio criterion of the form \SII/\Ha\ \textgreater~0.4--0.5 \citep{1973ApJ...180..725M, 1980A&AS...40...67D, 1984ApJ...281..658F,1997ApJS..112...49M,1997ApJS..108..261B,2010ApJ...710..964D,Lee_2014,2019A&A...628A..87V,2019SerAJ.198...13V,2018A&AT...30..379V,Lin_2020}. This criterion separates shock-ionisation from photoionisation in \ac{SNRs} from \HII\ regions and \ac{PNe} \citep{2010PASA...27..129F}. \ac{SNR} radiative shocks collisionally excite sulphur ions in the extended recombination region resulting in S$^+$, hence the larger contribution of \SII\ accounting for an increase of the \SII\ to \Ha\ ratio. In typical \HII\ regions, sulphur exists predominantly in the form of S$^{++}$, yielding low \SII\ to \Ha\ emission ratios. Ratios from narrow-band imaging are usually verified spectroscopically, since \NII\ lines at 6548 and 6584\AA\ can contaminate the \Ha\ images at an unknown and variable level. Spectroscopic observations of such emission nebulae also can provide other evidence of shock heating, such as strong \OI\ $\lambda$6300 emission, elevated \NII\ to \Ha\ with respect to \HII\ regions, or high \OIII\ electron temperatures, verifying the candidate as being a \ac{SNR} \citep{1981ApJ...247..879B, 1982ApJ...254...50B, 1990ApJS...72...61L, 1993ApJ...407..564S, 1997ApJS..108..261B}. Although somewhat biased as an isolated criterion, this method is proven and a good way of identifying ordinary radiatively cooling \ac{SNRs} in nearby galaxies. We note that young, Balmer-dominated \ac{SNRs} \citep{1980ApJ...235..186C} would be missed by this criterion.

The \ac{LMC} galaxy, the target of this study, lies towards the south ecliptic pole and is in one of the coldest parts of the radio sky, uncontaminated by Galactic foreground emission \citep{doi:10.1175/1520-0469(1991)048<0651:OTCOED>2.0.CO;2,2010MNRAS.405.1349R}. The \ac{LMC}'s position and its known distance of 50~kpc \citep{2008MNRAS.390.1762D} makes the \ac{LMC} arguably an ideal galaxy in which to study \ac{SNRs} in our Local Group of galaxies. \ac{LMC} objects can also be assumed to be located at approximately the same distance even with the tilt (inclination) of the \ac{LMC} at 23\D\ toward the line of sight \citep{2010A&A...520A..24S}. This tilt introduces $<$10~per cent of additional uncertainty in diameter estimates. 

Apart from the above mentioned \ac{LMC} \ac{SNR} survey papers, there are a number of studies focusing on particular \ac{LMC} \ac{SNRs}, such as N\,103B and N\,132D. Some recent studies include: \citet{2007MNRAS.378.1237B, 2009SerAJ.179...55C, 2008SerAJ.177...61C, 2010A&A...518A..35C, Crawford_2014, 18e685d5aef24c268b0ab9ad6dcb56d7, 2012A&A...539A..15G, 2017ApJS..230....2B, 2012A&A...540A..25D, 2014AJ....147..162D, 2013A&A...549A..99K, 2014ApJ...780...50B, 2016A&A...586A...4K, 2015A&A...583A.121K, 2015A&A...579A..63K, 2015A&A...573A..73K, 2015MNRAS.454..991R, 2014A&A...561A..76M, 2014A&A...567A.136W, 2014Ap&SS.351..207B, 2015PKAS...30..149B, 2013MNRAS.432.2177B, 2014MNRAS.439.1110B, 2014Ap&SS.351..207B, 2014MNRAS.440.3220B, 2012RMxAA..48...41B, 2012SerAJ.184...69B, 2012MNRAS.420.2588B, 2012SerAJ.185...25B, 2017ApJ...837...36L, 2017ApJ...847..122G, 2018ApJS..237...10D, 2019MNRAS.490.5494M, 2019Ap&SS.364..204A, 2019PhRvL.123d1101S, 2019AJ....158..149L,2020arXiv200404531P}.

In this paper, we present new optical narrow band imaging data for 19 \ac{LMC} objects. For 12 of these 19 objects we also have spectroscopic follow-up observations. The nebular lines detected include \OIII\ 4959/5007\AA, \OI\ 6300/6364\AA, \SII\ 6716/6731\AA, and hydrogen Balmer lines, \Ha, \Hb, \Hy\ and \Hd. Further, \NII\ 6548/6583\AA\ ratio (and line detection) are generally common in \ac{SNR} spectra, but in the \ac{MCs}, due to a low abundance of nitrogen \citep[see][]{1979AuJPh..32..123D, 1990ApJS...74...93R,2019AJ....157...50D} these lines are not necessarily detected, or if they are seen they are usually not as intense compared with the same lines of \ac{SNRs} in the \ac{MW} or in other galaxies. These optical lines are common for evolved SNRs, i.e., for remnants whose interiors are in the cooling phase while their shells are merging with the \ac{ISM}. Compared with old \ac{SNRs}, younger ones show many more lines in their optical spectra, although they do not share the same spectral classification such as the type of \ac{SN} explosion and the surrounding \ac{ISM}. Likewise, in the \ac{LMC} we notice the existence of non-radiative Balmer dominated and oxygen rich \ac{SNRs} \citep[see][]{1991ApJ...375..652S, 2011Ap&SS.331..521V,2018ApJ...853L..32S,2018NatAs...2..465V} like others discovered in the Galaxy or other galaxies but, somewhat surprisingly, not in M\,33 \citep{2010ApJS..187..495L,Lin_2020}.

The layout of this paper is as follows: In Section~\ref{sec:optical imaging} we describe the observations and imaging techniques. In Section~\ref{sec:results} we present our results on the 19 \ac{LMC} objects studied and in Section~\ref{sec:discussion} we investigate the true nature of these objects. Finally, in Section~\ref{sec:conclusion} we summarise our findings.

\section{Optical and X-ray Observations}
\label{sec:optical imaging} 

\subsection{\ac{MCELS} Observations}

We used images from the Magellanic Cloud Emission Line Survey \citep[\ac{MCELS};][]{1999IAUS..190...28S}. These images were taken at the UM/CTIO (University of Michigan) Curtis Schmidt telescope at Cerro Tololo Inter-American Observatory (CTIO). The detector, a Tek $2048 \times 2048$ CCD with 24~$\mu$m pixels, had a scale of $2.3^{\prime\prime}$ per pixel and a resulting angular resolution of approximately $4.6^{\prime\prime}$. The narrowband images were taken with filters centered on the \OIII\ (${\lambda}_{\rm c}$ 5008\AA, FWHM=50\AA), \Ha+\NII\ (${\lambda}_{\rm c}$ 6563\AA, FWHM=30\AA), and \SII\ (${\lambda}_{\rm c}$ 6724\AA, FWHM=50\AA) emission-lines along with green (${\lambda}_{\rm c}$ 5130\AA, FWHM=155\AA) and red (${\lambda}_{\rm c}$ 6850\AA, FWHM=95\AA) continuum filters. The optical data were reduced using the IRAF\footnote{\ac{IRAF} is distributed by the \ac{NOAO}, which is operated by the Association of Universities for Research in Astronomy (AURA) under a cooperative agreement with the \ac{NSF}.} software package for bias subtraction and flat-field correction. The astrometry was derived from stars in the Two Micron All Sky Survey (2MASS) J-band catalog \citep{2006AJ....131.1163S}. The data were flux-calibrated using observations of spectro-photometric standard stars \citep{1994PASP..106..566H,1992PASP..104..533H} and then continuum subtracted. More details about \ac{MCELS} observations can be found in \citet{2019ApJ...887...66P}\footnote{Data can be found in: ftp://ftp.ctio.noao.edu/pub/points/MCELS/LMC/}. 

In Table~\ref{tab:snrtable} we list all 19 objects studied in this paper. In Col.~2 we list the date of the spectroscopic observations for objects listed in Table~\ref{tab:snrflux}; source name and its central position (RA and DEC) are listed in Cols.~3, 4 and 5; source extent as major and minor axis/diameter are listed in Col.~6 (in arcsec) while in Col.~7 we show the average of major and minor axes converted to parsecs for a distance of 50~kpc; the position angle (PA; Col.~8) is measured from north to east. \SII/\Ha\ from \ac{MCELS} in Col.~9 represent the average measured value within the \ac{SNR} candidate extent after subtracting local noise. Only values $>5\sigma$ of the local noise were used and \SII\ value provided in Col.~9 is the sum of the 6716\AA\ and 6731\AA\ lines. In Col.~10 we show the number of massive OB stars found within $\sim$100~pc radius as well as within the object's extent as measured in (Col.~6).

\subsection{WiFeS}
\label{wifes}

The Wide-Field Spectrograph \citep[WiFeS;][]{Dopita2007, 2010Ap&SS.327..245D} was used to obtain the integral field spectra of 12 \ac{MCELS} selected objects (see Section~\ref{sec:3.1} for more details). This integral field unit (IFU) is mounted on the Australia National University (ANU) 2.3-m telescope at the Siding Spring Observatory (SSO). The spectrograph is an image slicer, consisting of a combination of 25 $\times$ 1\arcsec\ wide adjacent slits each 38\arcsec\ in length to yield an effective 25\arcsec $\times$ 38\arcsec\ field of view on the sky. 

We performed our WiFeS spectral observation between October~21 and 23, 2017 as well as on September~7, 2019 using the RT560 dichroic as a beam splitter to send light to both the blue and red arms of the spectrograph. However, only the September~7, 2019 observation was done under excellent atmospheric (photometric) conditions. Each observation was 20--50~minutes in length and we used Nod-and-Shuffle techniques as per \citet{Dopita2007}. In the blue arm, the medium resolution grating with 708~lines~mm$^{-1}$ (B3000; $\sim$51.7~\AA~mm$^{-1}$) was used and for the red arm we used the higher resolution grating of 1200~lines~mm$^{-1}$ (R7000; $\sim$29.0~\AA~mm$^{-1}$). According to \citet[][Table~2]{2010Ap&SS.327..245D}, the R7000 grating actually achieves R$\sim$6800 ($\sim$45~km~s$^{-1}$) and the B3000 grating is actually R $\sim$2900 ($\sim$105~km~s$^{-1}$). At \Ha, this means that R7000 grating has a resolution of just under 1~\AA\ (0.965) (Figure~\ref{fig:A5}) and at \Hb\ B3000 has a resolution of about 1.67~\AA.

Our choice of gratings provided overlapping blue and red spectra with a coverage from ${\sim}$3400~to~7000~\AA. \citet{Dopita2007} and \citet{2010Ap&SS.327..245D} presented the on-telescope end-to-end transmission of the WiFeS spectrograph, including the telescope, atmosphere and detectors. According to this work, the transmission of the R7000 grating (which we used for the red arm observations) at the wavelength of the \SII\ sulphur lines is 12.5~per~cent lower compared to the transmission at the wavelength of \Ha. This transmission correction was applied to all of our relative flux observations of the \SII\ lines to \Ha.

After the data reduction of the WiFeS observations, which was performed with the PyWiFeS pipeline (Version 0.7.0 e.g. \citealt{2014Ap&SS...349..617C}), the final product is a 3D spectral data cube with R.A., Dec and $\lambda$ as the third dimension. Using a circular or square aperture, summed 1D spectra were extracted from the cube (see example in \citealt{2018MNRAS...479.4432}). For our sample of obtained cubes, we used circular apertures between 5\arcsec\ and 10\arcsec\ positioned at the brightest parts of the \ac{SNR} filaments to extract 1D spectra (see example in Figure~\ref{fig:A6} (bottom right)). We used the \ac{IRAF} splot routine to subtract the sky. The same results were achieved using Starlink Splat routine. \OI\ spectral lines at 6300/6363~\AA\ are frequently present in the spectrum of \ac{SNRs} and are good \ac{SNR} indicators especially in their young phase. However, they are also present in the spectrum of the night sky. Therefore, distinguishing whether such lines belong to the object or they are from the night sky is of essential importance. To resolve this problem, spectra with the presence of these lines were first checked on 2D data and then their wavelengths were measured. If the wavelengths were exactly 6300/6360~\AA, then the lines definitely belong to the night sky and are deleted from the 1D spectra. If the lines were shifted for some 5--6~\AA\ (our usual shift for the \ac{LMC}) its assumed they originate from the given object, and therefore, the lines are kept in the 1D spectrum. We present details in Section~\ref{sec:notes} for all 12 spectroscopically observed objects. Also, Tables~\ref{tab:snrtable} and \ref{tab:snrflux} provide aperture positions, relative intensities, and ratios of observed lines.

We note that the intrinsically fainter blue flux of three sources produced very weak blue spectra, so that these three candidates only have data for the red part of spectrum (see Table~\ref{tab:snrflux}). The blue spectra of other candidates should be taken with the understanding that we have not applied individual reddening corrections. 
For one object, J0548--6941, we have determined physical fluxes of the observed lines (see Table~\ref{tab:snrflux}). The fluxes were determined by calibration against the spectrophotometeric standard star LTT1020.

Some line ratios are set by atomic physics. Specifically, the \mbox{$^1$D$_2$ $\rightarrow ^{3}$P} doublet ratios \OIII\ 5007/4959\AA\ and \NII\ 6583/6548\AA\ must both be $\sim$2.9 \citep{1989Msngr..58...44A}, which makes measuring the flux of the fainter blue lines (\OIII4959\AA\ and \NII6548\AA) difficult in several cases (e.g. source J0542--7104 (Figure~\ref{fig:A18}) where \OIII4959\AA\ is hardly above the background fluctuations). To avoid adding noise, we measure only the brighter red component of the doublets, and multiply by (1+1/3) to calculate \NII/H$\alpha$ and \OIII/H$\alpha$ ratios.

\subsection{{\it XMM-Newton} data}
 \label{sec:XMM}
To search for X-ray emission from our candidate \ac{SNRs} we created an up-to-date mosaic of all {\it XMM-Newton} \citep{2001A&A...365L...1J} observations available in the \ac{LMC} area. This included all archival observations that were public up to July~2020, as well as recent observations from projects with author involvement centred within 4\degr\ around R.A.(J2000)=05$^h$22$^m$00$^s$ and Dec(J2000)=--68\degr\ 30\arcmin\ 00\arcsec\ \citep[for more detailed descriptions see][]{2016A&A...585A.162M,2013A&A...558A...3S,2012A&A...545A.128H}. Observations were processed with \textit{XMM-Newton} science analysis software \texttt{SAS} version 17.0.0. We combined the data from the European Photon Imaging Camera, EPIC-pn and EPIC-MOS \citep{2001A&A...365L..18S,2001A&A...365L..27T} and created images in the 0.3--0.7~keV, 0.7--1.1~keV and 1.1--4.2~keV energy bands which were combined with an RGB image. From EPIC-pn we used only single- and double-pixel events (\texttt{PATTERN $\leq$ 4}), and from EPIC-MOS all single- to quadruple-pixel events (\texttt{PATTERN $\leq$ 12}). The X-ray properties of the detected \ac{SNRs} as revealed by our {\it XMM-Newton} mosaic will be discussed in future papers including (Kavanagh~et~al.~2020).

\subsection{Magellanic Clouds Photometric Survey}
 \label{sec:MCPS}
It is very important to understand the type of stellar environment the progenitor of these objects came from. Because of that, we make use of data from the Magellanic Clouds Photometric Survey \citep[MCPS;][]{2004AJ....128.1606Z} in order to construct colour-magnitude diagrams. This way we identify blue stars more massive than $\sim$8\,M$_{\sun}$ within a 100~pc (6.9\arcmin) radius of objects selected in this study. This allows us to see the prevalence of early-type stars close to the remnant candidates.

\section{Results}
\label{sec:results}

\begin{figure*}
  \begin{center}
\resizebox{1\linewidth}{!}{\includegraphics[trim = 90 0 130 0,clip]{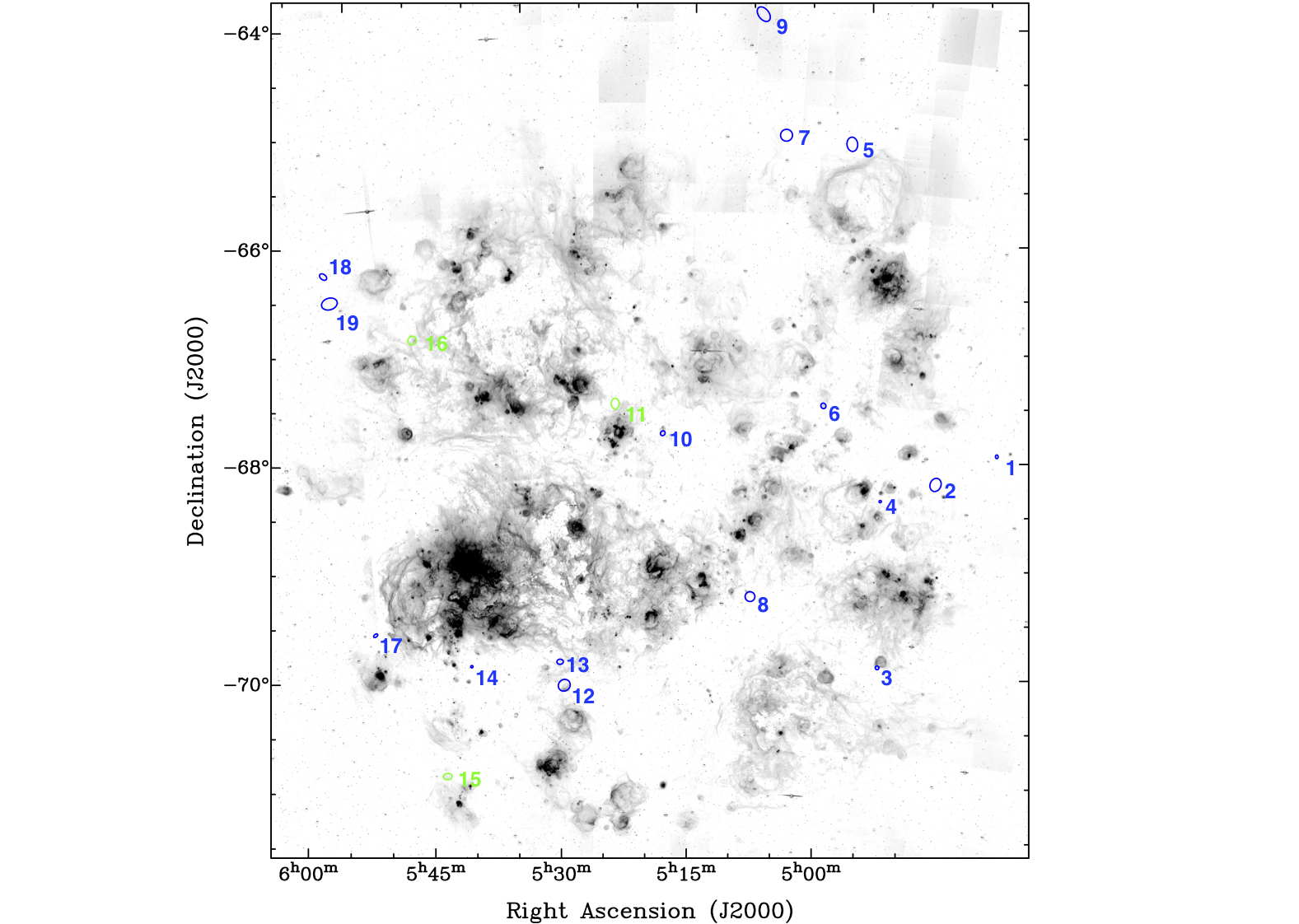}}
   \caption{The positions of 16 new \ac{SNR} candidates in the \ac{LMC} are marked in blue. Previously classified X-ray \ac{SNR} --- MCSNR~J0541--6659 and here classified MCSNR~0522--6740 and MCSNR~J0542--7104 are marked in green colour with corresponding numbers as in Table~\ref{tab:snrtable} (Col.~1). Background (grey scale) image is \ac{MCELS} \Ha.}
   \label{fig:1new}
  \end{center}
 \end{figure*}

From careful examination of the \ac{MCELS} images, we select 19 \ac{LMC} objects and classify them as potential new \ac{SNR} candidates. We base this classification on their morphological characteristics as well as their \SII/\Ha\ ratios (see Table~\ref{tab:snrtable} and Figure~\ref{fig:1new} for their positions across the \ac{LMC} field). We include in our sample MCSNR~J0541--6659, which was previously classified as an \ac{SNR} by \citet{2012A&A...539A..15G} using X-ray observations (Figure~\ref{fig:A17}) but never confirmed spectroscopically.

\begin{table*}
	\centering
	\caption{The main characteristics of the 19 \ac{LMC} objects in this study.
	}
	\begin{tabular}{ccccccccccc} 
		\hline
	No. & Obs. & Name    & RA (J2000) & DEC (J2000)             & D$_{maj} \times$ D$_{min}$ & D$_{av}$ & PA   & \ac{MCELS} & OB stars & Figure  \\
		& Date &         & (h m s)    &  (\D\ \arcmin\ \arcsec) & (\arcsec)                  & (pc)     & (\D) & \SII/\Ha   & No.      & No.     \\
		(1) & (2) & (3) & (4) & (5) & (6) & (7) & (8) & (9) & (10) & (11)  \\
		\hline
	\noalign{\smallskip}
1  & 22/10/2017& 0444--6758 & 04 44 27.8 & --67 58 13.1 & $124\times92$  & 25.9  & 80  & 0.7 &  1/0  & \ref{fig:A1}  \\
2  & 23/10/2017& 0450--6818 & 04 50 12.4 & --68 18 04.8 & $454\times366$ & 98.8  & 105 & 0.8 &  1/1  & \ref{fig:A2}  \\
3  & not obs.  & 0454--7003 & 04 54 19.1 & --70 03 30.5 & $129\times121$ & 30.3  & 80  & 0.8 &  18/0 & \ref{fig:A3}  \\
4  & 21/10/2017& 0455--6830 & 04 55 36.8 & --68 30 34.6 & $80\times78$   & 19.1  & 170 & 0.9 &  18/0 & \ref{fig:A4}  \\
5  & 22/10/2017& 0500--6512 & 05 00 58.5 & --65 12 15.3 & $480\times360$ & 100.8 & 80  & 0.8 &  0/0  & \ref{fig:A5}  \\
6  & 22/10/2017& 0502--6739 & 05 02 02.5 & --67 39 31.3 & $190\times168$ & 43.3  & 60  & 0.6 &  18/2 & \ref{fig:A6}  \\
7  & not obs.  & 0506--6509 & 05 06 51.7 & --65 09 16.6 & $402\times398$ & 97.0  & 170 & 0.6 &  ---  & \ref{fig:A7}  \\
8  & 22/10/2017& 0508--6928 & 05 08 31.8 & --69 28 29.8 & $330\times330$ & 80.0  & 0   & 0.7 &  7/2  & \ref{fig:A8}  \\
9  & 22/10/2017& 0509--6402 & 05 09 16.1 & --64 02 11.3 & $556\times350$ & 106.9 & 50  & 0.7 &  ---  & \ref{fig:A9} \\
10 & 23/10/2017& 0517--6757 & 05 17 51.1 & --67 57 46.0 & $170\times150$ & 38.7  & 130 & 0.7 &  17/0 & \ref{fig:A11} \\
11 & not obs.  & 0522--6740 & 05 22 32.4 & --67 40 56.0 & $360\times270$ & 75.6  & 90  & 1.0 &  7/1  & \ref{fig:A12} \\
12 & not obs.  & 0528--7017 & 05 28 46.0 & --70 17 56.8 & $416\times380$ & 96.4  & 140 & 0.9 &  13/4 & \ref{fig:A13} \\
13 & not obs.  & 0529--7004 & 05 29 05.9 & --70 04 40.6 & $216\times174$ & 47.0  & 0   & 1.0 &  12/0 & \ref{fig:A14}\\
14 & not obs.  & 0538--7004 & 05 38 47.2 & --70 04 15.8 & $80\times80$   & 19.4  & 0   & 0.8 &  2/0  & \ref{fig:A15} \\
15 & 23/10/2017& 0541--6659 & 05 41 51.5 & --66 59 02.8 & $300\times272$ & 69.2  & 0   & 0.5 &  18/2 & \ref{fig:A16} \\
16 & 23/10/2017& 0542--7104 & 05 42 37.9 & --71 04 13.6 & $300\times210$ & 60.8  & 0   & 0.7 &  1/0  & \ref{fig:A18} \\
17 & 07/09/2019& 0548--6941 & 05 48 49.1 & --69 41 18.3 & $156\times95$  & 29.5  & 150 & 0.6 &  4/0  & \ref{fig:A19} \\
18 & not obs.  & 0549--6618 & 05 49 30.1 & --66 18 15.9 & $270\times180$ & 53.4  & 45  & 1.0 &  0/0  & \ref{fig:A20} \\
19 & 23/10/2017& 0549--6633 & 05 49 14.5 & --66 33 43.1 & $540\times390$ & 111.2 & 170 & 1.4 &  0/0  & \ref{fig:A21} \\
	\noalign{\smallskip}
		\hline
	\end{tabular}
 \label{tab:snrtable}
\end{table*} 

Figures~\ref{fig:A1}, \ref{fig:A3}, \ref{fig:A4}, \ref{fig:A5}, \ref{fig:A6}, \ref{fig:A8}, \ref{fig:A9}, \ref{fig:A11}, \ref{fig:A16}, \ref{fig:A18}, \ref{fig:A19} and \ref{fig:A21} show the 12 spectroscopically studied \ac{LMC} objects with their associated WiFeS spectra and the \ac{MCELS} images (namely: \Ha, \SII\, \OIII\ and the ratio \SII/\Ha). 

In Figures~\ref{fig:A2}, \ref{fig:A7}, \ref{fig:A12}, \ref{fig:A13}, \ref{fig:A14}, \ref{fig:A15} and \ref{fig:A20} we show seven additional \ac{MCELS} objects, which we propose as new \ac{LMC} \ac{SNRs} or \ac{SNR} candidates. However, we have not yet obtained any spectroscopic follow-up observations of these objects. Three of these seven objects are detected in X-rays which additionally support their classification as \ac{SNRs}. Because of the lack of spectroscopic confirmation as well as the lack of confirmed detections at other frequencies, we assign a lower confidence classification (as \ac{SNR} candidates) for the four remaining sources in this sample of seven objects.

\subsection{Optical Identification of Newly Selected \ac{LMC} objects}
 \label{sec:3.1}

It is well established that the most reliable \ac{SNR} diagnostic of optical spectral observations is the ratio of the \SII/\Ha\ lines. This confirms the presence of radiative shock(s) and a ratio of $>$0.4 suggests an \ac{SNR} origin. However, we note that \SII/\Ha\ can also reach 0.4 (and higher) in ionisation fronts as well as in diffuse gas ionised by a diluted UV flux. A good example is NGC\,7793 where the optical search for \ac{SNRs} was severely complicated by extensive diffuse \SII/\Ha\ photoionised gas throughout the galaxy and rises well above the value of 0.4 \citep{1997ApJS..108..261B}. The \SII/\Ha\ ratio alone is thus not sufficient to robustly confirm an \ac{SNR} classification as X-ray and/or radio confirmation is also needed. All discrete objects in Table~\ref{tab:snrflux} satisfy these optical criteria indicating their possible \ac{SNR} nature. The shell structure of the observed sample, clearly seen on \ac{MCELS} images in \Ha, \SII\ and \OIII\ light, also supports the fact that these objects are excellent \ac{SNR} candidates. 

Additionally, the existence of \OIII\ emission that is localised to a thin shell is another indicator that the source is indeed of \ac{SNR} origin as it traces out the radiative cooling zone. But if it is extended, especially in the inner part of the nebula, it is more typical of a high (photo-)ionisation nebula, and argues against an \ac{SNR} interpretation.

\begin{table*}
	\centering
	\caption{
	The relative fluxes and most important line ratios (in the red arm) of observed emission lines for 12 \ac{LMC} objects. To extract spectra from WiFeS cube we used apertures between 5 and 10 arcsec. LDL (Column~15) stands for ``low density limit''. Also, J0548--6941 is the only source with physical (calibrated) flux values for which the spectroscopic standard star LTT1020 was used. All other objects are with relative flux (counts). All fluxes for J0548--6941 in units of erg~cm$^{-2}$~s$^{-1}$\AA$^{-1}$. Errors in flux estimates for blue end of the spectra are in order of 17--19~per~cent, and for red arm observations 23--25~per~cent what can be expected for lower signal-to-noise for the higher resolution spectra, while dispersion errors are as in \citet{Dopita2007,2010Ap&SS.327..245D,2014Ap&SS...349..617C}. \NII\ and \SII\ used in Columns~13 and 14 for the ratios relative to H$\alpha$ are the sum of the doublet lines. The \SII 6548\AA\ line flux is derived from that of the 6563\AA\ line (see Sect.\,\ref{wifes}).	}
	\label{tab:snrflux}
	{
	\setlength\tabcolsep{1pt}
	\begin{tabular}{ccccccccccccccc}
		\hline
		Object & \Hd\ & \Hy\ & \Hb\ & \FeII\ & \OIII\ & \OI\ & \Ha\ & \NII\ & \SII\ & \SII\ & \SII/\SII\ & \NII/\Ha\ & \SII/\Ha & Elect. den.\\
	    Name & 4101\AA & 4342\AA & 4861\AA & 4890\AA & 5007\AA & 6300\AA & 6563\AA & 6583\AA & 6716\AA & 6731\AA & 6716\AA/6731\AA&&&(cm$^{-3}$) \\
		(1) & (2) & (3) & (4) & (5) & (6) & (7) & (8) & (9) & (10) & (11) & (12) & (13) & (14) & (15)  \\
		\hline\noalign{\smallskip}
		J0444--6758~~ &      &      &  83  & & 1085 &     & 1103  & 277  & 446  & 286  & 1.5 & 0.33 & 0.67 & LDL\\
		J0450--6818~~ &      &      &      & &      & 302 & 1101  & 212  & 523  & 364  & 1.4 & 0.26& 0.80 & 26.5 \\
		J0455--6830~~ &      &      & 100  & & 580  &     & 1824  & 624  & 1192 & 838  & 1.4 & 0.46& 1.10 & 26.5 \\
		J0500--6512~~ &      &      &      & &      &     & 1111  & 152  & 540  & 359  & 1.5 & 0.18 & 0.81 & LDL \\
		J0502--6739~~ &      & 588  & 1563 & &      & 180 & 5196  & 536  & 1692 & 1180 & 1.4 & 0.14 & 0.55 & 26.5 \\
		J0508--6928~~ & 631  & 1270 & 4260 & &      & 533 & 22952 & 5073 & 9906 & 6628 & 1.5 & 0.29 & 0.72 & LDL\\
	    J0509--6402~~ &      &      &      & &      & 280 & 1107  & 224  & 461  & 373  & 1.2 & 0.27 & 0.75 & 29.8 \\
		J0517--6757~~ &      & 82   & 1420 & & 2078 & 488 & 6284  & 1531 & 2447 & 1871 & 1.3 & 0.32 & 0.69 & 26.5\\
		J0541--6659~~ & 1281 & 3805 & 12957& & 5944 &     & 19692 & 3045 & 5672 & 4120 & 1.4 & 0.21 & 0.50 &26.5 \\
		J0542--7104~~ &      &      & 30   & &  101 & 331 & 3688  & 113  & 1835 & 1198 & 1.5 & 0.04 & 0.82 & LDL \\
        J0548--6941~~ & \scriptsize{2.5$\times$10$^{-16}$} & & \scriptsize{1.1$\times$10$^{-15}$~~} & \scriptsize{3.1$\times$10$^{-16}$~~} &  \scriptsize{1.5$\times$10$^{-15}$} & & \scriptsize{1.3$\times$10$^{-15}$~~} & \scriptsize{2.5$\times$10$^{-16}$~~} &  \scriptsize{3.7$\times10^{-16}$~~} & \scriptsize{2.4$\times$10$^{-16}$} & 1.5 & 0.26 & 0.47 & LDL\\
        J0549--6633~~ &      &      &      & &      & 731 & 1989  & 400  & 1618 & 1129 &1.4& 0.27 &1.38&26.5\\
		\hline
		\end{tabular}
	}
\end{table*}

\begin{figure}
 \begin{center}
  \resizebox{0.99\linewidth}{!}{\includegraphics[trim = 0 0 0 0]{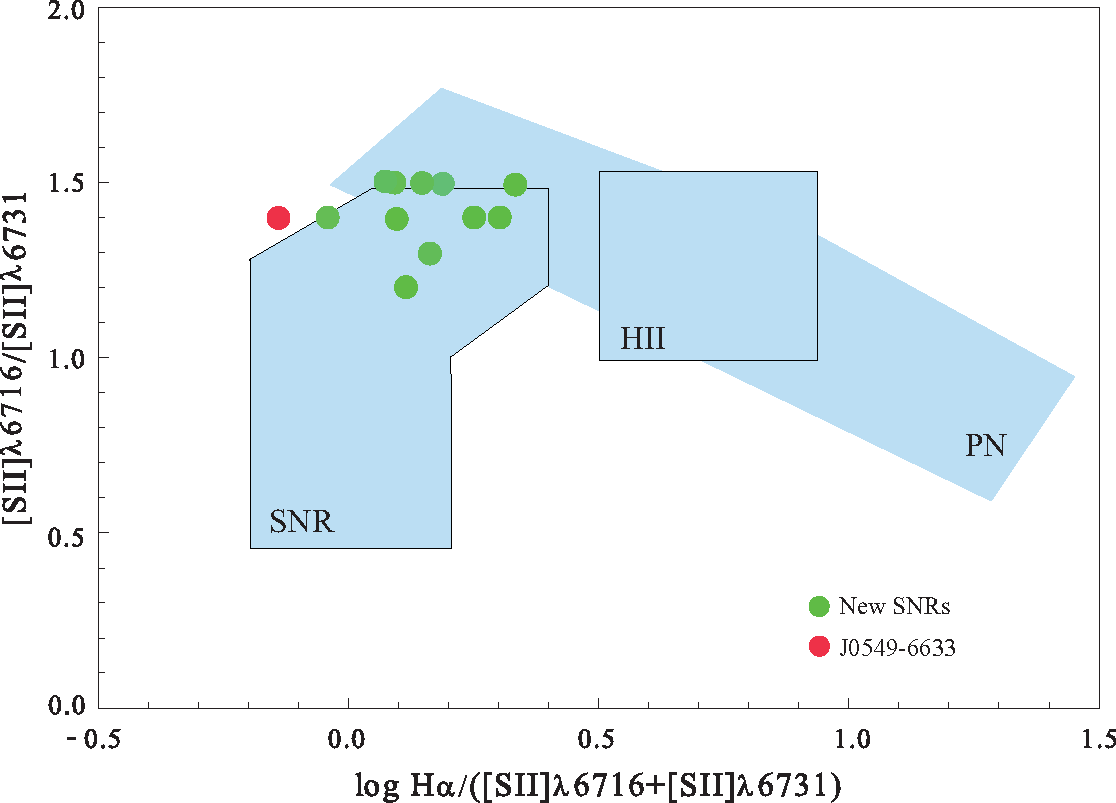}}
  \caption{Ratio of \SII$\lambda$6716 and \SII$\lambda$6731 lines versus log~\Ha/\SII$\lambda$6716+\SII$\lambda$6731 for the sample of our 12 spectroscopically observed \ac{LMC} objects presented in this paper. 11 objects from Table~\ref{tab:snrflux} fit in the designated area of \ac{SNRs} while the object marked as a red dot (J0549--6633; Section~\ref{sec:0549-6633}) is slightly out. Noticeably, all these objects occupy the area suggested for older remnants as predicted by \citet{1977A&A...60.147}.}
  \label{fig:2new}
 \end{center}
\end{figure}

In Figure~\ref{fig:2new} we show the ratio of the \SII$\lambda$6716 and \SII$\lambda$6731 lines versus log~(\Ha/(\SII$\lambda$6716+\SII$\lambda$6731)) for the sample of our 12 spectroscopically studied \ac{LMC} objects (following \citet{1977A&A...60.147}). This diagram maps the locations in this graph that would be occupied by \HII~regions, \ac{SNRs} and \ac{PNe}. This diagram is based on Galactic \ac{SNRs}, \HII~regions and \ac{PNe} and is representative of nearby galaxies such as the \ac{LMC}.

Using our high resolution spectroscopic observations, we find one of the initially selected \ac{SNR} candidates having an elevated \SII/\Ha\ ratio which is just out of the range expected for \ac{SNRs} (J0549--6633; marked in red in Figure~\ref{fig:2new}). We still classify this object, though, as a somewhat more likely \ac{SNR} candidate then a (super)bubble (see Appendix~\ref{sec:0549-6633}). The rest of our sample nicely fits in the group of older \ac{SNRs} because of the higher ratio of sulphur lines -- just as it is predicted by \citet{1977A&A...60.147}. This is in agreement with the findings of \citet{2018ApJ...855..140L} for M\,33 \ac{SNRs} where some very large shells with elevated \SII\ emission were seen. Recently, \citet{2020MNRAS.tmp.2593F} also found a new large optically visible but radio and X-ray faint Galactic \ac{SNR}. 

\citet{1977A&A...60.147} also use a diagram of \Ha/\NII\ vs. \Ha/\SII\ (in log scale, their Fig.~2). \HII~regions have smaller \NII/\Ha\ ratios than \ac{SNR} which adds another tool to separate these two group of sources. A very low value for our candidates J0502--6739 (looks more like \HII\ region) and J0542--7104 casts some doubts on their classification as \ac{SNR} candidates from the spectroscopic point of view. Certainly, \citet{1977A&A...60.147} estimates were for \ac{MW} objects, and as we mentioned earlier (see Section~\ref{sec:intro}) the \ac{LMC} has nitrogen deficiency. 

In Table~\ref{tab:snrflux} we also show the ratio of \NII/\Ha\ lines with a median (for this sample) of $\sim$0.27 (SD=0.10). This is a somewhat lower than expected ratio compared to Galactic \ac{SNR}s, but it is acceptable because of the overall lower nitrogen abundance of the \ac{LMC} compared to the \ac{MW} (see also \citealt{2015MNRAS.454..991R}). Depending on the local abundance in the Galaxy (see example in \citealt{2018MNRAS...479.4432}), this ratio for \ac{SNRs} could vary by as much as a factor of $\sim$2 (or even more). In the blue part of the spectrum, we detected the oxygen line \OIII$\lambda$5007 in six \ac{SNR} candidates (MCSNR~J0541--6659, J0444--6758, J0455--6830, J0517--6757, MCSNR~J0542--7104 and J0548--6941). The presence of this line suggests that the shock velocity is between 80~\kms\ and 140~\kms\ because a speed of over 140~\kms\ would produce more \Hb\ emission than the oxygen line which becomes relatively weak. However, spectra of these six putative remnants (and their relative intensities as listed in Table~\ref{tab:snrflux}) show a rather strong \OIII$\lambda$5007~line. The exception is again MCSNR~J0541--6659 where \Hb\ is much stronger than the usually strong \OIII$\lambda$5007~line. The last column in Table~\ref{tab:snrflux} presents electron densities for the given sample calculated on the basis of the \SII$\lambda$6716/\SII$\lambda$6731 line ratios with an assumed electron temperature of 10\,000\,K. Assessment of values in this column suggests an overall older \ac{SNR} population including five objects that are at a low electron density (e.g. out of the electron density function; marked with LDL (low density limit) in the last column of Table~\ref{tab:snrflux}).

While these spectroscopic results suggest that our newly selected objects are \ac{SNRs}, we cannot take that as the final classification. Apart from a confirmation at other frequencies \citep{1998A&AS..130..421F,2017ApJS..230....2B}, a question arises from the fact that the shell structure and the ratio of sulphur lines with \Ha\ (e.g. the shock) are not seen only in \ac{SNRs}. For example, a \ac{WR} star nebula is also a source of nebulously where shocks are observed as they are released from a \ac{WR} star. However, the sizes of \ac{WR} shells are certainly smaller compared to \ac{SNRs}. For our sample of sources studied here, the indication of a possible \ac{WR} star nebula can be also rejected as none of our sources can be found in the recent comprehensive survey of \ac{WR} stars in \ac{LMC} \citep{2018Apj...863..181N}.

\citet{1977ApJ...212..390L} demonstrated that some \ac{LMC} shell nebulae would have \SII\ (6716, 6731\AA) strengths of the same order as \Ha\ as well as expansion velocities of $\sim$30~\kms. He argued that the ionisation is radiative and that the additional heating required to explain the strong \SII\ is furnished by shocks. In summary, it is challenging to determine whether these shocks are driven by \ac{SN} blasts or by stellar winds. A number of examples can be seen across the \ac{LMC} in objects such as N\,185, N\,186 and N\,70 \citep{2002AJ....123..255O,2014AJ....147..162D}. However, some of the objects \citep[for example N\,9;][]{2017ApJS..230....2B} from the \citet{1977ApJ...212..390L} sample were later confirmed as a true \ac{SNRs}. Essentially, if any of these large shells are driven by stellar winds, one would expect to find some massive (OB) stars in the vicinity of these objects to create those stellar winds. This would be a most straightforward way to distinguish one from the other.

Similar to \ac{SNRs} and \ac{WR} star nebula, (super)bubbles also show shell structure as well as shocked radiative spectra. But in most cases, inside the shell, the material is far less dense (particularly seen in \Ha\ light) and propelled by the fast stellar winds because of the loss of energy from a massive star (see \citealt{2008IAUS..250..341C}). (Super)bubbles are also usually associated with a massive stellar population. \ac{MCELS} images of the whole sample (except MCSNR~J0542--7104; see later discussion) do not show any obvious dense material inside the shells. Thus, for our sample of objects, a classification as a (super)bubble is somewhat unlikely. One, though, should always consider a possible confusion with other nearby objects such as \HII\ regions \citep{2007MNRAS.378.1237B}. 

As radio-continuum objects, \ac{SNRs} are well known by their non thermal emission seen at different radio frequencies. Our search for radio emission from the new optically selected sample of \ac{SNR} candidates failed to detect any such emission. We searched all present-day radio surveys including the latest and most sensitive \ac{ASKAP} survey of the \ac{LMC} at 888~MHz (Filipovi\'c et al., in prep). Existing X-ray surveys of the \ac{LMC} such as from the {\it R\"Ontgen SATellite (ROSAT)} \citep{1998A&AS..127..119F} and {\it XMM-Newton} \citep{2016A&A...585A.162M}, with their noticeably limited area coverage compared to radio and optical data, likewise did not show any signs of \ac{SNR}-like emission except for MCSNR~J0542--7104 ([HP99]~1235; Section~\ref{sec:0542-7104}) as reported in Kavanagh~et~al.~(2020; in~prep.), J0454--7003 (Section~\ref{sec:J0454--7003}), J0529--7004 ([HP]~1077; Section~\ref{sec:J0529--7004}) and MCSR~J0522--6740 (Section~\ref{sec:J0522--6740}). 

We did not find any infrared emission from any object in our sample \citep[][also see;]{2015ApJ...799...50L}. \ac{SNRs} are often connected with infrared emission from the time of a \ac{SN} explosion where it is suggested that the dust grains are formed within the expanding and cooling ejecta (see more details in \citealt{2017hsn..book.2105W,2015ApJ...799...50L}). During the later \ac{SNR} phase, the source of infrared emission is shock-heated dust. Usually younger \ac{SNRs} emit in mid-infrared wavelengths even though some recent observations of the \ac{SMC} remnant and \ac{PWN} DEM\,S5 suggest otherwise \citep{2019MNRAS.486.2507A}. In older \ac{SNRs}, which is the most likely case for our sample studied here, the dominant radiation is optical (see our spectra) and UV radiation where cooled gas in a post-shock environment with low radiative shock speed (about 100~km~s$^{-1}$) is observed. This cooler emission is in the far infrared and could explain why we saw no infrared emission in our sample of evolved remnants.

From these 19 \ac{SNR} candidates, MCSNR~J0522--6740, MCSNR~J0541-6659 and MCSNR~J0542--7104 are confirmed \ac{SNRs} (Appendix~\ref{sec:J0522--6740}, \ref{sec:0541-6659} and \ref{sec:0542-7104}). The remaining 16 objects studied here are all excellent new \ac{SNR} candidates.

\section{Discussion}
\label{sec:discussion}

We found that our new \ac{LMC} \ac{SNRs} and \ac{SNR} candidates are larger in comparison to other confirmed \ac{LMC} \ac{SNRs} (see Table~\ref{tab:snrtable}) but only after excluding three smaller size objects: J0444--6758, J0455--6830 and J0538-7004. Those three were most likely previously undetected \ac{SNRs} due to the confusion with larger \HII\ regions. 

\begin{figure}
 \begin{center}
 \resizebox{1\linewidth}{!}{\includegraphics[trim = 0 15 20 20,clip]{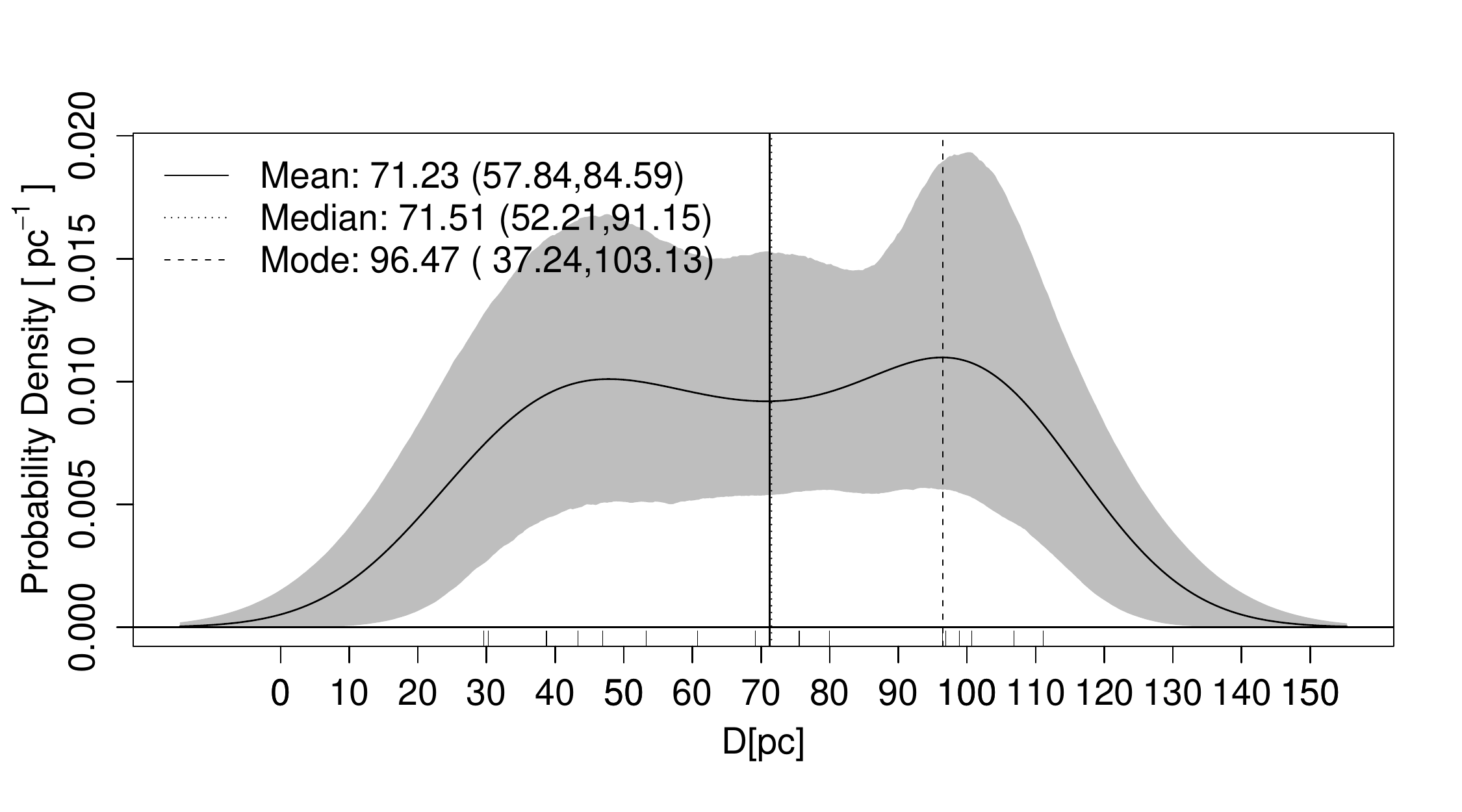}}
  \caption{Smoothed diameter distribution for the samples of 16 selected and here studied \ac{LMC} \ac{SNRs}. The smoothing procedure and 95~per~cent bootstrap confidence intervals are calculated as described in \citet{2019A&A...631A.127M}. The grey lines present a 95~per~cent confidence interval around the smoothed distribution. The obtained value for the} smoothing bandwidth $h$ is also designated on the plot.
  \label{fig:3}
 \end{center}
\end{figure}

To calculate the physical diameter distribution for these 16 \ac{SNRs} and \ac{SNR} candidates the kernel smoothing procedure described in \citet{2019A&A...631A.127M} is applied (Figure~\ref{fig:3}). For comparison, the same procedure is also applied to the sample of 59 \ac{SNRs} from \citet[][Fig.~8]{2017ApJS..230....2B}. It is evident that the sample diameters from this work have values almost two times larger (with correspondingly higher uncertainties) for the stated distribution parameters, while the smoothing bandwidth differs by about 15~per~cent. We find that the average size of the 16 \ac{SNRs} and \ac{SNR} candidates to be 71$\pm$14~pc (SD=27). \citet{2017ApJS..230....2B} found that the \ac{LMC} \ac{SNR} population exhibits a mean diameter of 39$\pm$4~pc for the earlier confirmed 59 \ac{LMC} \ac{SNR}. Our discovery of these 16 large \ac{LMC} \ac{SNR} candidates nicely agree with the \citet{2017ApJS..230....2B} prediction that the present sample with sizes D$>$40~pc is incomplete, leaving room for a future detection of mainly large (and older) \ac{LMC} \ac{SNRs} (such as these 16). We also note that our new \ac{SNR} size distribution is much closer to the \ac{SMC} sample \citep{2019A&A...631A.127M} which is smaller in number but noticeably more complete. The sample from this work shows indications of the secondary distribution peak at $\approx45$~pc, which corresponds to the peak of the distribution for the sample from \citet{2017ApJS..230....2B}.

As mentioned earlier, \citet{2020MNRAS.tmp.2593F} found a large, new optically visible but radio and X-ray faint Galactic \ac{SNR} -- G107.0+9.0 with an estimated size of $\sim$75 -- 100~pc, an advanced age of $\sim$100\,000~yrs and well above the Galactic plane (250--300~pc). However, most of its large shell is not as high in \SII/\Ha\ ratio as the objects studied here. Still, one selected location where \citet{2020MNRAS.tmp.2593F} obtained a spectrum does contain an elevated ratio, apparently from a relatively slow but radiative shock in an isolated emission knot. This provokes an interesting question: what would G107.0+9.0 look like if it were at the distance of the \ac{LMC}? To start with, if it is anywhere close to the 30~Doradus region it would most likely be indistinguishable from the local environment which is crowded with bright \HII\ region filaments. Or at best, we would see something similar to J0528--7017 which we study here. But, if G107.0+9.0 is positioned well outside the \ac{LMC} we would most likely see it to be similar to our J0450--6818, J0500--6512, J0506--6509, J0508--6928, J0522--6740, MCSNR J0542--7104 or J0549--6618.

As in \citet{2017ApJS..230....2B}, we investigate the spatial distribution and local environment of these 19 new \ac{LMC} \ac{SNRs} and \ac{SNR} candidates in relation to the \HI\ peak temperature map from \citet{1998ApJ...503..674K}. Most of these new \ac{SNR} candidates (15 out of 19) are positioned near or around the edges of the \ac{LMC}'s main body, in obviously less dense regions. 

As argued by \citet{2018ApJ...855..140L}, the majority of the objects with high \SII/\Ha\ line ratios ($>$0.4) are indeed \ac{SNRs}, but the distinction between \HII\ regions and \ac{SNRs} becomes far less obvious at low surface brightness, and additional criteria, such as X-ray or radio continuum detection, are needed. We also note that these outer regions of the \ac{LMC} lack good X-ray coverage, and therefore confirmation of the true nature of these objects cannot yet be made. The recent {\it XMM-Newton} detection of MCSNR~J0542--7104 (Kavanagh~et~al.~2020; in prep) and MCSNR~J0522--6740 which is classified here, show that the \ac{SNR} candidacy of our selected object sample is solid. Further, we report here the detection of diffuse soft X-ray emission from three of our objects with {\it XMM-Newton} (J0454--7003, MCSNR~J0522--6740 and J0529--7004). We note that in all three cases the X-ray emission is more centrally peaked. However, the short exposures and large off-axis angles of the observations do not allow a more detailed analysis of their faint X-ray emission. Despite the lack of deeper coverage we manage to classify one of these three sources (MCSNR~J0522--6740; see Appendix~\ref{sec:J0522--6740}) as a new bona-fide \ac{SNR}.

As discussed by \citet{2018SSRv..214...44L}, the environment and the surrounding medium influences \ac{SNR} morphology. More specifically, the progenitors of \ac{CC} \ac{SNe} have short main-sequence lives (3 -- 50~Myr) and their explosions are expected to occur within or near the dense media from which the massive stars formed. The mean \HI\ column density in the direction of the 59 confirmed and 15 candidate remnants, from \citet{2017ApJS..230....2B}, was estimated to be $\sim2\times10^{21}$~atoms~cm$^{-2}$ (with a SD=1$\times$10$^{21}$~atoms~cm$^{-2}$). Our sub-sample of 15 new \ac{SNR} objects (excluding three smaller \ac{SNR} candidates and bona-fide \ac{SNR} MCSNR~J0541--6659 as they are possibly embedded in \HII\ regions) show a marginally lower environmental density of \mbox{$\sim1.8\times10^{21}$~atoms~cm$^{-2}$} (SD=1.3$\times$10$^{21}$~atoms~cm$^{-2}$). Therefore, we find no significant difference in the \HI\ environment in which our new sub-sample of the \ac{LMC} \ac{SNR} candidates resides and evolves. 

\citet{2016A&A...585A.162M} argue that the relatively large population of the \ac{LMC} \ac{SNRs} could be a product of type~Ia \ac{SNe}, as we also argued for the distant candidate J0509--6402 studied here and new bona-fide \ac{SNR} MCSNR~J0542--7104, although it should be considered with caution. We emphasise that the \HI\ environment around \ac{SNR} candidates studied here is not vastly different from the previously established \ac{LMC} \ac{SNRs} sample as shown above. But the low number of massive (OB) stars within the boundaries of the \ac{SNR} candidates may favour an \ac{SN} type~Ia origin. However, this should be taken only as indirect evidence.

The larger physical size of our \ac{SNR} candidates sample could indicate their possibly older age and more evolved phase as also observed for a number of Galactic \ac{SNRs} by \citet{2008MNRAS.390.1037S}, \citet{2019MNRAS.486.4701F} and recently \citet{2020MNRAS.tmp.2593F}. However, one should be very careful comparing our results from the \ac{MCs} sample with \ac{MW} \ac{SNRs} because close to the Galactic plane the absorption is sufficient to absorb most X-ray emission for any but the young \ac{SNRs} (emitting above 1 or 2~keV). When comparing with other waveband surveys such as X-ray \textit{Chandra} or \textit{XMM-Newton} and \ac{ASKAP}/\ac{ATCA} radio images, we find that these new \ac{LMC} \ac{SNR} candidates are seen almost exclusively in the optical wavelengths. However, the exception are three confirmed \ac{SNRs} (MCSNR~J0522--6740, MCSNR~J0541--6659 and MCSNR~J0542--7104) as well as two candidates (J0454--7003 and J0529--7004) which can be detected in the X-ray surveys. This could potentially imply that we are discovering a previously unknown class of large and predominantly optically visible \ac{LMC} \ac{SNRs}. We suggest that these \ac{SNRs} are mainly residing in a very rarefied environment and are likely relatively old ($>$20~kyr). This would make them less visible to the present generation of radio and X-ray telescopes. 

Evidently, larger (and older) remnants seen mostly in optical wavebands are radio and X-ray quiet since they are in the last dissipation phase and they almost blend with the \ac{ISM}. Their emission in those domains are likely to cease because of radiative cooling and the decrease of their strong magnetic fields to the level of the galactic background. \ac{SNR} non-thermal emission ends because of this dissipation and therefore is not detectable \citep{2008MNRAS.390.1037S}. 

\ac{SNRs} in free expansion and Sedov phases of evolution (defined by the non-radiative shocks) can not be easily detected by the \SII/\Ha\ method. In these phases, \SII\ emission of \ac{SNR} is at best small because they are usually mixed within the local \ac{ISM} (or \HII\ regions). Even for the nearby galaxies such as the \ac{LMC}, 
this causes a selection effect that adversely affects the detection of objects such as \ac{SNRs}. In the early phases of their evolution, \ac{SNRs} are mainly detected by radio and X-ray observations and (for SN type~Ia) in optical bands only by Balmer lines \citep{Lin_2020}. \HII\ regions are also detected by Balmer lines and therefore to rely only on optical detection will not be sufficient especially for \ac{SNRs} in the free expansion and Sedov phases. In the radiative phases of evolution, \ac{SNRs} are usually not easily detected in radio and X-rays if they are distant and evolve in a low density environment. In general, this is one of the major detection challenges for distant extragalactic \ac{SNR} samples where the instrumental sensitivity selection effects dominates the construction of a complete \ac{SNR} sample.

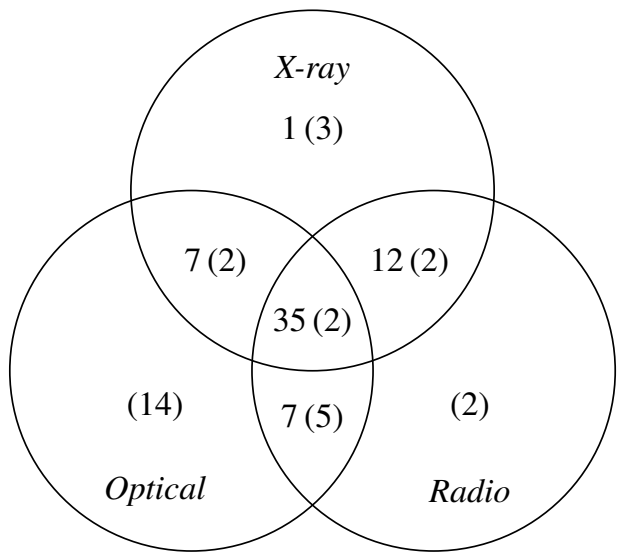
\begin{figure}
\centering
\def\firstcircle{(0,0) circle (1.5cm)}
\def\secondcircle{(1,1.5) circle (1.5cm)}
\def\thirdcircle{(0:2cm) circle (1.5cm)}
\resizebox{!}{0.3\textheight}{\begin{tikzpicture}
    \begin{scope}[shift={(3cm,-4cm)}, fill opacity=0.25]
        \fill[white] \firstcircle;
        \fill[white] \secondcircle;
        \fill[white] \thirdcircle;
        \draw \firstcircle node[below] {\textit{}};
        \draw \secondcircle node [above] {\textit{}};
        \draw \thirdcircle node [below] {\textit{}};
        \node[color=black, opacity=1] at (-0.3,-1) {\textit{Optical}};
        \node[color=black, opacity=1] at (2.3,-1) {\textit{Radio}};
        \node[color=black, opacity=1] at (1,2.5) {\textit{X-ray}};
        \node[color=black, opacity=1] at (1,.4) {{35\,(2)}}; 
        \node[color=black, opacity=1] at (1,2) {{1\,(3)}};	
        \node[color=black, opacity=1] at (1.8,0.9) {{12\,(2)}}; 
        \node[color=black, opacity=1] at (0.2,0.9) {{7\,(2)}}; 
        \node[color=black, opacity=1] at (-0.3,-0.3) {{(14)}};
        \node[color=black, opacity=1] at (2.3,-0.3) {{(2)}}; 
        \node[color=black, opacity=1] at (1,-0.4) {{7\,(5)}};
    \end{scope}
\end{tikzpicture}
}
\caption{A Venn diagram showing 62 confirmed \ac{LMC} \ac{SNRs} (including three from this work). We also show (in brackets) the previously known 14 \ac{SNR} candidates from \citet{2017ApJS..230....2B} and the here proposed 16 new \ac{SNR} candidates in different electromagnetic domains.}
 \label{fig:4}
\end{figure}

We follow (and update) the \citet{2017ApJS..230....2B} comparison of multi-frequency emission from known \ac{SNRs} (60; including one from \citet{2019MNRAS.490.5494M}) and \ac{SNR} candidates (32 including 18 from this work and 14 from \citet{2017ApJS..230....2B} as \citet{2019MNRAS.490.5494M} confirmed MCSNR~J0513--6731) in the \ac{LMC}. Also, we plot a Venn diagram (Figure~\ref{fig:4}) that summarises the number of \ac{SNRs} (and candidates) exhibiting emission in different electromagnetic domains. We emphasise that the lack of detected emission does not always mean that the remnant does not emit such radiation. Alternatively, it may indicate that the emission is below the sensitivity level of current surveys. Importantly, there are examples of \ac{SNRs} such as the \ac{SMC} \ac{SNR} HFPK\,334 \citep{2014AJ....148...99C,2019MNRAS.490.1202J,2019A&A...631A.127M}, SMC\,IKT\,23, MCSNR~J0528--6713 \citep{2010A&A...518A..35C} or the Galactic Vela~Jr \ac{SNR} \citep{2001AIPC..565..267F,2005AdSpR..35.1047S,Fukui_2017,2018ApJ...866...76M} that could not be identified in optical frequencies despite extensive searches.

Various optical, radio and X-ray \ac{SNR} detection methodologies have biases as can be seen in our Venn diagrams (Figure~\ref{fig:4} and \citep{2017ApJS..230....2B}). If there are no biases one would expect that all \ac{SNRs} from one sample would converge to common intersection of all three circles of the Venn diagrams, with zero \ac{SNRs} in different observational wavebands.

\section{Conclusions}
\label{sec:conclusion}

We study previously selected MCSNR~J0541--6659 and confirm its optical \ac{SNR} signature. We also classify previously unknown MCSNR~J0522--6740 and MCSNR~J0542--7104 as new \ac{LMC} \ac{SNRs}. Finally, 16 other objects studied here for the first time are good \ac{SNR} candidates that require further studies to confirm their real nature.

In total, this work adds two new bona-fide \ac{SNRs} to the list of 60 previously confirmed \ac{SNRs} and 16 new \ac{SNR} candidates to the list of 14 previously known in the \ac{LMC} as per \citet{2017ApJS..230....2B} and \citet{2019A&A...631A.127M}, respectively. We believe the reason as to why these \ac{SNR} candidates were not detected previously is due to the fact they are mainly positioned in the outer field of the \ac{LMC} where they can only be detected because of the high sensitivity of \ac{MCELS}. This could mean we are looking at an unknown (but predicted) class of large and only (at this stage) optically visible \ac{SNRs}. The 16 new \ac{SNRs} and \ac{SNR} candidates studied here have an average size of 71~pc which is almost a factor of 2 larger (71~pc vs. 39~pc) than those in \citet{2017ApJS..230....2B}. We suggest that this sample is older and perhaps in the last evolutionary phase of their lives.

J0509--6402 is a prime candidate of type~Ia \ac{SNR} situated 2\D\ north of the \ac{LMC}, in a field where low surface brightness stellar population from the \ac{LMC} extend much further than the main (gaseous) body \citep{2018ApJ...858L..21M}. Two other candidates (J0454--7003 and J0529--7004) were found that exhibit X-ray emission but only further studies can confirm their real nature.

\section*{Acknowledgements}
The \ac{MCELS} was funded through the support of the Dean B. McLaughlin fund at the University of Michigan and through \ac{NSF} grant 9540747. M.S.~acknowledges support by the Deutsche Forschungsgemeinschaft through the Heisenberg professor grants SA 2131/5-1 and 12-1. We thank You-Hua Chu for her comments and suggestions.
B.V. was funded by the Ministry of Education, Science and Technological Development of the Republic of Serbia through the contract number 451-03-68/2020-14/200002.
Lastly, we thank the referee for their very constructive and insightful comments and suggestions.

\section*{Data Availability}
The data underlying this article will be shared on reasonable request to the corresponding author.


\bibliographystyle{mnras}
\bibliography{References} 


 \appendix

 \section{Notes on individual objects}
\label{sec:notes}

Below, we present notes on each individual source from this sample. These notes also include the number of OB stars found in the vicinity of each. As explained in Section~\ref{sec:MCPS}, the significance of the number of OB stars is related to the possible classification of a \ac{SNR} as a likely \ac{CC} or a type~Ia origin. This, together with the global (galactic) and local morphological properties, can indicate the potential of each of these individual objects to have a \ac{SN} origin \citep{2011ApJ...732..114L} .

\begin{figure*}
 \begin{center}
 \resizebox{0.47\textwidth}{!}{\includegraphics[trim = 0 0 0 0]{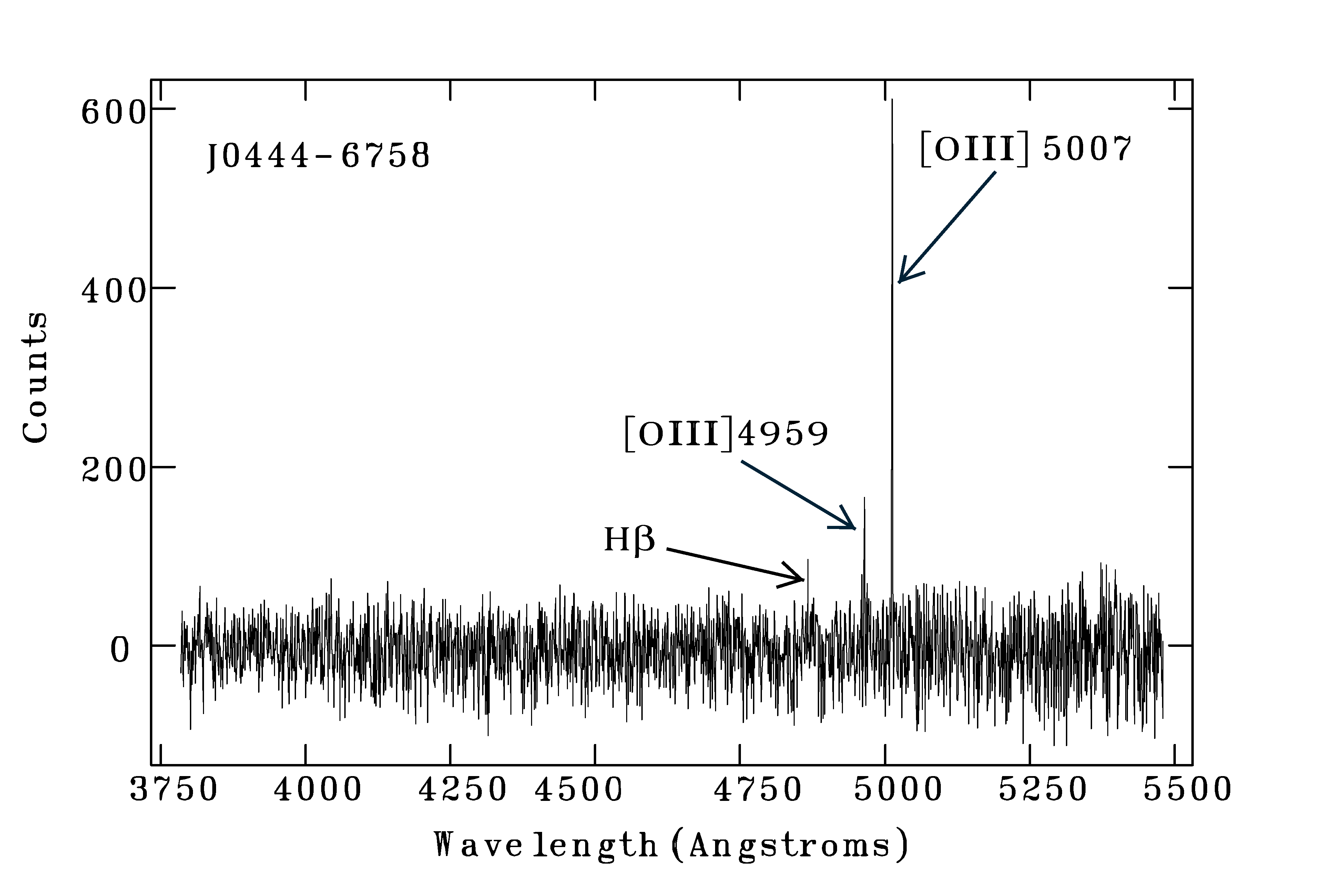}}
 \resizebox{0.47\textwidth}{!}{\includegraphics[trim = 0 0 0 0]{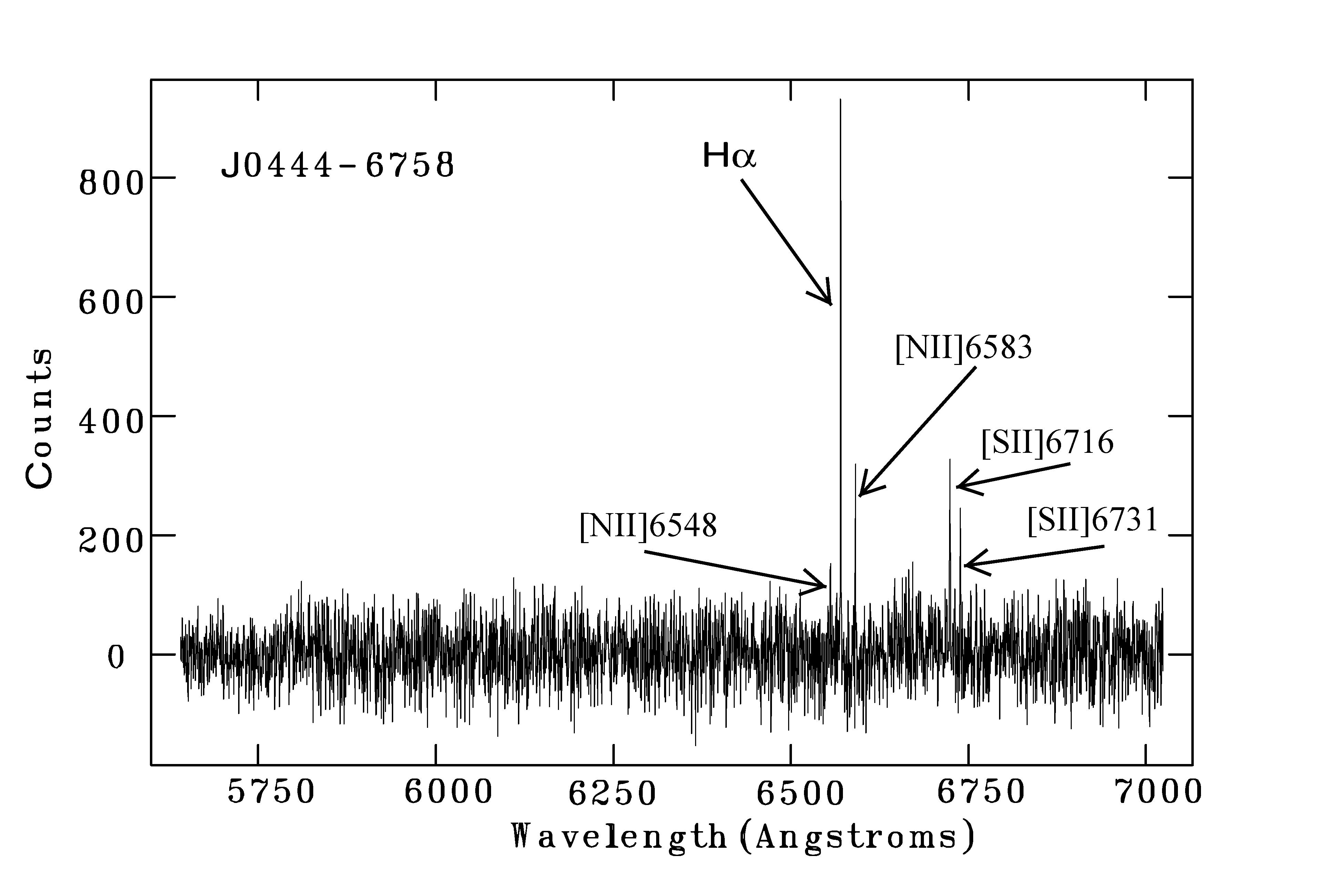}}
\resizebox{1\linewidth}{!}{\includegraphics[trim = 0 0 0 0]{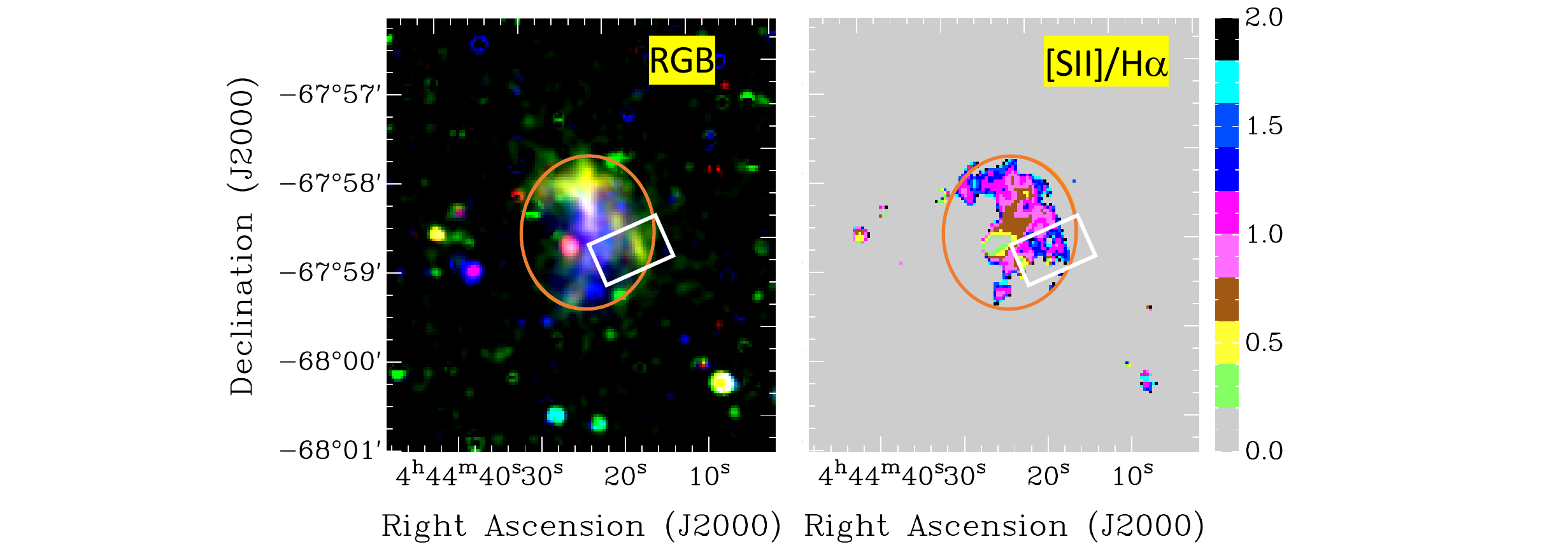}}
  \caption{J0444--6758: (Top) showing the spectra from both arms (left; blue, right; red) of the spectrograph; (Bottom) colour images produced from \ac{MCELS} data, where RGB corresponds to \Ha, \SII\ and \OIII\ while the ratio map is between \SII\ and \Ha. The rectangular box (white) represents an approximate position of the WiFeS slicer. The orange ellipse indicates the extent of the optical emission seen from the object.} 
  \label{fig:A1}
 \end{center}
\end{figure*}

\subsection{J0444--6758 (Figure~\ref{fig:A1})}
 \label{sec:J0444--6758}
This object's \Ha\ image shows patchy semi-circular diffuse emission with D=23.6~pc assuming the distance to the \ac{LMC} of 50~kpc. This remnant candidate, as can be seen in Figure~\ref{fig:A1}, is faint and it has a \SII/\Ha\ ratio of $\sim$0.67 indicating it to be likely an \ac{SNR} candidate. We searched the MCPS for the massive (OB) stars in and around this object's 100~pc boundaries and found that only one such star can be found at the far North position from the \ac{SNR} candidate. 

A bit of a caution for the classification of this object as an \ac{SNR} candidate since \OIII\ lines are quite prominent while at the same time we note an absence of the \OI\ line. The \ac{MCELS} \OIII\ image also shows emission concentrated in the centre rather than in a shell. Unfortunately, this area is presently poorly covered in X-ray surveys and further study is needed.

\begin{figure*}
 \begin{center}
 \resizebox{0.48\textwidth}{!}{\includegraphics[trim = 0 0 0 0]{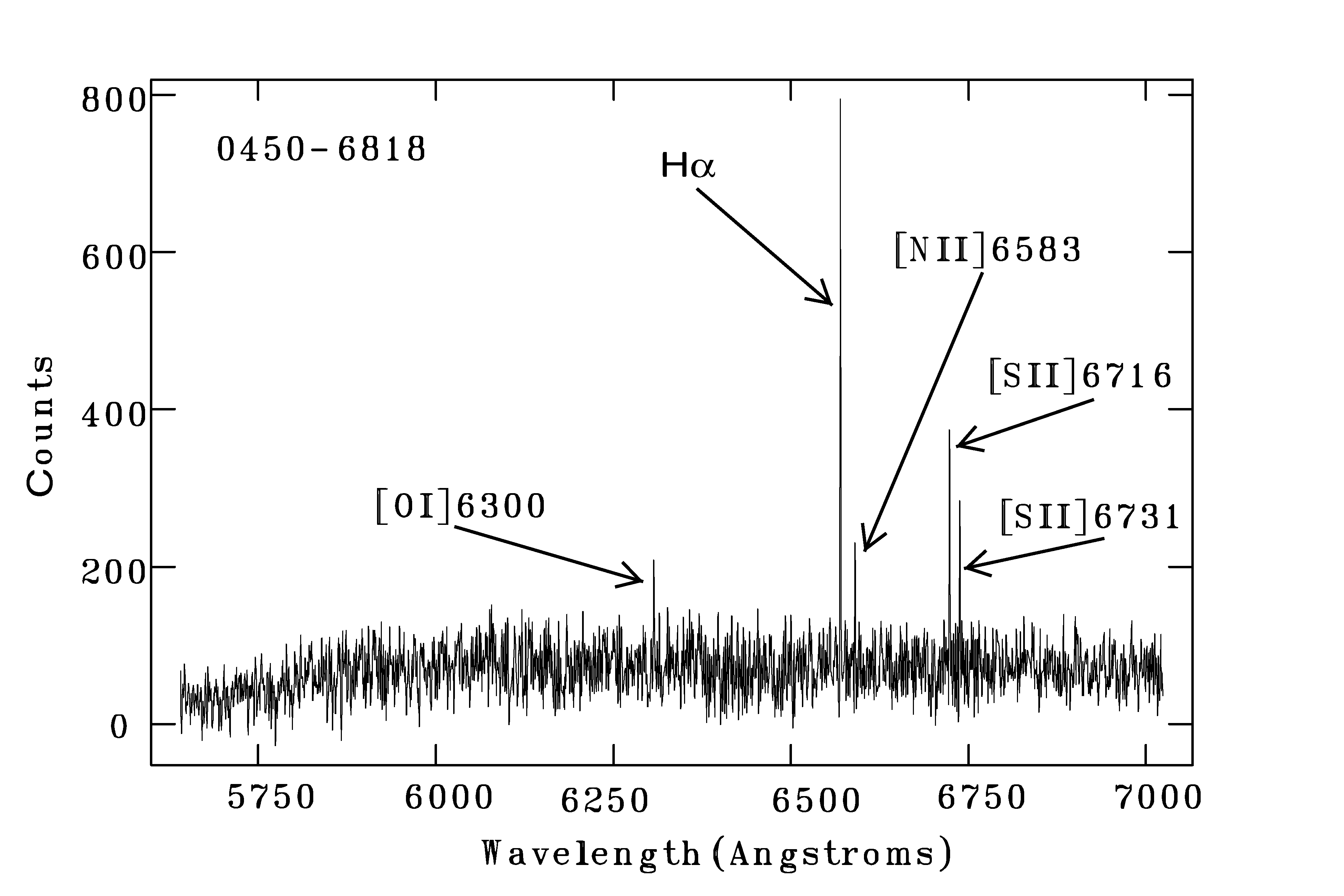}}
 \resizebox{0.50\linewidth}{!}{\includegraphics[trim = 150 0 140 0,clip]{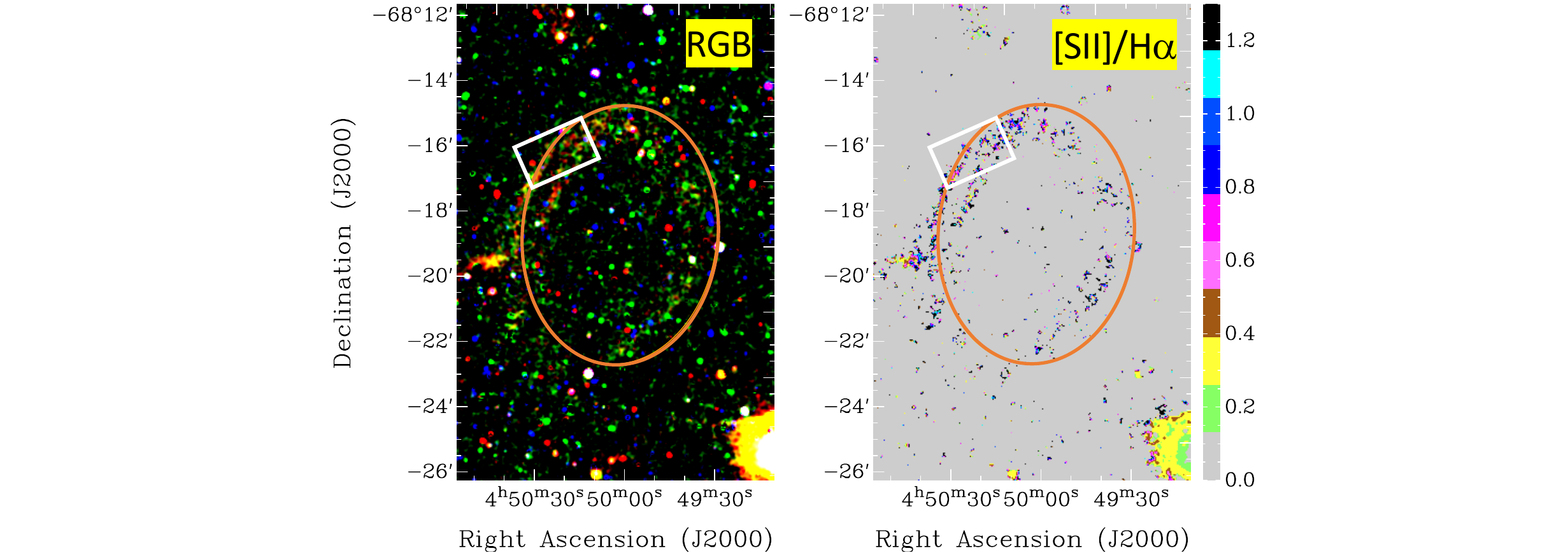}}
  \caption{J0450--6818: (Left) showing the spectra from one arm (red) of the spectrograph; (Middle and right) colour images produced from \ac{MCELS} data, where RGB corresponds to \Ha, \SII\ and \OIII\ while the ratio map is between \SII\ and \Ha. The rectangular box (white) represents an approximate position of the WiFeS slicer. The orange ellipse indicates the extent of the optical emission seen from the object.}
  \label{fig:A2}
 \end{center}
\end{figure*}

\subsection{J0450--6818 (Figure~\ref{fig:A2})}
 \label{sec:J0450--6818}
 
The \Ha\ image for this object shows two distinctly disjointed filaments, giving a bilateral and elliptical shell morphology. The emission lines are strongest in these areas. The \SII/\Ha\ ratio is $\sim$0.8 and has an elliptical diameter of about 116$\times$87~pc which satisfies one of the selection criteria for the larger diameter class of \ac{SNRs} known as ``evolved" \citep{1973ApJ...180..725M}. This remnant, as seen in Figure~\ref{fig:A2}, was also detected in \OI\ which boosts the \ac{SNR} confirmation as it gives evidence of shock heating. As for the J0444--6758 \ac{SNR} candidate, the stellar content of this object's area shows the presence of only one nearby massive OB star.

\begin{figure*}
 \begin{center}
 \resizebox{1\linewidth}{!}{\includegraphics[trim = 0 20 0 0]{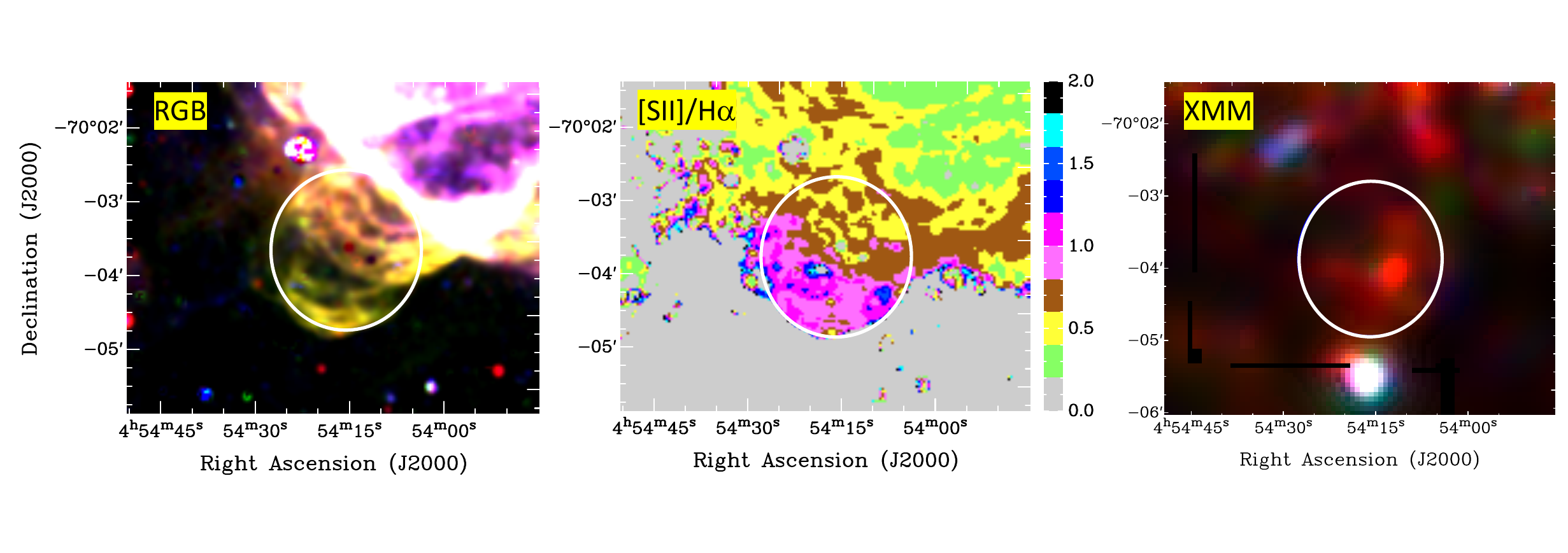}}
  \caption{J0454--7003: colour images produced from \ac{MCELS} data, where RGB corresponds to \Ha, \SII\ and \OIII\ (left) while the ratio map is between \SII\ and \Ha\ (middle). {\it XMM-Newton} EPIC RGB (R=0.3--0.7~keV band, G=0.7--1.1~keV band and B=1.1--4.2~keV band) image from the area of J0454--7003 is shown in the right panel. The white circle indicates the extent of the optical emission seen from the object. The line-shaped features in the left and bottom hand side of the image are smoothing artefacts, caused by borders of adjacent observations, which introduce steps in exposure.
}
  \label{fig:A3}
 \end{center}
\end{figure*}

\subsection{J0454--7003 (Figure~\ref{fig:A3})}
 \label{sec:J0454--7003}
This circular object is positioned at the outer south-east boundary of the large superbubble DEM\,L25 \citep{2002AJ....123..255O}. With a diameter of 30~pc (Figure~\ref{fig:A3}) and no optical spectra available for this object, we use \ac{MCELS} images and estimate that the overall \SII/\Ha\ ratio of $>$0.8 which warrants further investigation of this object as an \ac{SNR} candidate. For the larger DEM\,L25 shell, the \OIII/\Ha\ ratio is also higher in the interaction zone (ranging around 0.8), which together with the high \SII/\Ha\ ratio suggest a ram pressure-induced shock. We note that there is also a possible filamentary shell (``blow-out'') from the much larger and brighter DEM\,L25 shell. There are a number of nearby OB stars in projection to J0454--7003, and their stellar content was classified by \citet{1996ApJ...465..231O}. 

Very weak and soft ($<$0.7~keV) X-ray emission is seen around the centre of the ellipse marking the optical emission (Figure~\ref{fig:A3} (right)). Clearly extended soft X-ray emission is detected from the nearby DEM\,L25 shell. Unfortunately, the exposure of $\sim$3.8\,ks (EPIC-pn) is too short for a more quantitative analysis of the X-ray emission from J0454--7003. 

\begin{figure*}
 \begin{center}
 \resizebox{0.47\textwidth}{!}{\includegraphics[trim = 0 0 0 0]{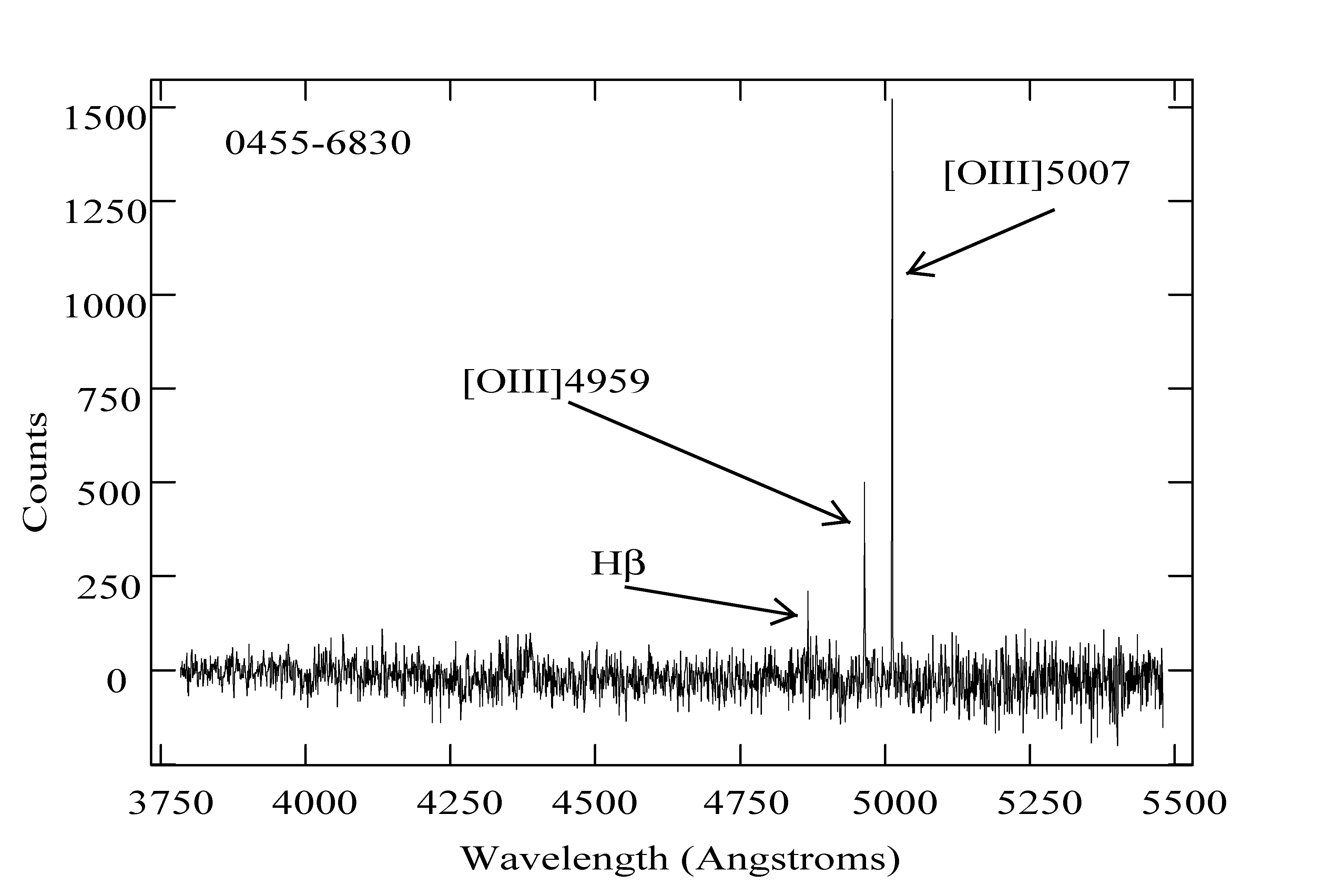}}
 \resizebox{0.47\textwidth}{!}{\includegraphics[trim = 0 0 0 0]{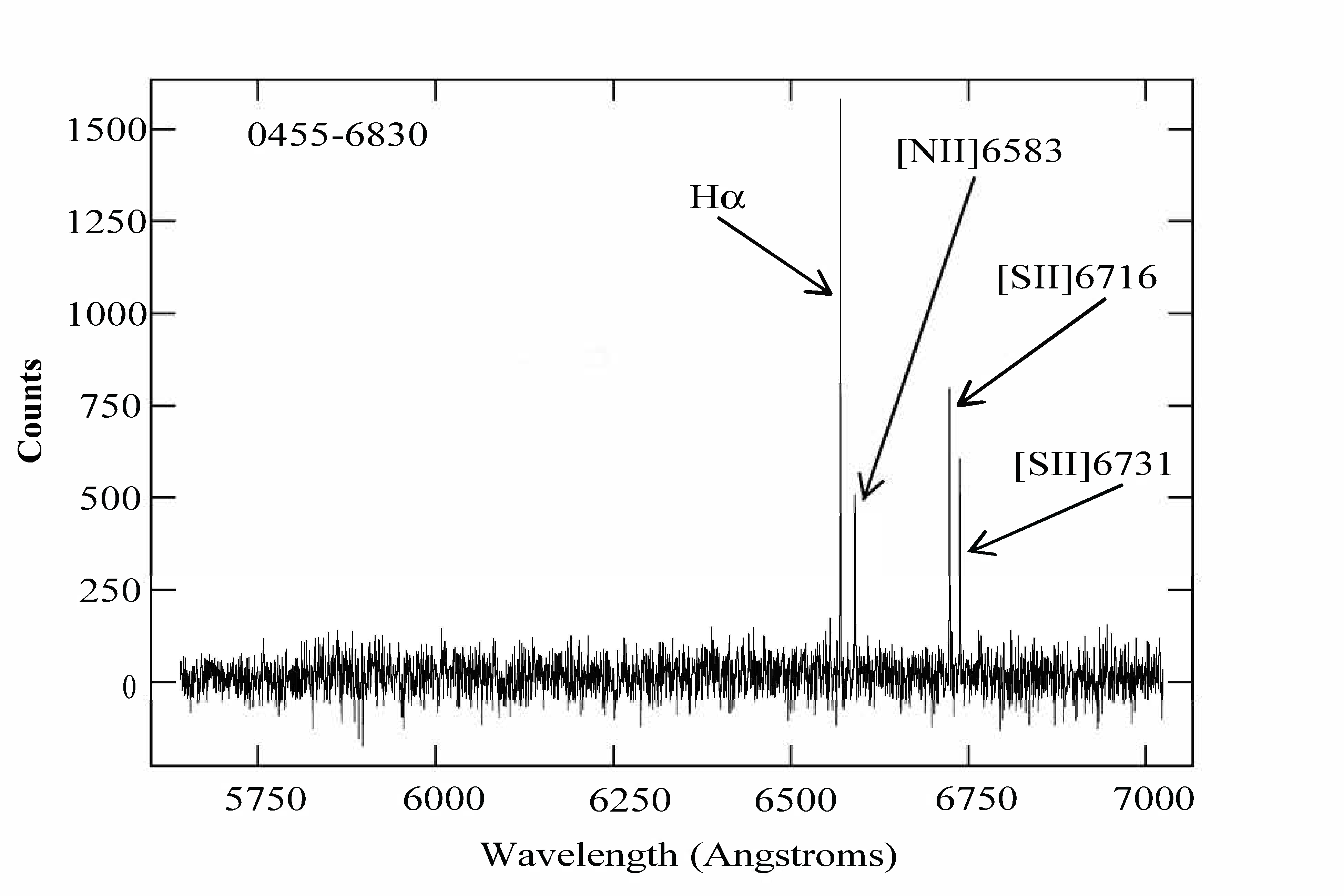}}
 \resizebox{1\linewidth}{!}{\includegraphics[trim = 0 0 0 0]{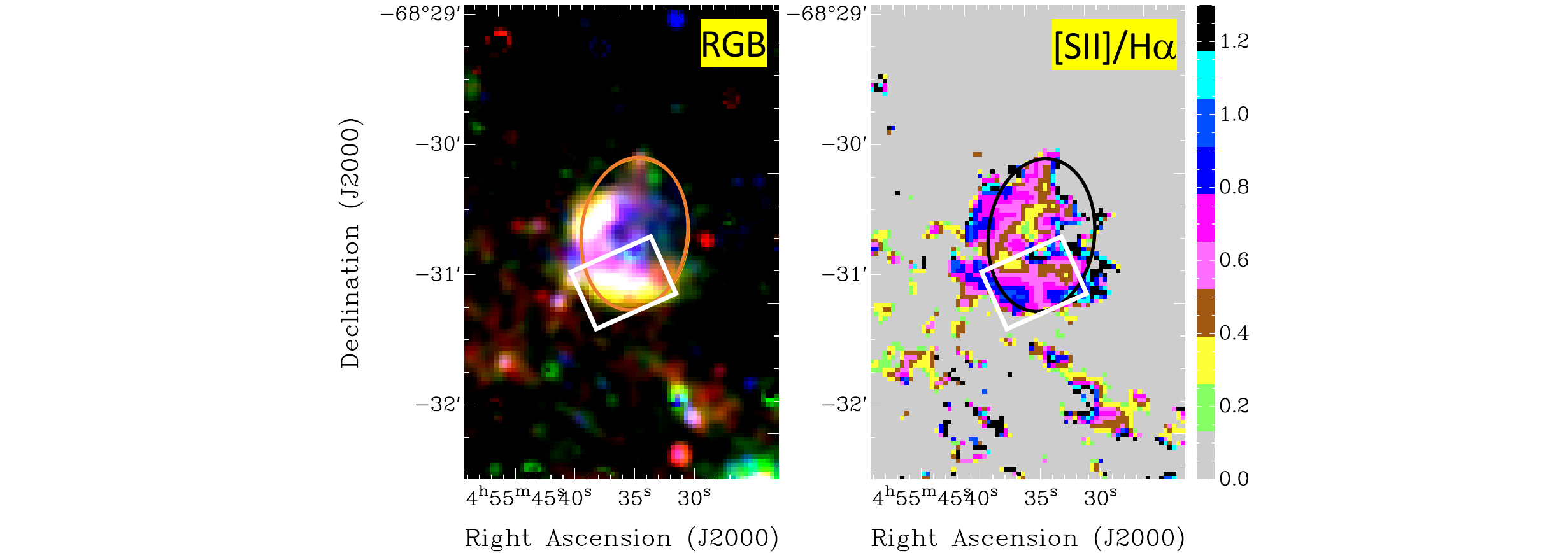}}
  \caption{J0455--6830: (Top) showing the spectra from both arms (left; blue, right; red) of the spectrograph; (Bottom) colour images produced from \ac{MCELS} data, where RGB corresponds to \Ha, \SII\ and \OIII\ while the ratio map is between \SII\ and \Ha. The rectangular box (white) represents an approximate position of the WiFeS slicer. The orange/black ellipse indicates the extent of the optical emission seen from the object.}
  \label{fig:A4}
 \end{center}
\end{figure*}

\subsection{J0455--6830 (Figure~\ref{fig:A4})}
 \label{sec:J0455--6830}

The \ac{MCELS} \Ha\ image of this \ac{SNR} candidate does not show an obvious shell morphology with diffuse and scattered emission across the suggested boundaries (Figure~\ref{fig:A4}). This \ac{SNR} candidate has the smallest diameter (D$_{av}$=17~pc) of our sample and the \SII/\Ha\ ratio of $\sim$1.1 is one of the sample's highest.

As in J0450--6818, this \ac{SNR} candidate also exhibits prominent (bright) \OIII\ lines while at the same time we did not detect an \OI\ line. The \ac{MCELS} \OIII\ image also shows distinctive emission that, in this case, follows the \Ha\ and \SII\ emission. Finally, one would expect that such a small sized \ac{SNR} would be bright in X-ray emission. Unfortunately, this area is presently poorly covered in all X-ray surveys.

We also searched MCPS for massive (OB) stars near this object (100~pc radius) and found 18 very distant OB stars -- none within the boundaries of this proposed \ac{SNR} candidate. 

\begin{figure*}
 \begin{center}
 \resizebox{0.47\textwidth}{!}{\includegraphics[trim = 0 0 0 0]{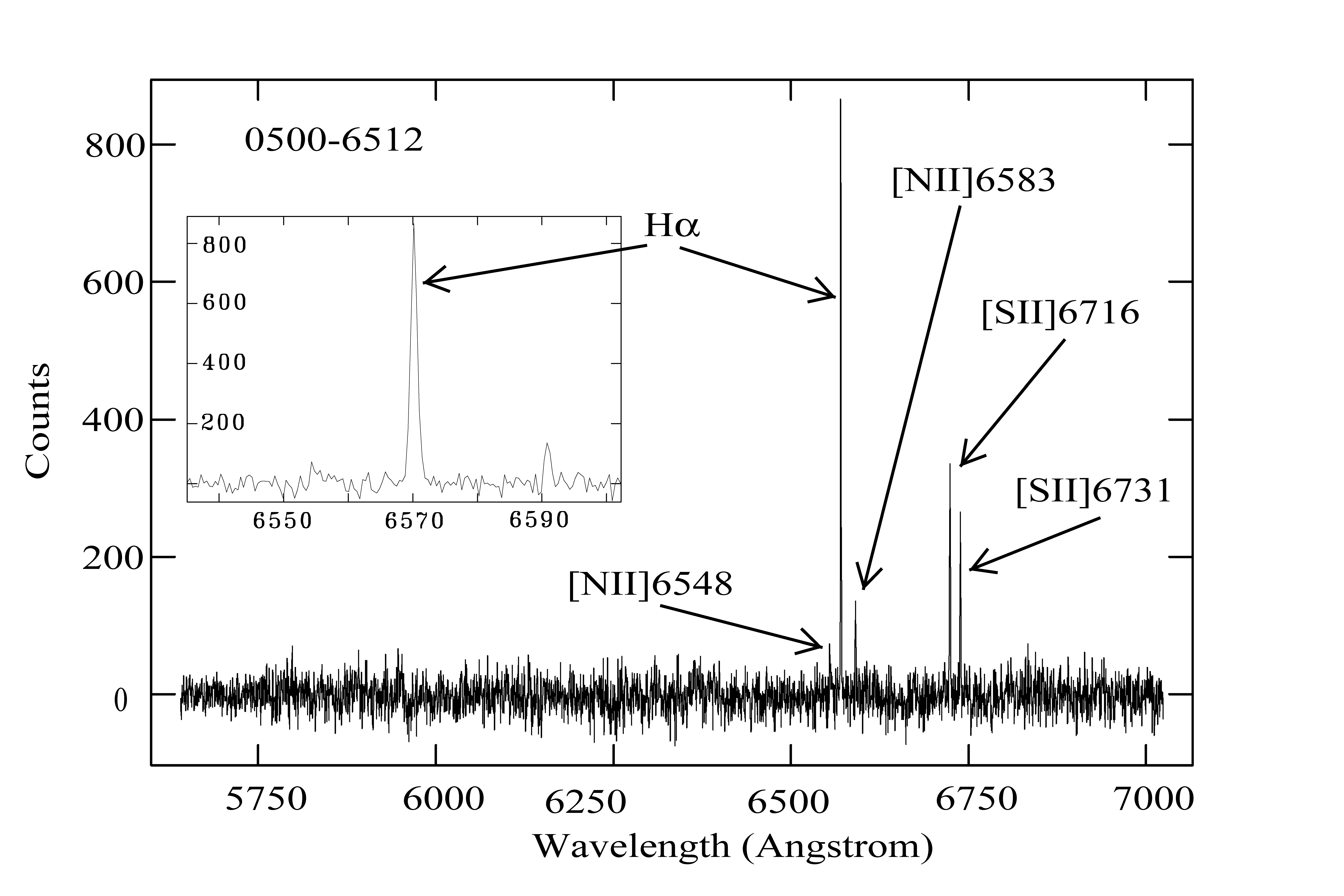}}
 \resizebox{0.5\linewidth}{!}{\includegraphics[trim = 150 0 140 0,clip]{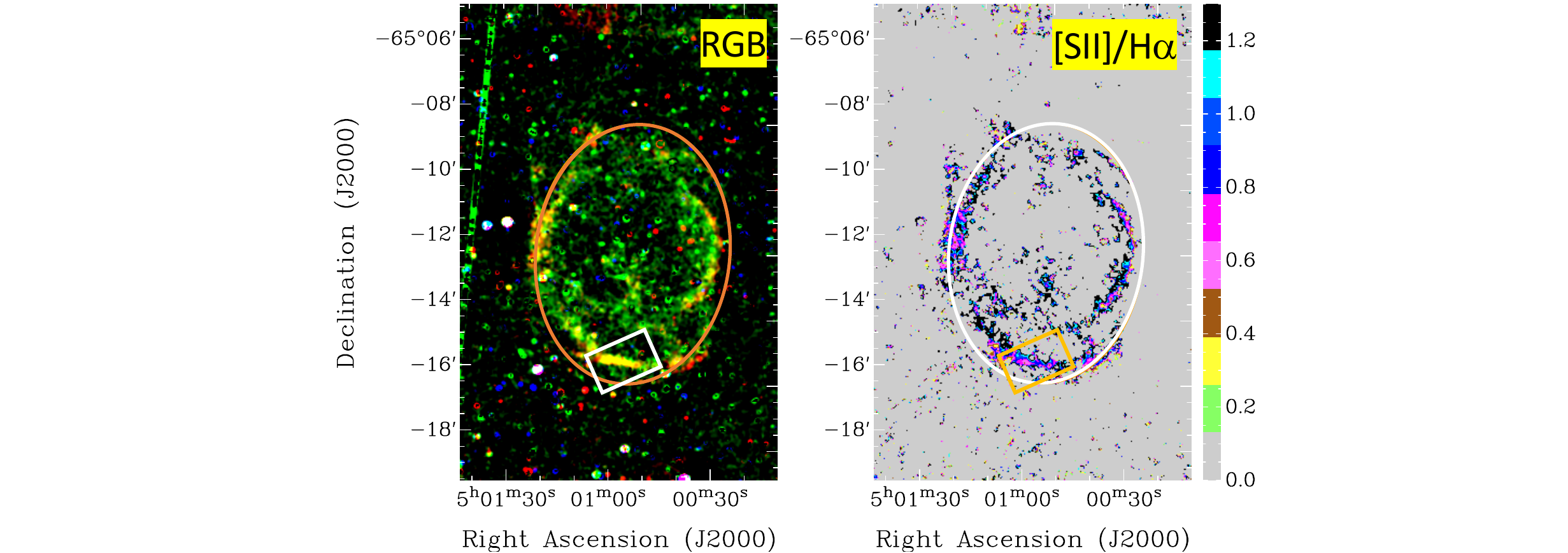}}
  \caption{J0500--6512: (Left) showing the spectra from one arm (red) of the spectrograph where the inset figure represents the ``zoomed-in'' \Ha\ line. This \Ha\ line is detected with the broadening of 1.20~\AA\ while the instrument width is $\sim$1~\AA\ ($\sim$45~km~s$^{-1}$); (Middle and right) colour images produced from \ac{MCELS} data, where RGB corresponds to \Ha, \SII\ and \OIII\ while the ratio map is between \SII\ and \Ha. The rectangular box (white/orange) represents an approximate position of the WiFeS slicer. The orange/white ellipse indicates the extent of the optical emission seen from the object.}
  \label{fig:A5}
 \end{center}
\end{figure*}

\subsection{J0500--6512 (Figure~\ref{fig:A5})}
 \label{sec:J0500--6512}

The \Ha\ image shows a looped filamentary structure (Figure~\ref{fig:A5}). This \ac{SNR} candidate has a complex (double-shell) morphology, with the shells overlapping in the South. We estimate that this is similar in size to the J0450--6818 candidate. The \SII/\Ha\ ratio is $\sim$0.81, typical of \ac{SNRs}. The limited variety of line intensities are not surprising as the weaker lines are lost in the noise.

\begin{figure*}
 \begin{center}
 \resizebox{0.47\textwidth}{!}{\includegraphics[trim = 0 0 0 0]{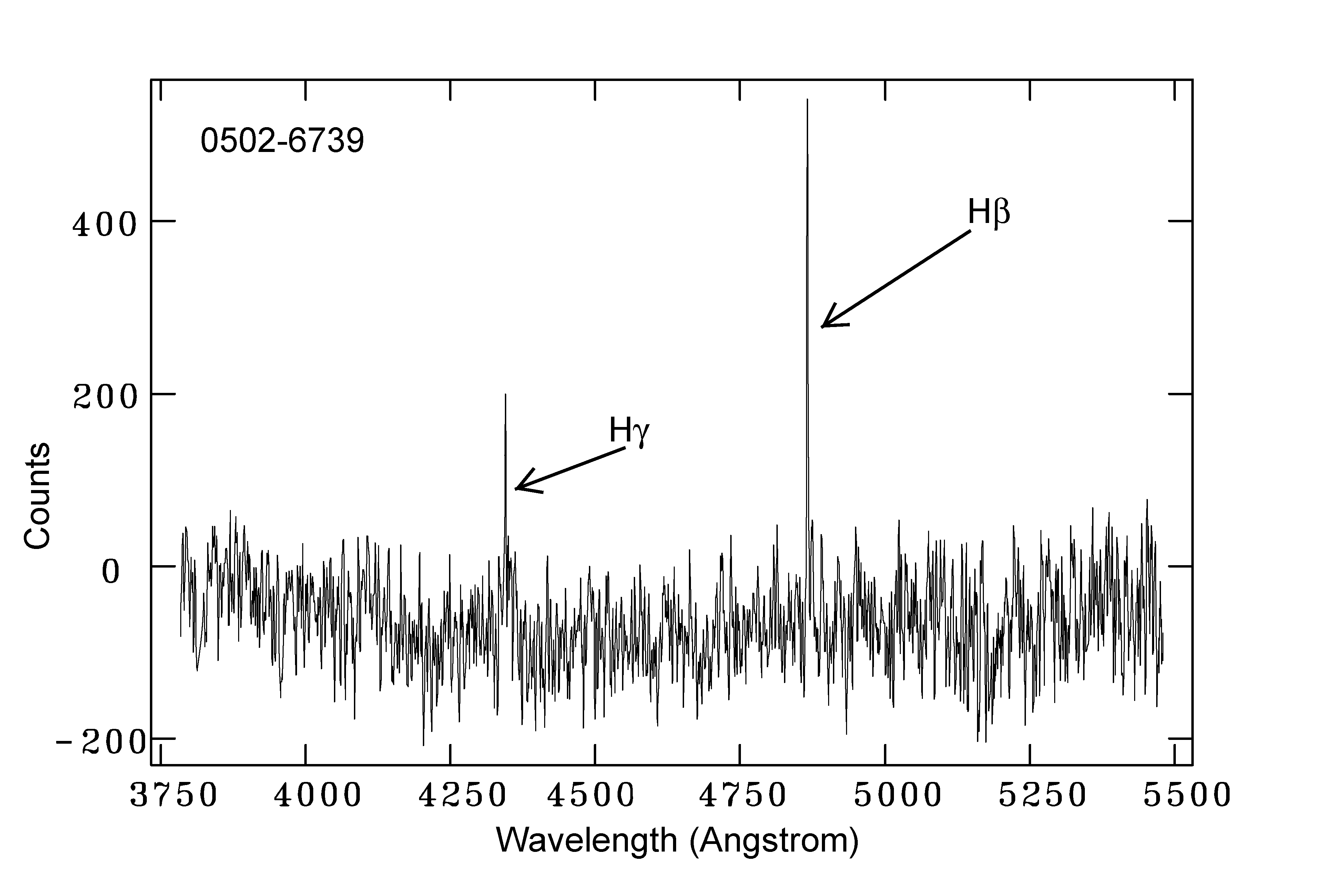}}
 \resizebox{0.47\textwidth}{!}{\includegraphics[trim = 0 0 0 0]{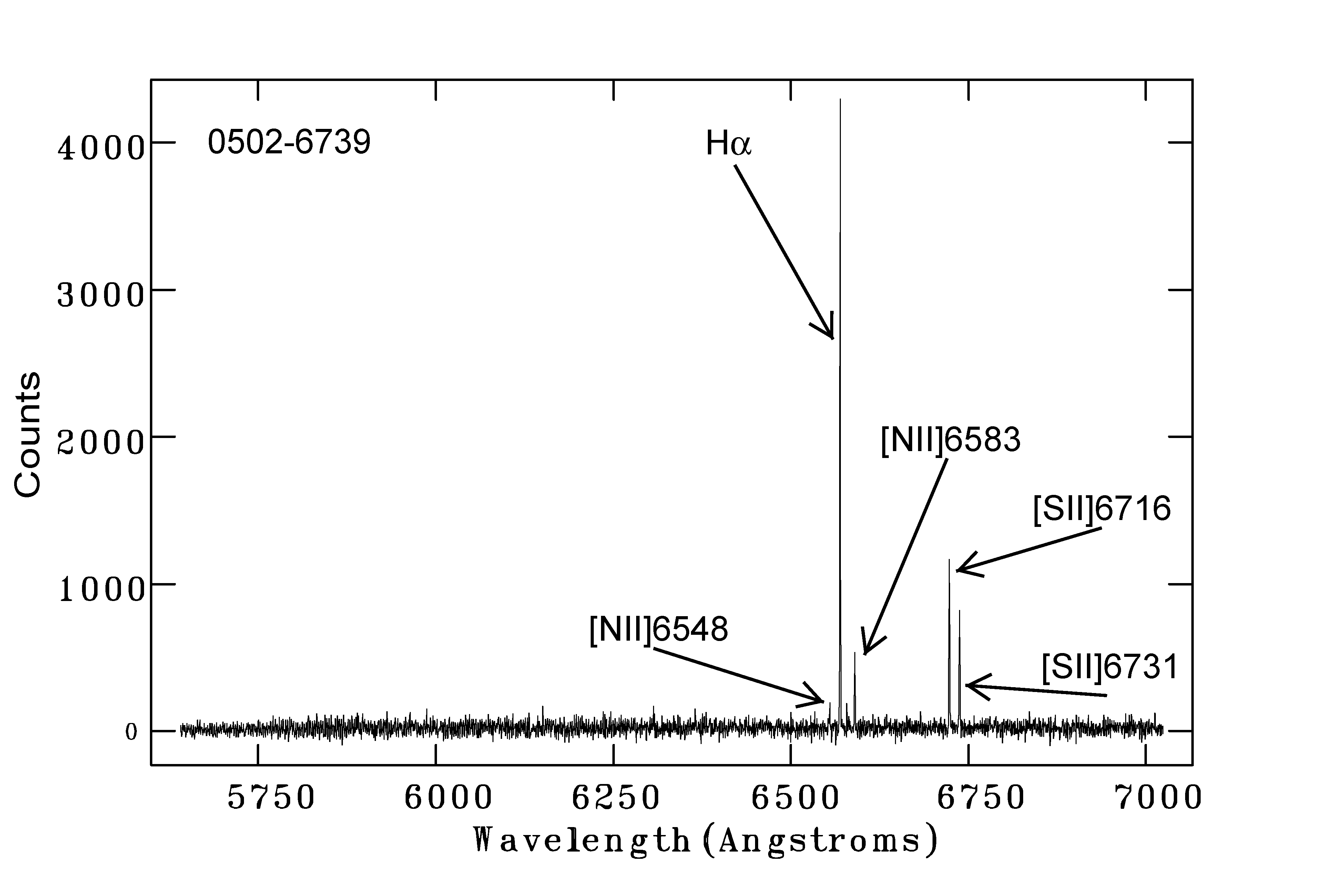}}
 \resizebox{1\linewidth}{!}{\includegraphics[trim = 0 0 0 0]{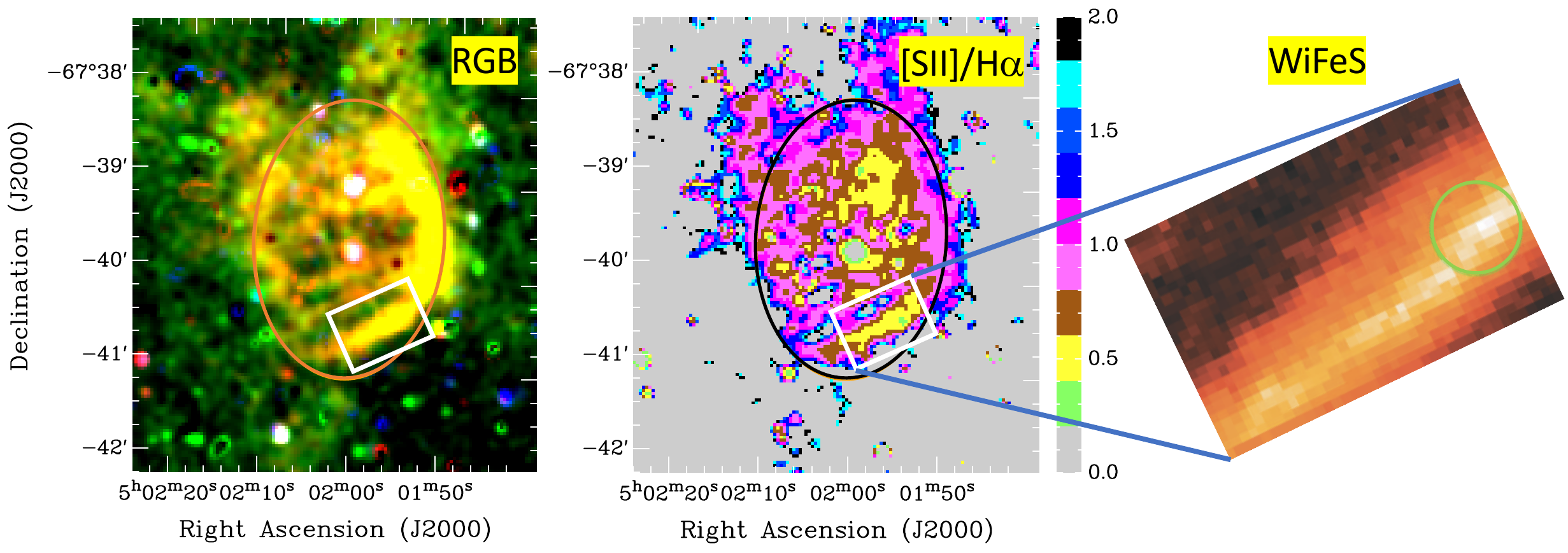}}
  \caption{J0502--6739: (Top) showing the spectra from both arms (left; blue, right; red) of the spectrograph; (Bottom left and right) colour images produced from \ac{MCELS} data, where RGB corresponds to \Ha, \SII\ and \OIII\ while the ratio map is between \SII\ and \Ha. 
  The orange/black ellipse indicates the extent of the optical emission seen from the object. (bottom right) image of 25$\times$38~arcsec (as in the white rectangular box) field of view of WiFeS spectrograph slicer which consists of 3152 slices (see Sect.~\ref{wifes}). The strongest part of \Ha\ emission of J0502--6739 is detected on slice number 2116. The green circle represents the position of the highest flux within a 10~arcsec aperture where the 1D spectrum of all spectral lines is extracted from the cube.}
  \label{fig:A6}
 \end{center}
\end{figure*}

\subsection{J0502--6739 (Figure~\ref{fig:A6})}
 \label{sec:J0502--6739}
 
This \ac{SNR} candidate structure has an average diameter of 43~pc and shows diffuse structureless emission which is slightly elongated (Figure~\ref{fig:A6}). Although the \SII/\Ha\ ratio is relatively low at $\sim$0.55, we still take this as an \ac{SNR} candidate but we note possible contamination from the nearby \HII\ region.

As for other objects in this sample, we searched for massive stars and found the richest population of OB stars (18 with 2 inside the object boundaries) among the sample studied here.

\begin{figure*}
 \begin{center}
 \resizebox{1\linewidth}{!}{\includegraphics[trim = 0 0 0 0]{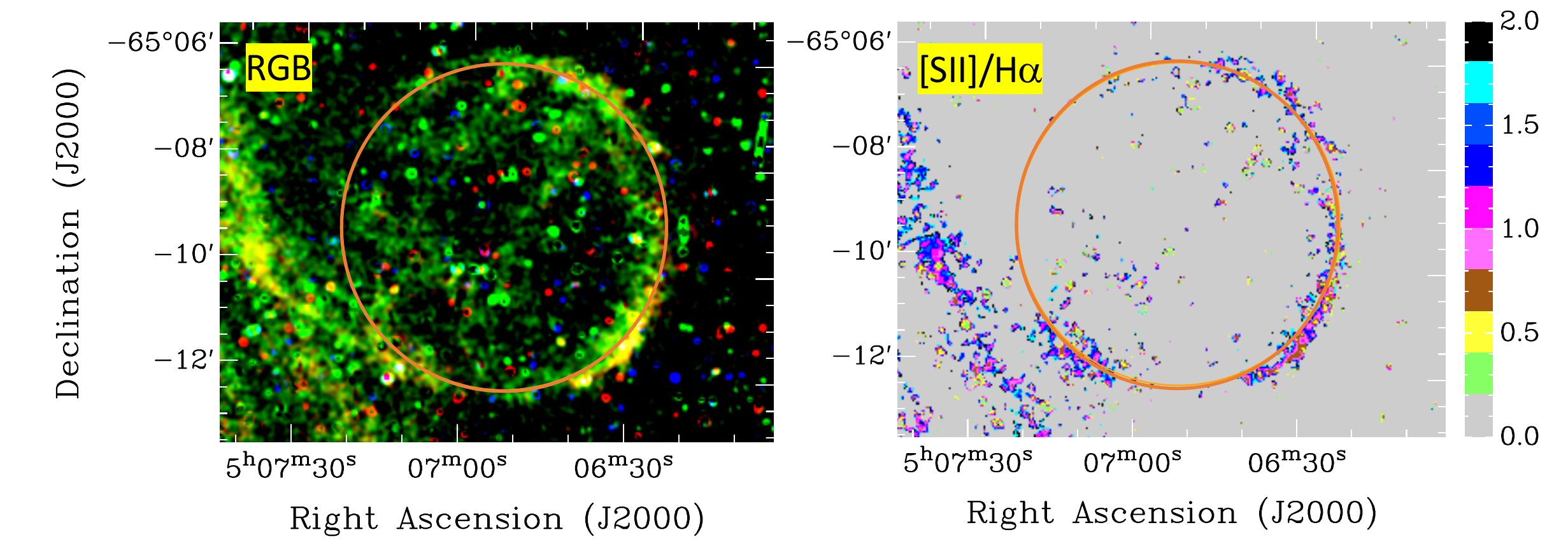}}
  \caption{J0506--6509: colour images produced from \ac{MCELS} data, where RGB corresponds to \Ha, \SII\ and \OIII\ while the ratio map is between \SII\ and \Ha. The orange circle indicates the extent of the optical emission seen from the object.}
  \label{fig:A7}
 \end{center}
\end{figure*}

\subsection{J0506--6509 (Figure~\ref{fig:A7})}
 \label{sec:J0506--6509}
The candidacy for the \ac{SNR} nature of this object is prompted by its spherical (circular) shell morphology with a diameter of 90~pc (Figure~\ref{fig:A7}). While there are no spectra available for this candidate, we use \ac{MCELS} images and estimate that the overall \SII/\Ha\ ratio is $\sim$0.6, which warrants further investigation of this object as an \ac{SNR} candidate. We note that no nearby massive stars could be found.

\begin{figure*}
 \begin{center}
 \resizebox{0.47\textwidth}{!}{\includegraphics[trim = 0 0 0 0]{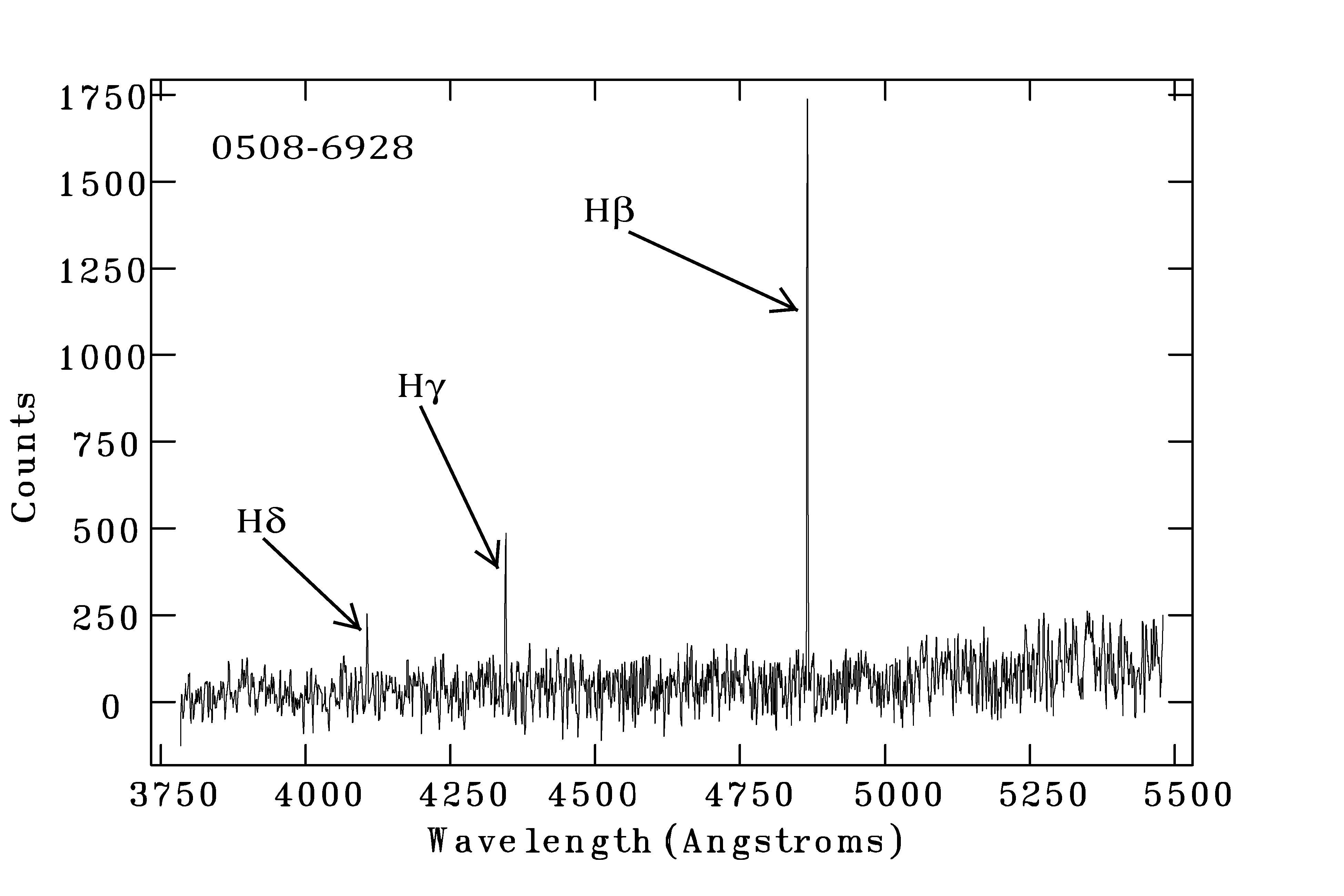}}
 \resizebox{0.47\textwidth}{!}{\includegraphics[trim = 0 0 0 0]{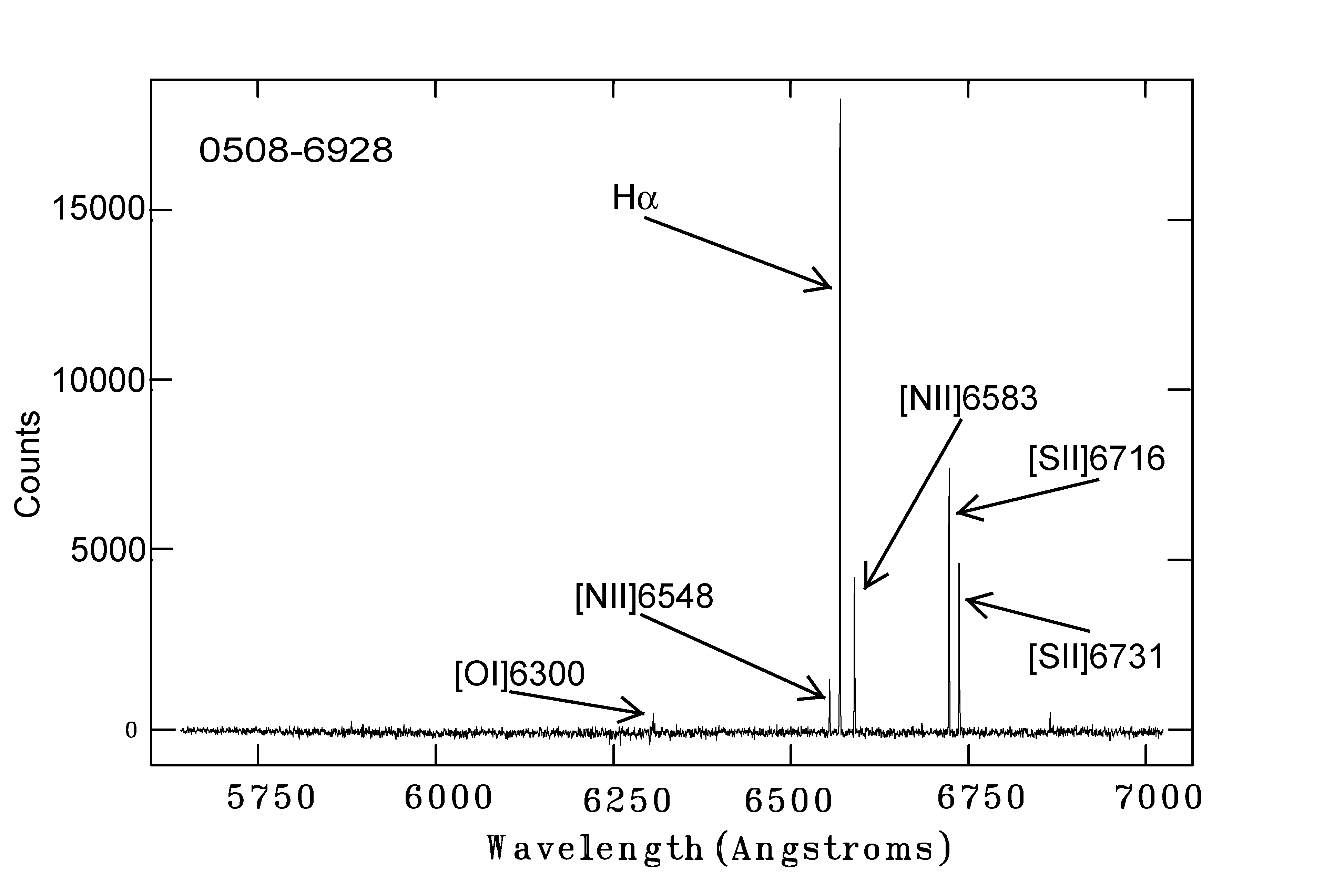}}
 \resizebox{1\linewidth}{!}{\includegraphics[trim = 0 0 0 0]{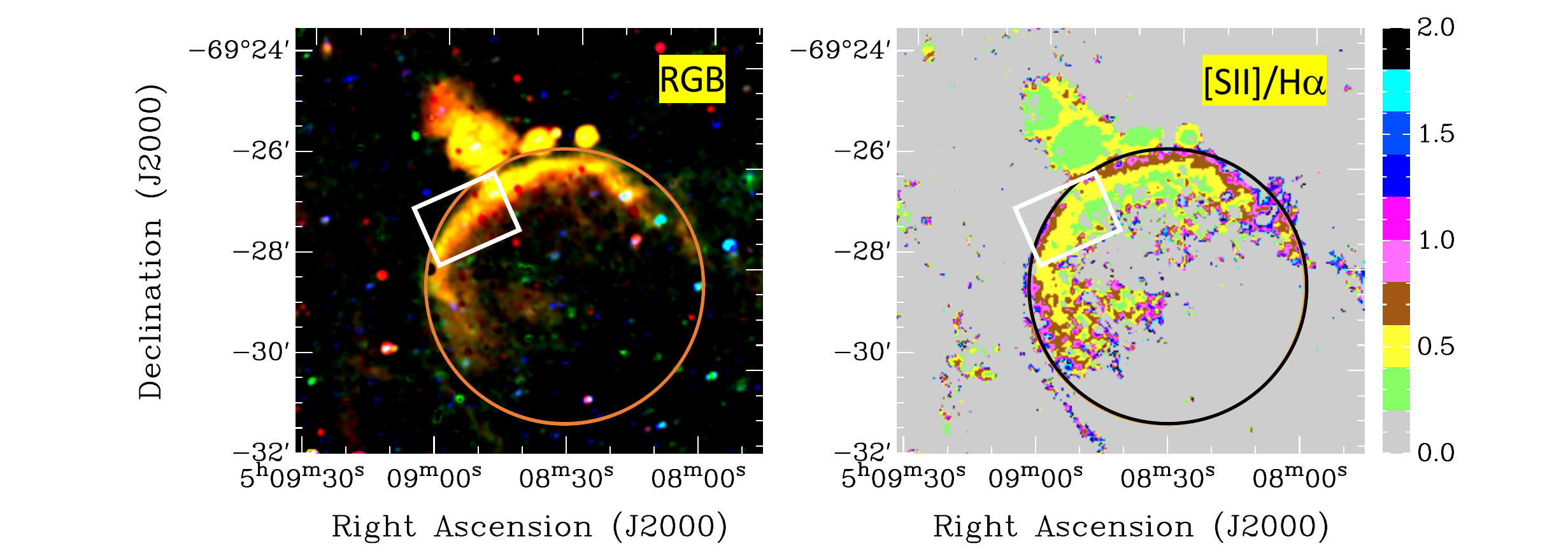}}
  \caption{J0508--6928: (Top) showing the spectra from both arms (left; blue, right; red) of the spectrograph; (Bottom) colour images produced from \ac{MCELS} data, where RGB corresponds to \Ha, \SII\ and \OIII\ while the ratio map is between \SII\ and \Ha. The rectangular box (white) represents an approximate position of the WiFeS slicer. The orange/black circle indicates the extent of the optical emission seen from the object.}
  \label{fig:A8}
 \end{center}
\end{figure*}

\subsection{J0508--6928 (Figure~\ref{fig:A8})}
 \label{sec:J0508--6928}
This \ac{SNR} candidate has an obvious half-shell (horse-shoe) morphology in the top-half, expanding a little past half-way (Figure~\ref{fig:A8}). The \SII/\Ha\ ratio is $\sim$0.72. This remnant also has the highest peak in \Ha\ in our sample but given the strength of \Hb\ in this object, we note the stunning lack of any \OIII\ emission. We find 2 massive OB stars inside the object bounds, 3 just north and 2 relatively close on the Eastern side of this \ac{SNR} candidate.

\begin{figure*}
 \begin{center}
 \resizebox{0.4425\textwidth}{!}{\includegraphics[trim = 0 0 0 0]{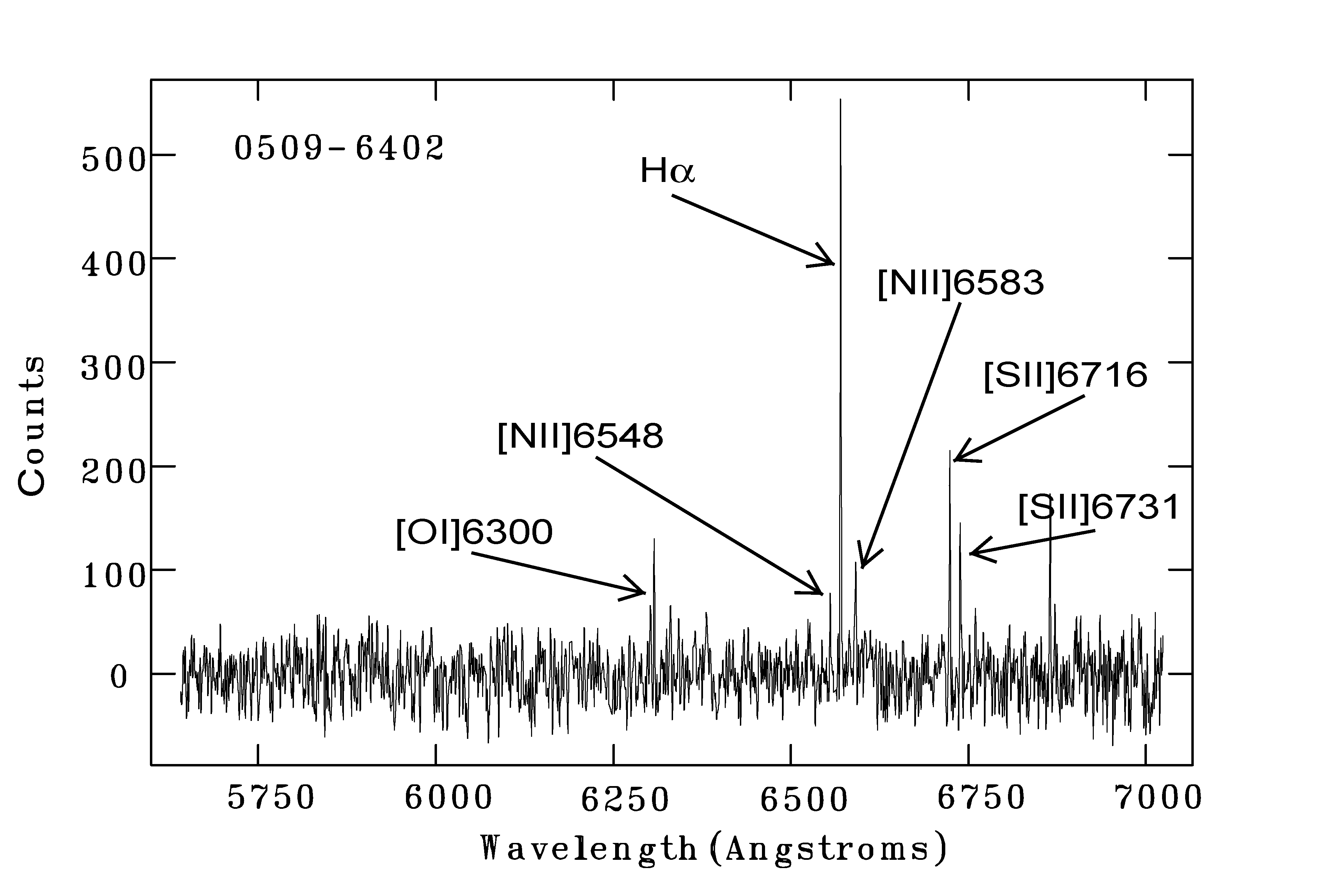}}
 \resizebox{0.55\linewidth}{!}{\includegraphics[trim = 120 0 120 0,clip]{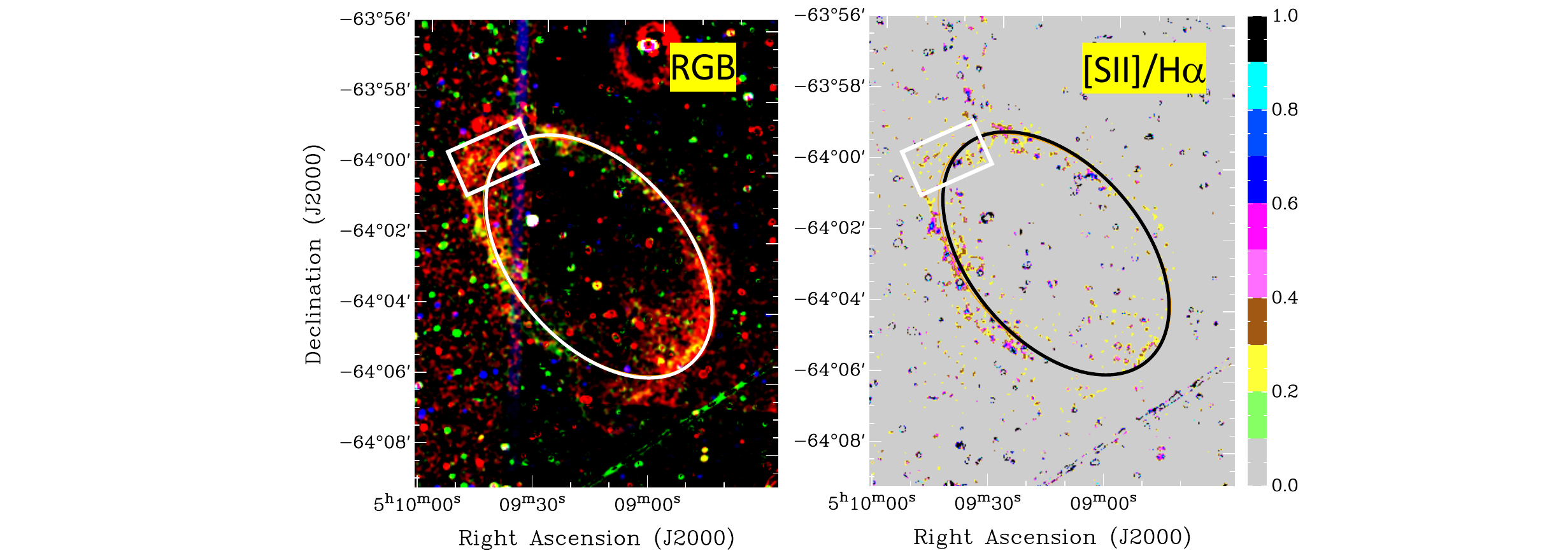}}
  \caption{J0509--6402: (Left) showing the spectra from one arm (red) of the spectrograph; (Middle and right) colour images produced from \ac{MCELS} data, where RGB corresponds to \Ha, \SII\ and \OIII\ while the ratio map is between \SII\ and \Ha. The rectangular box (white) represents an approximate position of the WiFeS slicer. The white/orange ellipse indicates the extent of the optical emission seen from the object.}
  \label{fig:A9}
 \end{center}
\end{figure*}

\begin{figure}
 \begin{center}
 \resizebox{0.99\linewidth}{!}{\includegraphics[angle=-90,trim = 0 0 0 0,clip]{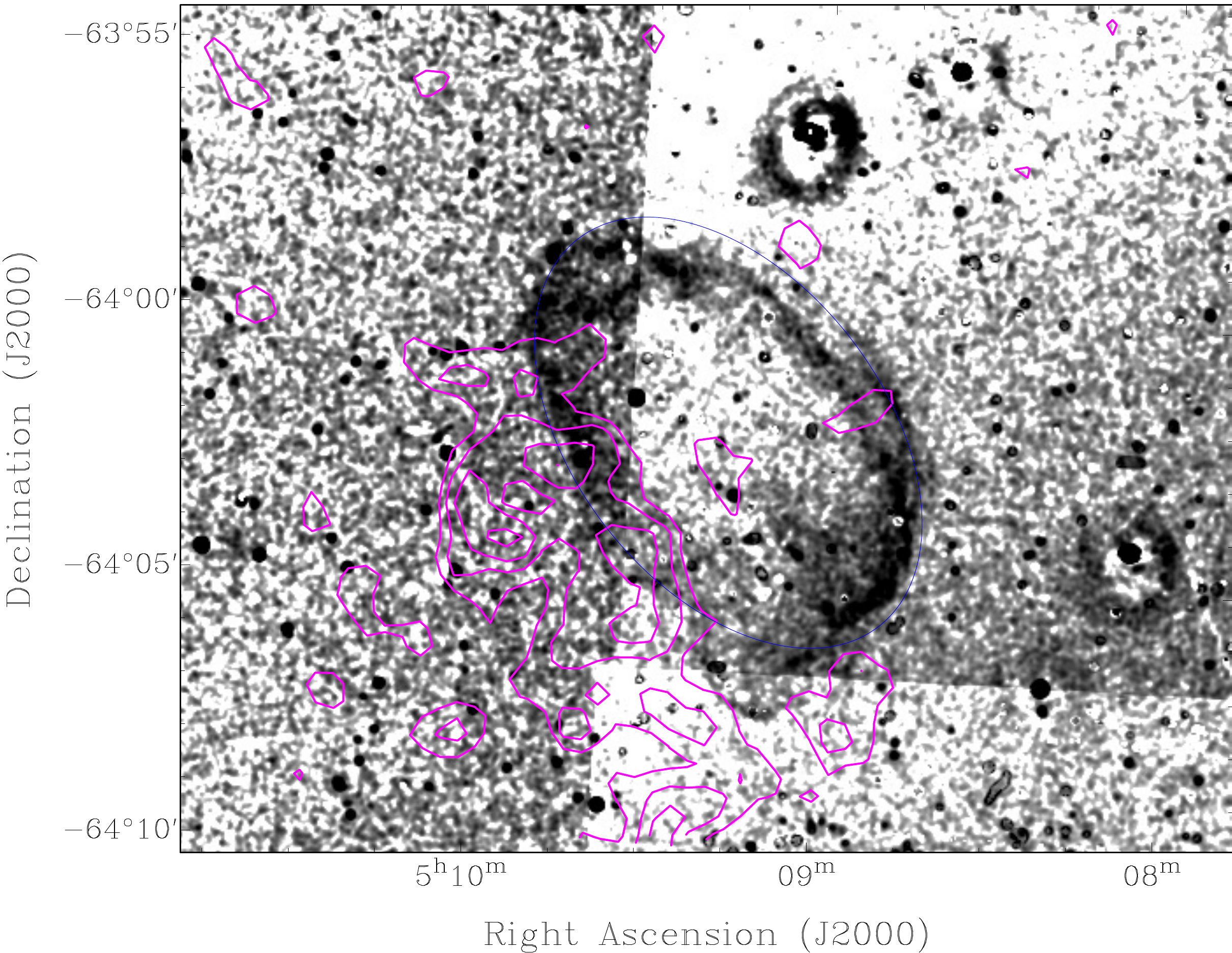}}
  \caption{ \Ha\ image of J0509--6402 overploted with \HI\ contours (6, 7, 8 and 9$\times10^{20}$~atoms~cm$^{-2}$. The blue ellipse indicates the extent of the optical emission seen from the object.}
  \label{fig:A9-2}
 \end{center}
\end{figure}

\subsection{J0509--6402 (Figures~\ref{fig:A9} and \ref{fig:A9-2})}
 \label{sec:J0509--6402}

This \ac{SNR} candidate lies 2\D\ north of the \ac{LMC} main body, in a field where we did not expect detection of any \ac{LMC} \ac{SNRs}. The \ac{MCELS} \Ha\ and \SII\ images shows a typical (shell-like) \ac{SNR} morphology but somewhat elongated (Figure~\ref{fig:A9}). At the same time, we report no \OIII\ emission in the \ac{MCELS} image. Due to the size of the J0509--6402 shell $\sim$8\arcmin~$\times$~5\arcmin\ ($\sim$116$\times$83~pc), the \ac{SNR} candidate cannot be morphologically classified as a young \ac{SNR}. If confirmed as an \ac{SNR}, this object would be certainly evolving in a much more rarefied environment than the rest of the known \ac{LMC} \ac{SNRs}. However, as can be seen in Figure~\ref{fig:A9-2}, the eastern side of J0509--6402 might be interacting with one the northernmost \HI\ spurs and perhaps create so called cloud-cloud interaction as seen in a number of other \ac{MCs} \ac{SNRs} and shells \citep{2018ApJ...867....7S,2019ApJ...873...40S,2019ApJ...881...85S,2020arXiv200707900S}. Sensitive X-ray and molecular gas observations are needed to confirm this possibility.

The optical spectra confirms shocks typical of \ac{SNRs} which can be further confirmed using the ratio of shock sensitive \SII\ lines with \Ha\, e.g. \SII/\Ha\ ratio of 0.75 (see Table~\ref{tab:snrflux}) which differentiates \ac{SNRs} from \HII\ regions and planetary nebulae. The most prominent line in its spectrum is \Ha, with the presence of \OI\ at 6300\AA, \NII\ at 6548\AA\ and 6583\AA\ as well as \SII\ lines at 6717\AA\ and 6731\AA. All these lines are always associated with spectra of \ac{SNRs} except the \OI\ line which is rarely seen. We estimate that this \ac{SNR} has an electron density of $\sim$30~cm$^{-3}$ (for a temperature of 10\,000~K) as a result of the ratio of \SII\ lines. This indicates that \ac{SNR} candidate J0509--6402 is most likely of a mature age. We also suggest that it might be more likely a result of a type~Ia \ac{SN} event because of its large distance from the main body of the \ac{LMC} where massive stars are rarely found. However, one cannot completely disregard a possible core-collapse \ac{SN} event as a very massive star could be ejected from the \ac{LMC} main body and survive ``traveling'' some $\sim$2~kpc (assuming a distance to the \ac{LMC} of 50~kpc). While no such \ac{SN} explosion is directly confirmed (so far) we suggest that only a small fraction of such \ac{SNRs} would exist in any galaxy. In principle, this scenario could be applied to any \ac{CC} type of \ac{SN} explosion unless a given \ac{SNR} doesn't currently sit near a known OB star.

As this object is so far north of the main body of the \ac{LMC} we could not search for massive stars since the area is not covered in \citet{2004AJ....128.1606Z}.

\begin{figure*}
 \begin{center}
 \resizebox{0.47\textwidth}{!}{\includegraphics[trim = 0 0 0 0]{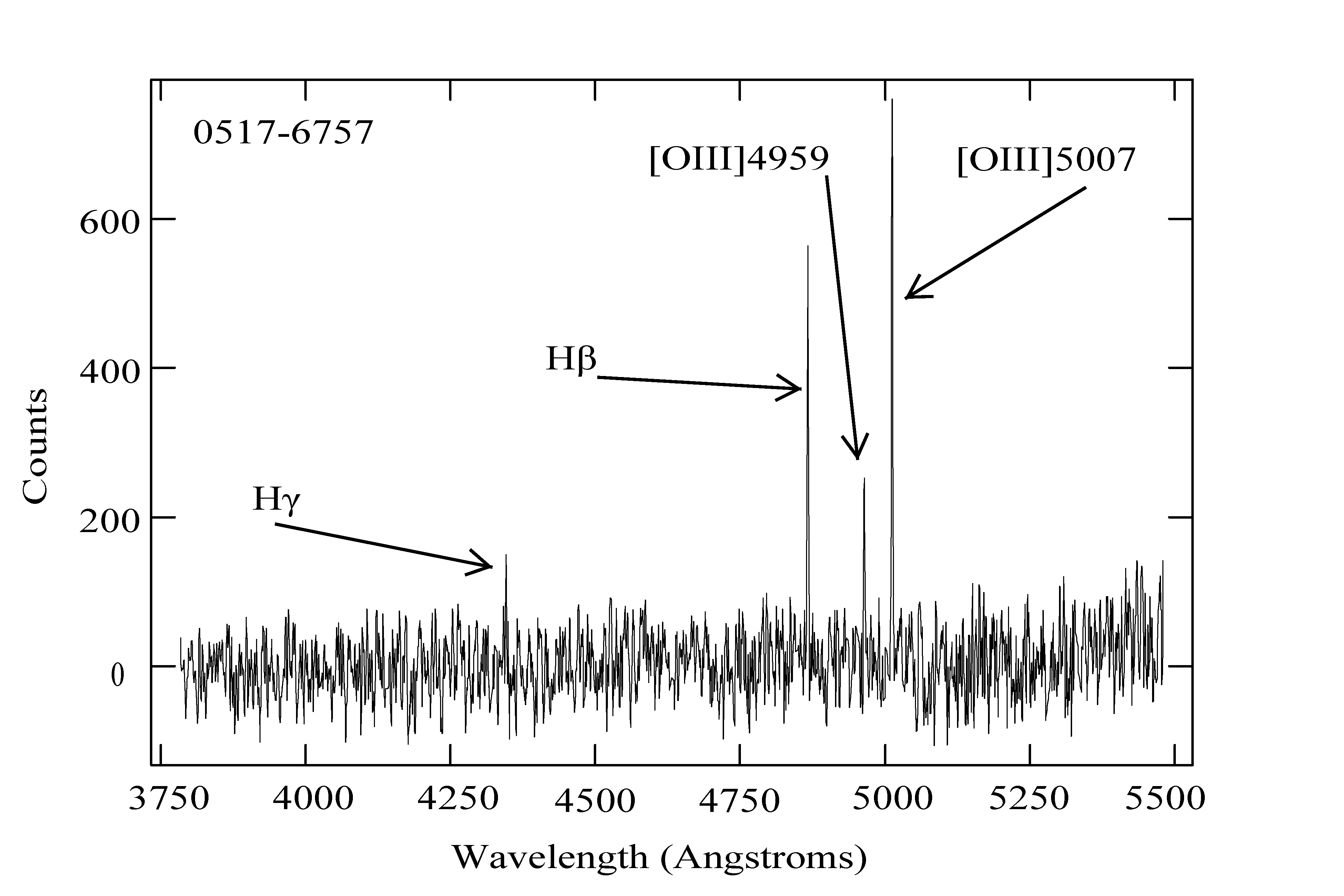}}
 \resizebox{0.47\textwidth}{!}{\includegraphics[trim = 0 0 0 0]{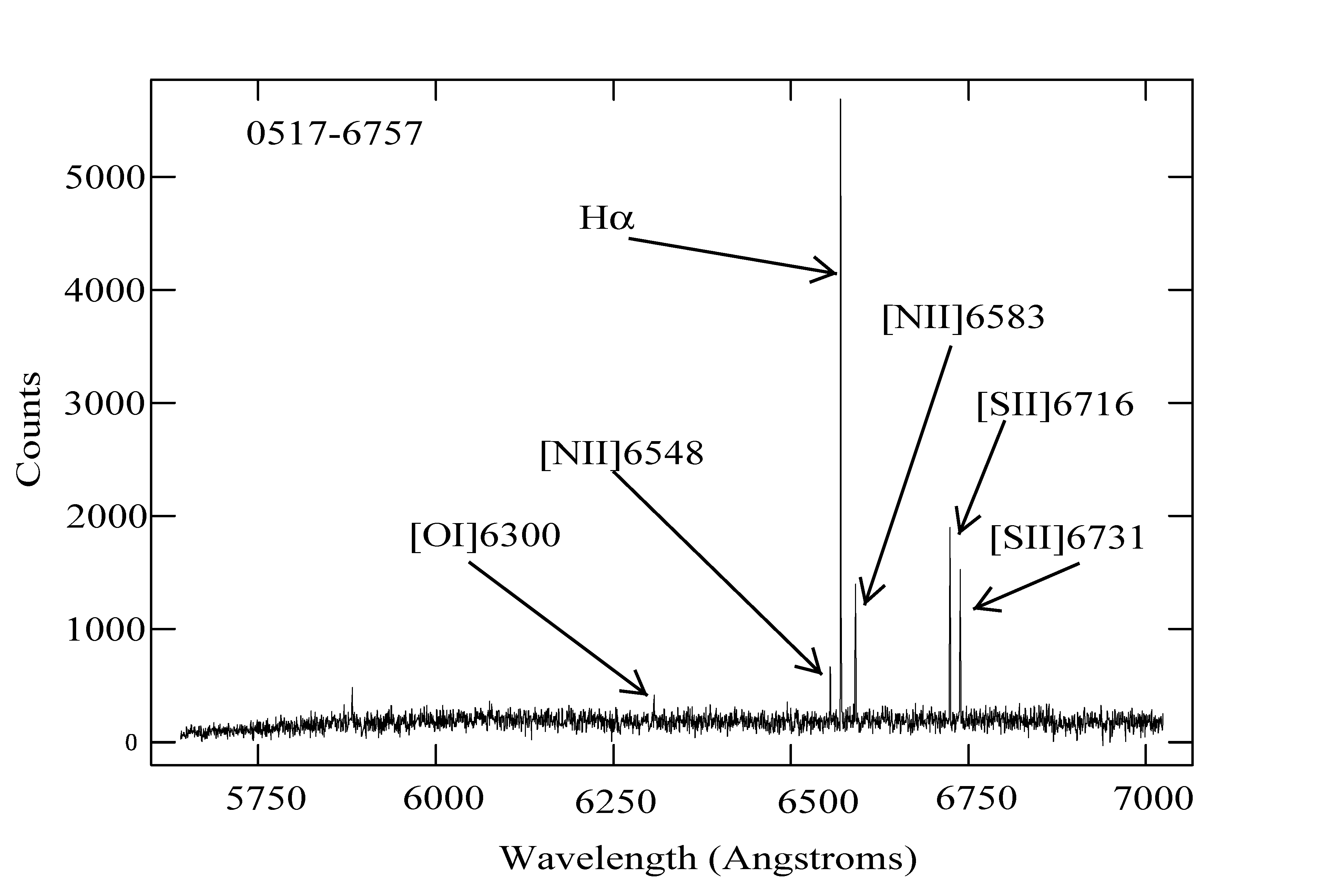}}
 \resizebox{1\linewidth}{!}{\includegraphics[trim = 0 0 0 0]{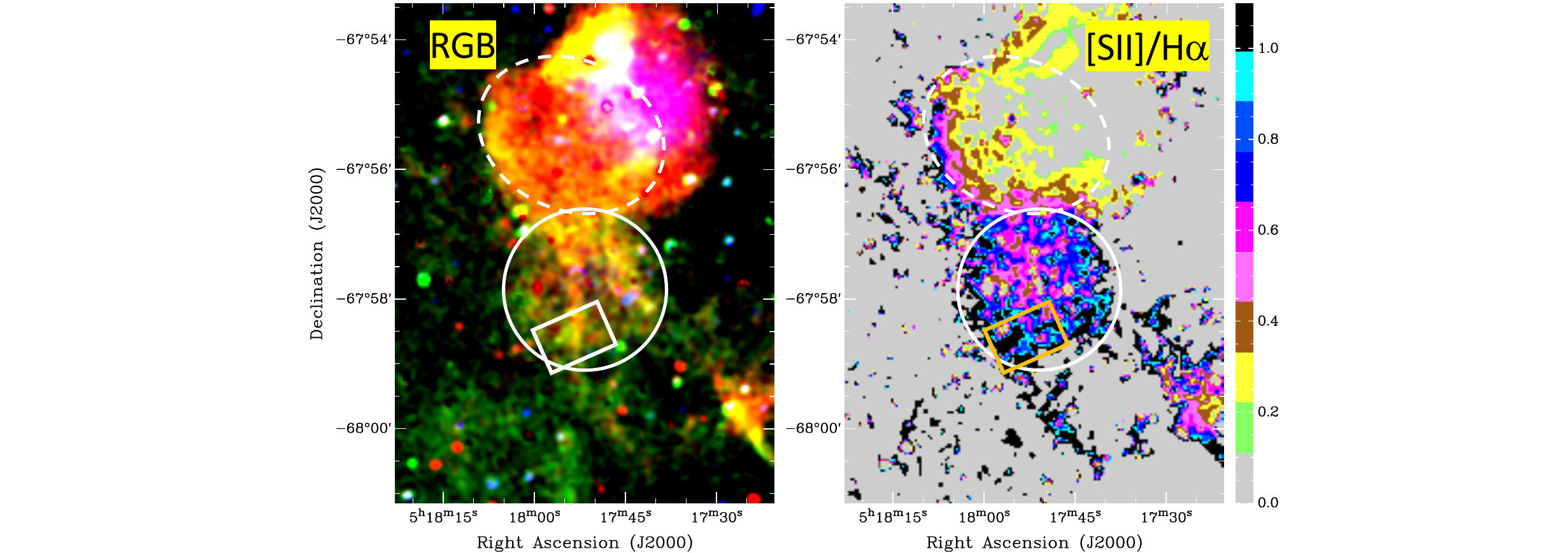}}
  \caption{J0517--6757: (Top) showing the spectra from both arms (left: blue, right: red) of the spectrograph; (Bottom) colour images produced from \ac{MCELS} data, where RGB corresponds to \Ha, \SII\ and \OIII\ while the ratio map is between \SII\ and \Ha. The rectangular box (white/orange) represents an approximate position of the WiFeS slicer. The solid white circle indicates the extent of the optical emission seen from the object while the dashed ellipse indicates possible adjacent \ac{SNR} candidate that we will investigate in our future studies.}
  \label{fig:A11}
 \end{center}
\end{figure*}

\subsection{J0517--6757 (Figure~\ref{fig:A11})}
 \label{sec:J0517--6757}
 
The \ac{MCELS} \Ha\ image (Figure~\ref{fig:A11}) of this \ac{SNR} candidate shows a faint, thin half-shell of emission towards the south of the object. It also shows line intensities for \Hy\ and \Hb\ in its spectrum. J0517--6757 is surrounded with a number of \HII\ regions, the most prominent of which is located just north of its suggested boundaries. Although unconfirmed, given that its \SII/\Ha\ ratio is $\sim$0.69 and a somewhat typical \ac{LMC} \ac{SNR} size of D=39~pc, we nominate this object to be an \ac{SNR} candidate. While we find some 17 OB stars in a 100~pc radius, none of these lie within the object's defined shell.

We also note an adjacent semi-loop somewhat north of our \ac{SNR} candidate J0517--6757 (see dashed ellipse in Figure~\ref{fig:A11}). With an elevated \SII/\Ha\ ratio of 0.6 and diameter of $\sim$150~arcsec ($\sim$36~pc), it will be further studied in our future follow up studies. 

\begin{figure*}
 \begin{center}
\resizebox{1\linewidth}{!}{\includegraphics[trim = 0 0 0 0]{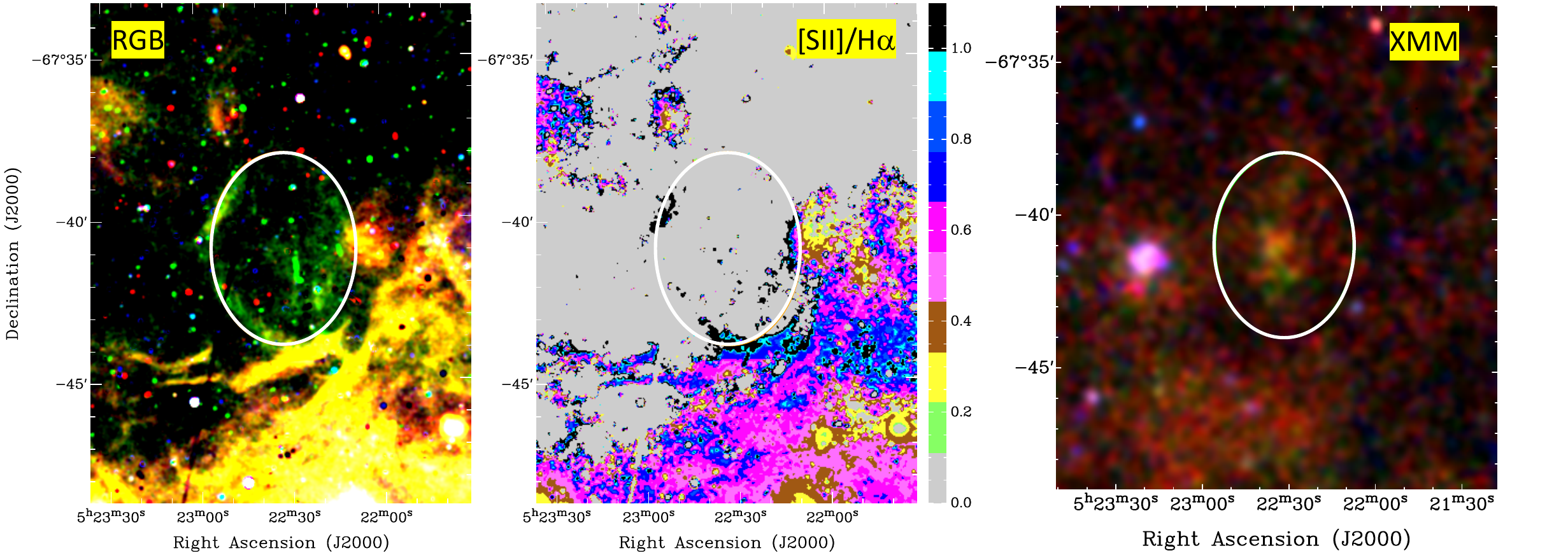}}
  \caption{MCSNR~J0522--6740: colour images produced from \ac{MCELS} data, where RGB corresponds to \Ha, \SII\ and \OIII\ (left) while the ratio map is between \SII\ and \Ha\ (middle). {\it XMM-Newton} EPIC RGB (R=0.3--0.7~keV band, G=0.7--1.1~keV band and B=1.1--4.2~keV band) image from the area of MCSNR~J0522--6740 is shown in the right panel. The white ellipse indicates the location of the optical shell.}
  \label{fig:A12}
 \end{center}
\end{figure*}

\subsection{MCSNR~J0522--6740 (Figure~\ref{fig:A12})}
 \label{sec:J0522--6740}
 
This source shows a diffuse elliptical shell morphology (D=87$\times$65~pc) which is indicative of an evolved \ac{SNR} (Figure~\ref{fig:A12} (left and middle)). It is found just outside the northern boundary of the giant \HII\ complex LHA~120-N\,44 \citep{1956ApJS....2..315H}. While there are no optical spectra available for this object, the source is most prominent in the light of \SII, and we estimate from the \ac{MCELS} images a very high \SII/\Ha\ ratio of $\sim$1.0. There are several isolated filaments where the \SII/\Ha\ ratio is elevated ($>$0.4). But, we couldn't connect any of these with possible new \ac{SNR} candidates. Seven massive OB stars are found in the vicinity, but only one within the bounds of this \ac{SNR} candidate.

Although the contour plot is located in a region with low X-ray exposure ($\sim$10\,ks EPIC-pn combining two observations), soft extended X-ray emission is significantly detected at the centre of the optical shell (Figure~\ref{fig:A12}, right), most prominently in the 0.7--1.1~keV band. We analysed the X-ray spectrum of MCSNR~J0522--6740 accumulated over the whole optical shell. The spectrum is thermal and reproduced by the emission of an optically-thin collisional-ionisation equilibrium plasma at \ac{LMC} abundance (about half-solar). The best-fit electronic temperature is $kT_e = 0.3 \pm 0.02$~keV. This is typical for evolved \ac{SNRs}, lending strong support to its confirmation as a bona-fide \ac{SNR}. The amount of \ac{LMC} neutral gas in front of the source is low (best-fit $N_H = 0$, and less than $10^{21}$~cm$^{-2}$ at the 90\,\% confidence level), but the total \ac{LMC} line-of-sight integrated \HI\ column at this position is only $1.6 \times 10^{21}$~cm$^{-2}$ \citep{2003ApJS..148..473K}. The 0.3--8~keV luminosity of MCSNR~J0522--6740 is $L_X = 1.7 \times 10^{34}$~erg~s$^{-1}$, potentially ranking this source among the 10~per~cent faintest (so far) \ac{LMC} \ac{SNRs} in X-rays.

We investigate whether the prominence of the medium 0.7--1.1~keV X-ray band could be due to the presence of elevated iron abundance of \ac{SN} ejecta origin, as was seen previously in several \ac{LMC} and \ac{SMC} \ac{SNRs} \citep[e.g.][]{2006ApJ...652.1259B,2014A&A...561A..76M,2016A&A...586A...4K,2019A&A...631A.127M}. When letting the oxygen and iron abundance free, fits with elevated Fe abundance are favoured despite large uncertainties (the upper value of Fe abundance is unconstrained), but the statistical improvement is moderate ($\Delta \chi ^2 = 4.95$ for 2 degrees of freedom, i.e. less than $2\sigma$). Exploring the parameter space of oxygen vs. iron abundance, the Fe/O ratio (by number) is $> 3$ at the 90\,\% confidence level. This is more than twice the average \ac{LMC} value of 1.4. We consider this enhancement to be indicative of a possible detection of iron ejecta. However, this evidence remains marginal given the limited exposure time and spectral resolution, and the intrinsic faintness of MCSNR~J0522--6740.

Based on above findings, we suggest that MCSNR~J0522--6740 is a new \ac{LMC} \ac{SNR}.

\begin{figure*}
 \begin{center}
 \resizebox{1\linewidth}{!}{\includegraphics[trim = 0 0 0 0]{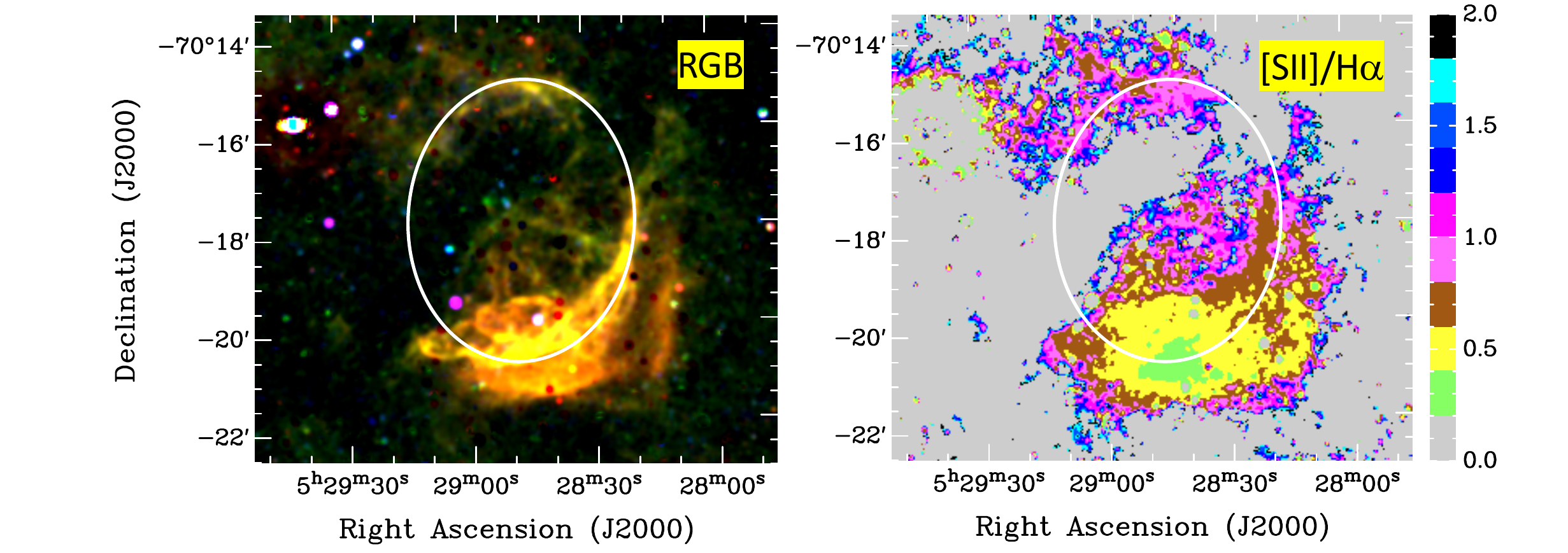}}
  \caption{J0528--7017: Colour images produced from \ac{MCELS} data, where RGB corresponds to \Ha, \SII\ and \OIII\ while the ratio map is between \SII\ and \Ha. The white ellipse indicates the location of the optical shell.}
  \label{fig:A13}
 \end{center}
\end{figure*}

\subsection{J0528--7017 (Figure~\ref{fig:A13})}
 \label{sec:J0528--7017}
The \ac{SNR} nature of J0528--7017 is drawn from its spherical (circular) shell morphology with a diameter (D=81~pc; Figure~\ref{fig:A13}) while \ac{MCELS} images indicates an overall \SII/\Ha\ ratio of $\sim$0.9. We searched MCPS images for massive (OB) stars near this object within a 100~pc radius and found 13 very distant OB stars from which 4 are within the boundaries of this proposed \ac{SNR} candidate.

\begin{figure*}
 \begin{center}
  \resizebox{1\linewidth}{!}{\includegraphics[trim = 0 0 0 0]{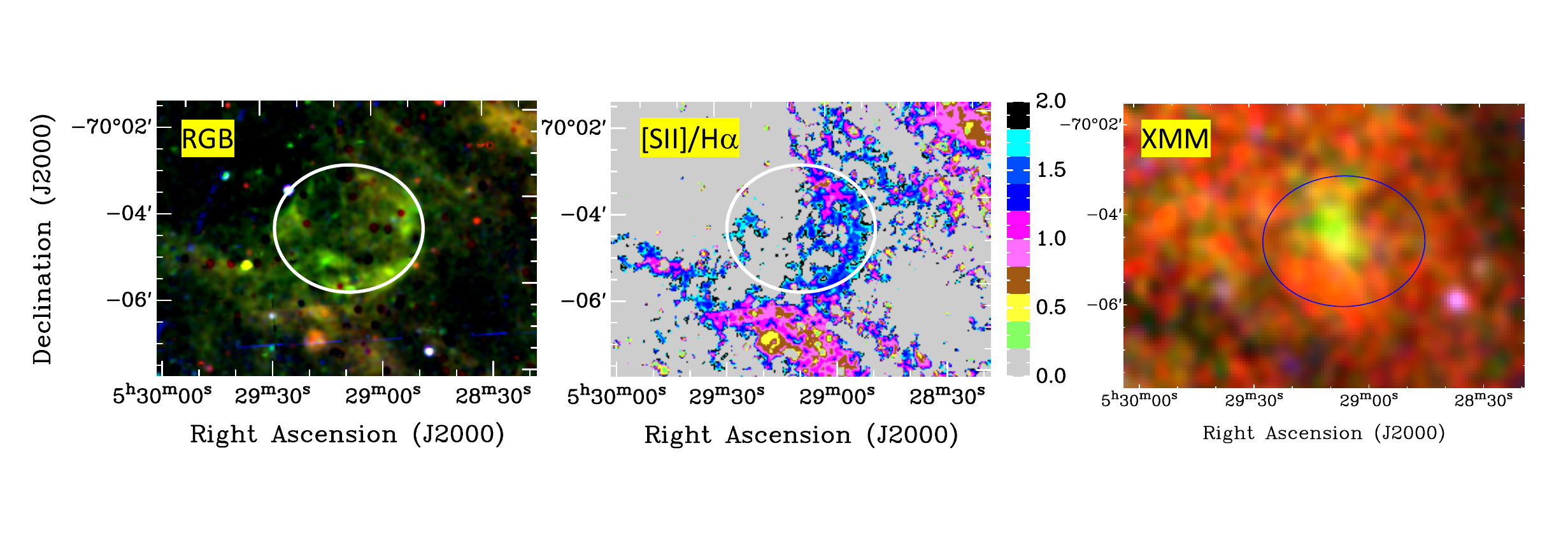}}
  \caption{J0529--7004: colour images produced from \ac{MCELS} data, where RGB corresponds to \Ha, \SII\ and \OIII\ (left) while the ratio map is between \SII\ and \Ha\ (middle). {\it XMM-Newton} EPIC RGB (R=0.3--0.7~keV band, G=0.7--1.1~keV band and B=1.1--4.2~keV band) image from the area of J0529--7004 (right). The white/blue ellipse indicates the location of the optical shell.}
  \label{fig:A14}
 \end{center}
\end{figure*}

\subsection{J0529--7004 (Figure~\ref{fig:A14})}
 \label{sec:J0529--7004}
The candidacy for the \ac{SNR} nature of J0529--7004 is also suggested by its spherical (circular) shell morphology with a diameter of 47~pc (Figure~\ref{fig:A14}). While there are no spectra available for this candidate, we use \ac{MCELS} images and found that the overall \SII/\Ha\ ratio of $\sim$1.0 warrants further investigation of this object as an \ac{SNR} candidate. We note that no massive stars could be found within the shell of this object but in a near vicinity (100~pc radius) we found 12~OB stars in the MCPS.

J0529-7004 is covered by two XMM-Newton observations, but both are at the very edge of the EPIC field of view. This results in an uneven exposure with 7.5\,ks in the eastern and 25\,ks in the western half of the region with optical emission. This candidate is located in an area of strong diffuse emission, but peaked soft X-ray emission is detected from J0529--7004 right at the centre of the optical emission. The position of the {\it ROSAT} source [HP99]~1077 is consistent with the bright spot seen in the EPIC image. The position of the {\it ROSAT} X-ray source source [HP99]~1077, which is about 18.5\arcsec\ away from the centre of J0529--7004, is consistent with the bright spot seen in the EPIC image.

\begin{figure*}
 \begin{center}
 \resizebox{1\linewidth}{!}{\includegraphics[trim = 0 0 0 0]{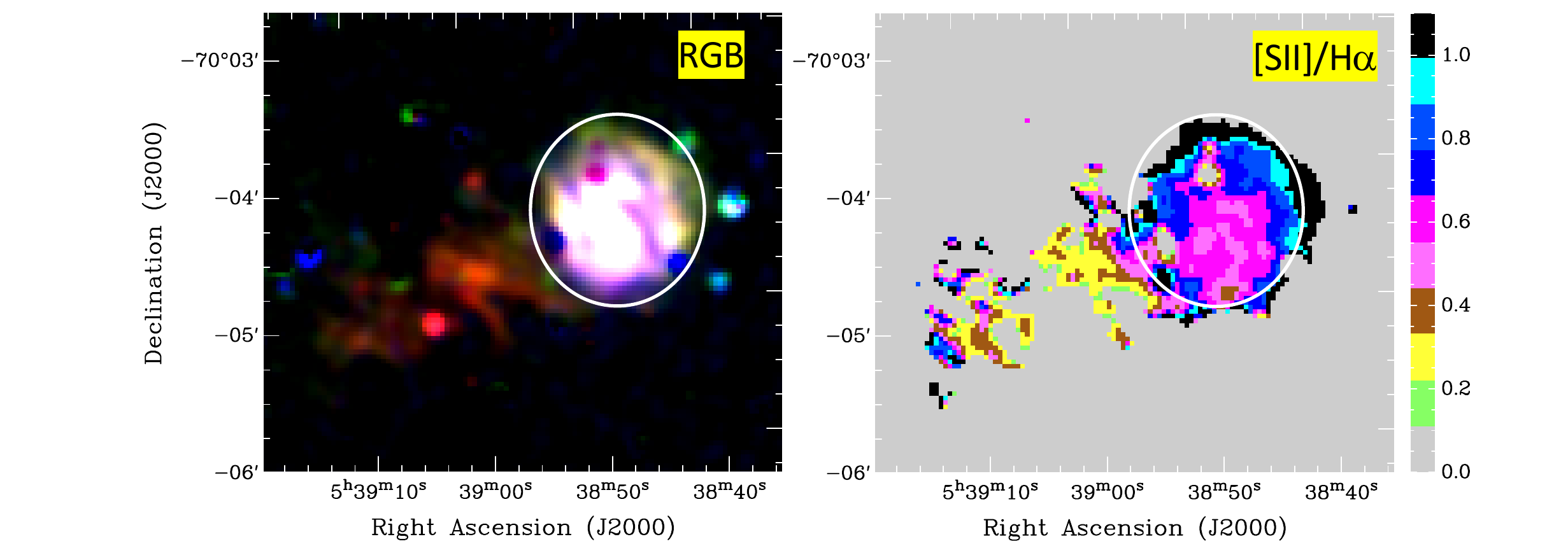}}
  \caption{J0538--7004: colour images produced from \ac{MCELS} data, where RGB corresponds to \Ha, \SII\ and \OIII\ while the ratio map is between \SII\ and \Ha. The white circle indicates the location of the optical shell.}
  \label{fig:A15}
 \end{center}
\end{figure*}

\subsection{J0538--7004 (Figure~\ref{fig:A15})}
 \label{sec:J0538--7004}
 
This \ac{SNR} candidate shows a partial filled-in shell, concentrating to the North-West (Figure~\ref{fig:A15}). There are no spectra available for this candidate but the \SII/\Ha\ ratio of $\sim$0.8 that we estimate from the \ac{MCELS} images is indicative of an \ac{SNR} nature. However, this is the second smallest \ac{SNR} candidate in our 19 object strong sample with D$_{av}$=19.4~pc. This would strongly argue for the presence of radio and X-ray emission which we don't see in any of our surveys despite the good coverage. Only two distant OB stars are found within the 100~pc from the object centre. As for the J0444--6758 (Section~\ref{sec:J0444--6758}), we see a bright central \OIII\ emission which is indicative of an \HII\ region. Therefore, this object is in a group of low confidence \ac{SNR} candidates in this sample.

\begin{figure*}
 \begin{center}
 \resizebox{0.47\textwidth}{!}{\includegraphics[trim = 0 0 0 0]{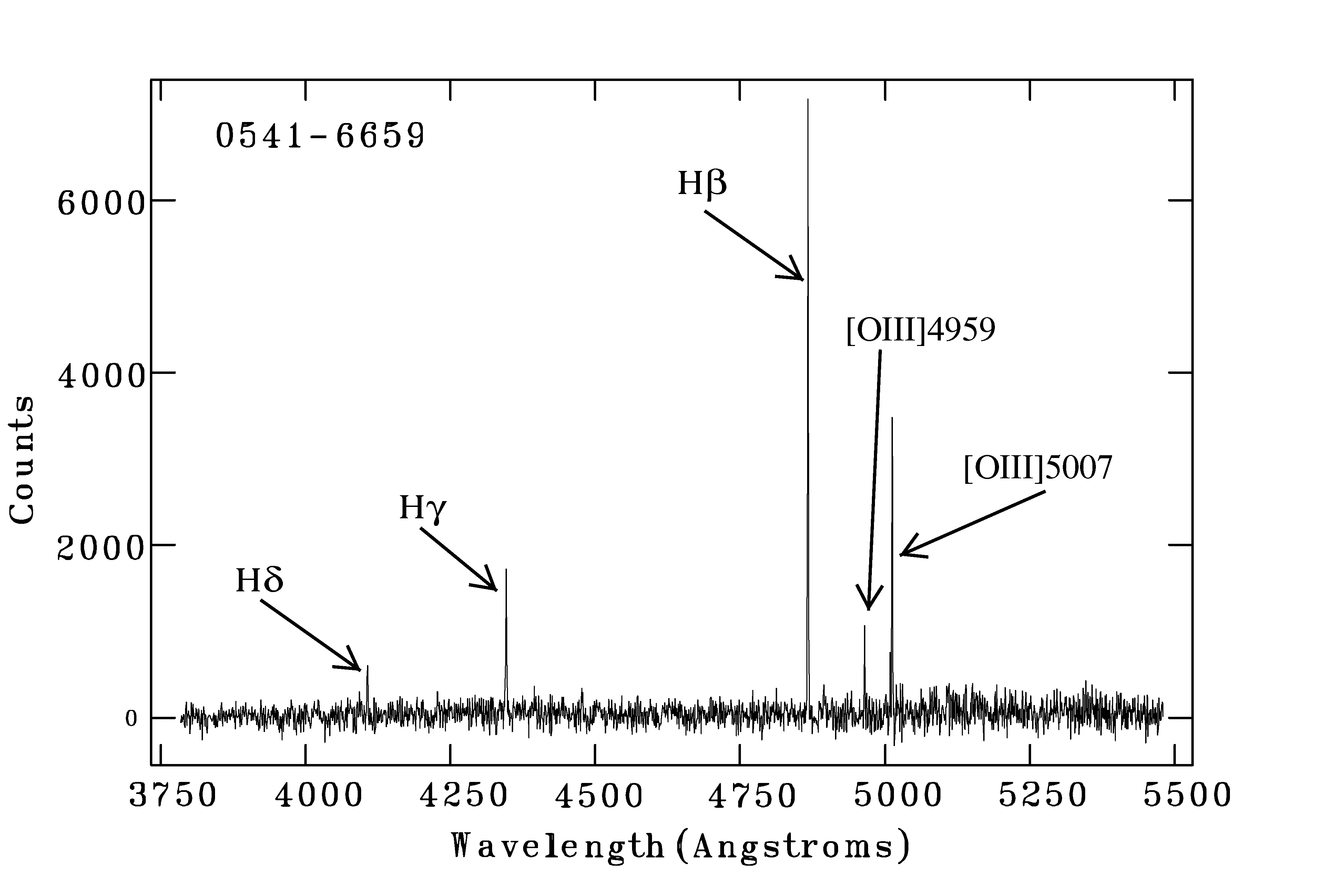}}
 \resizebox{0.47\textwidth}{!}{\includegraphics[trim = 0 0 0 0]{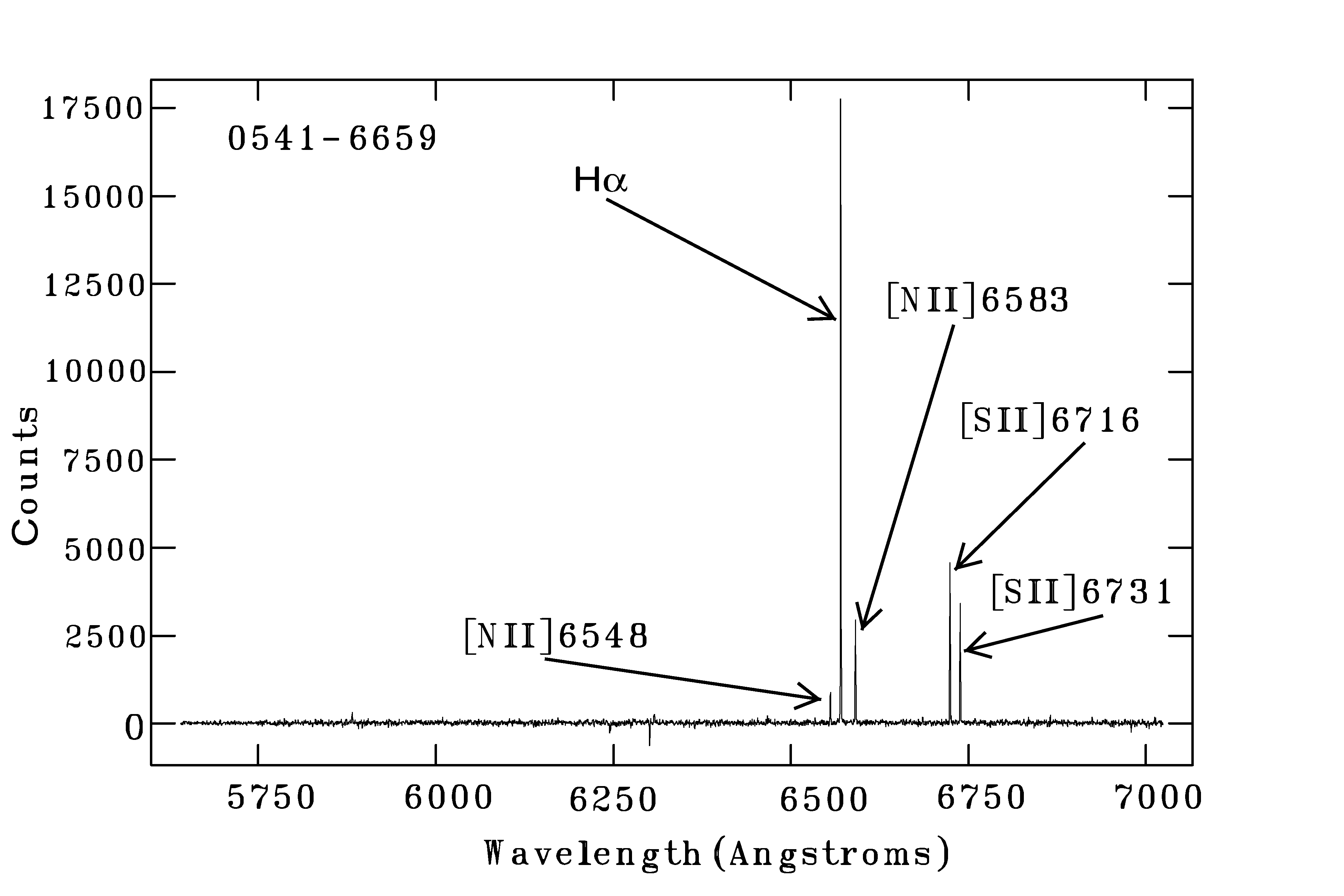}}
\resizebox{1\linewidth}{!}{\includegraphics[trim = 0 0 0 0]{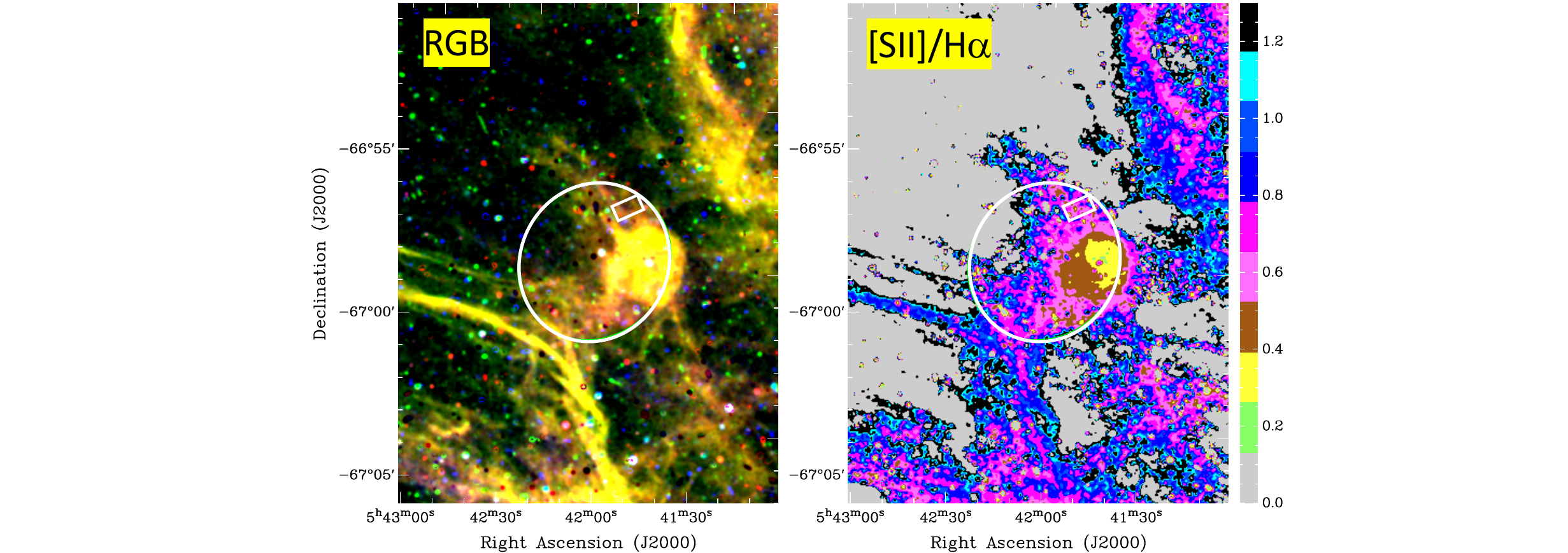}}
  \caption{MCSNR~J0541--6659: (Top) showing the spectra from both arms (left; blue, right; red) of the spectrograph; (Bottom) colour images produced from \ac{MCELS} data, where RGB corresponds to \Ha, \SII\ and \OIII\ while the ratio map is between \SII\ and \Ha. The rectangular box (white) represents an approximate position of the WiFeS slicer. The white circle indicates the location of the optical shell.}
  \label{fig:A16}
 \end{center}
\end{figure*}

\begin{figure*}
\centering
\includegraphics[width=.48\textwidth]{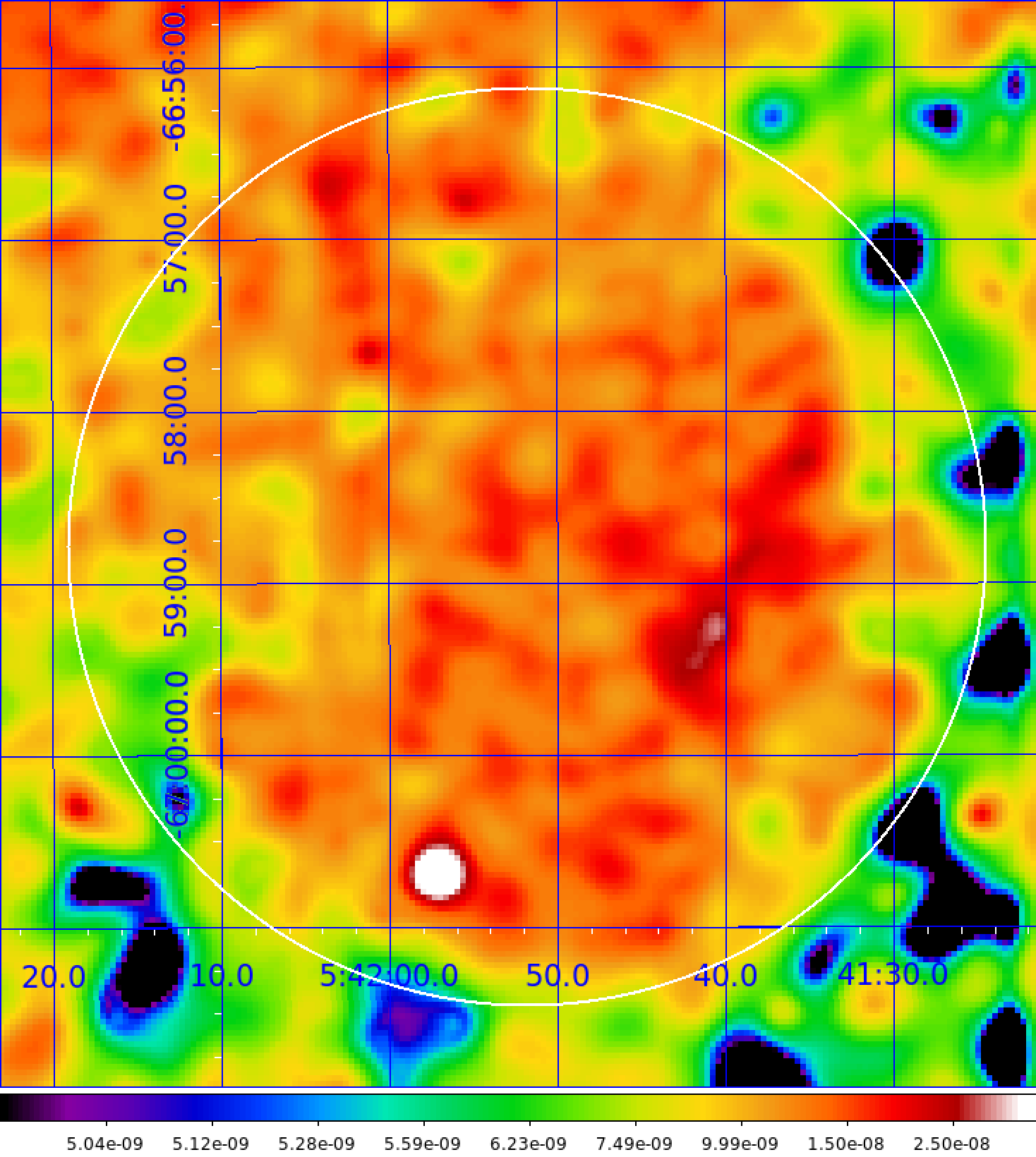}
\includegraphics[width=.48\textwidth]{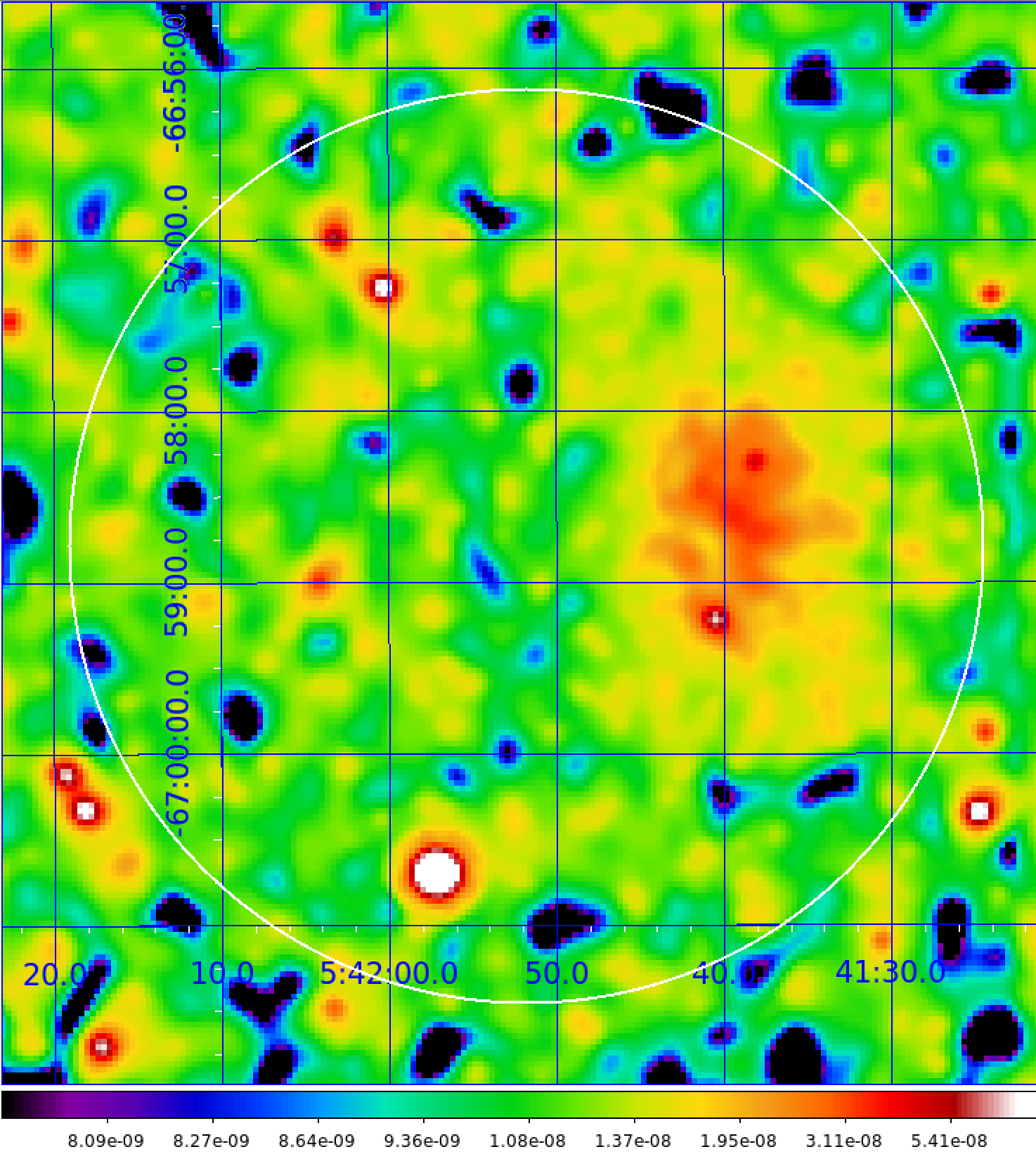}
\caption{\label{merged} {\it Chandra} ACIS images of MCSNR~J0541--6659 created from merged data of observations 16754 and 17675 (left: soft-band image, 0.3--1.0~keV; right: hard-band image, 1.0--8.0~keV). The images have been binned with a bin size of 4 pixels, smoothed adaptively, and exposure-corrected. The shell of the \ac{SNR} is indicated with a white circle.}
  \label{fig:A17}
\end{figure*}

\subsection{MCSNR~J0541--6659 (Figures~\ref{fig:A16} and \ref{fig:A17})}
\label{sec:0541-6659}

This source was previously suggested as an \ac{SNR} in the \ac{LMC} based on its initial X-ray detection \citep{2012A&A...539A..15G}. It was named [HP99]~456 in the {\it ROSAT}~PSPC survey of the \ac{LMC} \citep[][hereafter HP99]{1999A&AS..139..277H}. West of the X-ray source, there is an extended source in the optical bands. Here, for the first time, we present optical spectra of this object (see Figure~\ref{fig:A16}). It has some diffuse emission towards the top left and relatively low \SII/\Ha\ ratio of $\sim$0.4. We note a large peak in \Ha\ and \Hb\ which coupled with its \OIII\ value, gives a low \OIII/\Hb\ ratio. Therefore, this \ac{SNR} is possibly located next to an \HII\ region.

We obtained data from a 50~ks observation with the {\it Chandra} X-ray Observatory, using the ACIS-S array in VFAINT mode. The \ac{SNR} was observed at the aimpoint and was fully covered by the S3 chip. The observation was split into two parts with the same configuration: one observation performed on June~18,~2015 (ObsID 17675) with a net exposure time of 22~ks and another one performed 9 days later on June~27, 2015 (ObsID 16754) with a net exposure time of 29~ks. We reprocessed the data using CIAO Version~4.7 and CALDB Version 4.6.7. Using the reprocessed files, we created images in two bands: soft band (0.3--1.0~keV) and hard band (1.0--8.0~keV). We then smoothed the images adaptively using the tool {\tt dmimgadapt} with a Gaussian kernel. We also merged the event files of the two observations and created images.

The {\it Chandra} images show shell-like soft X-ray emission and confirm the \ac{SNR} identified with {\it XMM-Newton} data (Figure\,\ref{merged}). There is some additional faint soft diffuse emission extending to the northeast, which might be caused by acceleration of gas along a negative density gradient (possible ``blow out''). One should also consider that the \ac{SNR} has blown out as we can see open optical filaments in that direction in the \ac{MCELS} images.

Above 1~keV, extended emission is only visible in the western part, as had been already discovered with {\it XMM-Newton}. Inside the hard extended emission, which has an extent of $\sim$80\arcsec $\times$~100\arcsec, the superb spatial resolution of {\it Chandra} allowed us to detect two point sources: \ac{PSN} and \ac{PSS}. These sources were detected by applying the source detection procedure {\tt wavdetect} on the images of the different bands for both observations. For \ac{PSN}, we only obtained $\sim20$ net counts. The source is hard and only detected in the hard-band image. Source \ac{PSS} has $\sim60$ net counts and is detected both in the soft and the hard-band images. However, there is no indication that one of these sources might be a pulsar. In fact, according to \citet{2016A&A...585A.162M}, the stellar population in the region around the \ac{SNR} is rather old with an age of $>5 \times 10^7$ years for the majority of the stars \citep[based on the Magellanic Clouds Photometric Survey, MCPS,][]{2004AJ....128.1606Z}, which indicates that most of massive stars have ended their lives \citep{1993A&AS...98..523S}. In fact, a number of OB stars can be found within the 100~pc of the \ac{SNR} centre from which only two are within its boundaries.

\begin{figure*}
 \begin{center}
  \resizebox{0.47\textwidth}{!}{\includegraphics[trim = 0 0 0 0]{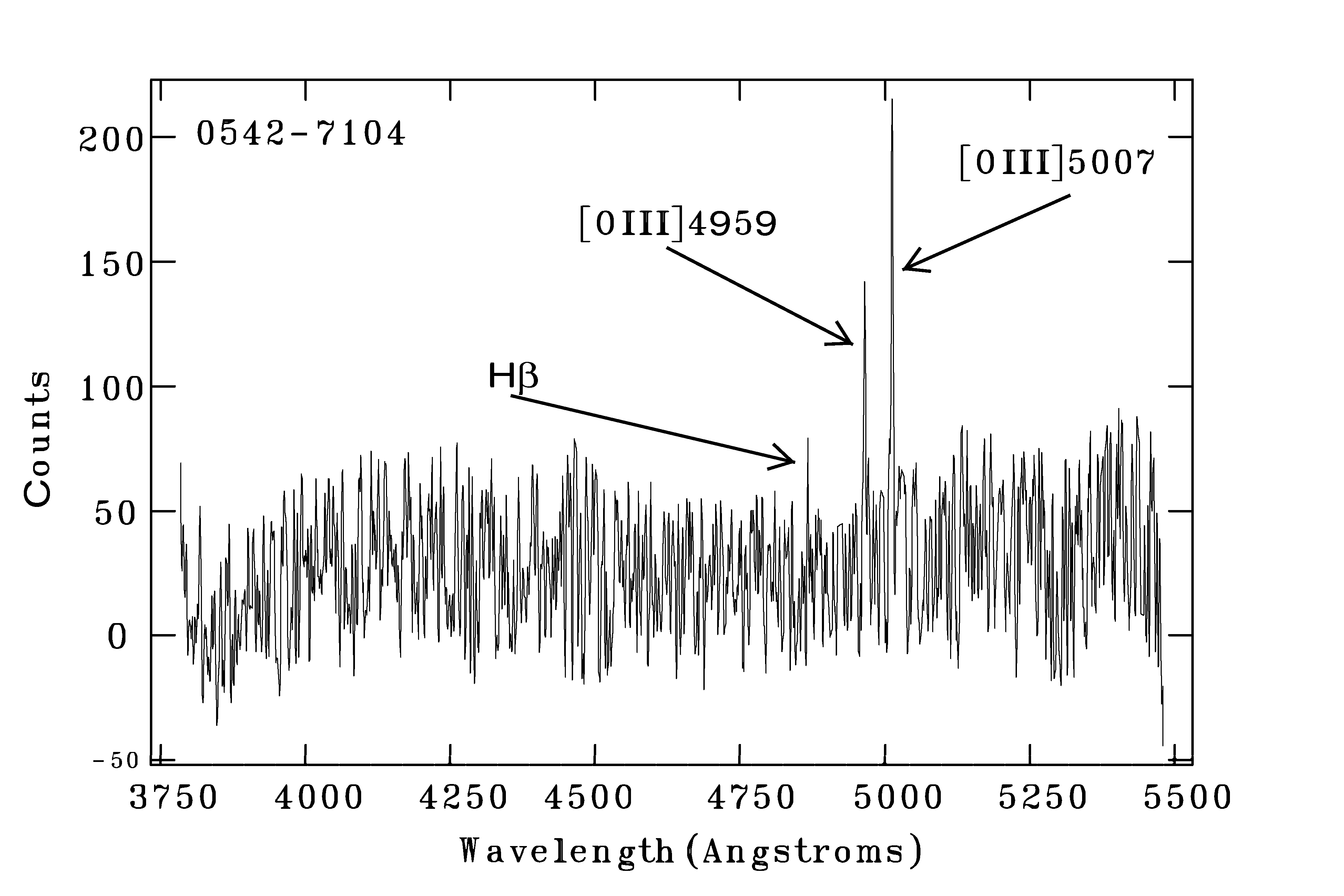}}
  \resizebox{0.47\textwidth}{!}{\includegraphics[trim = 0 0 0 0]{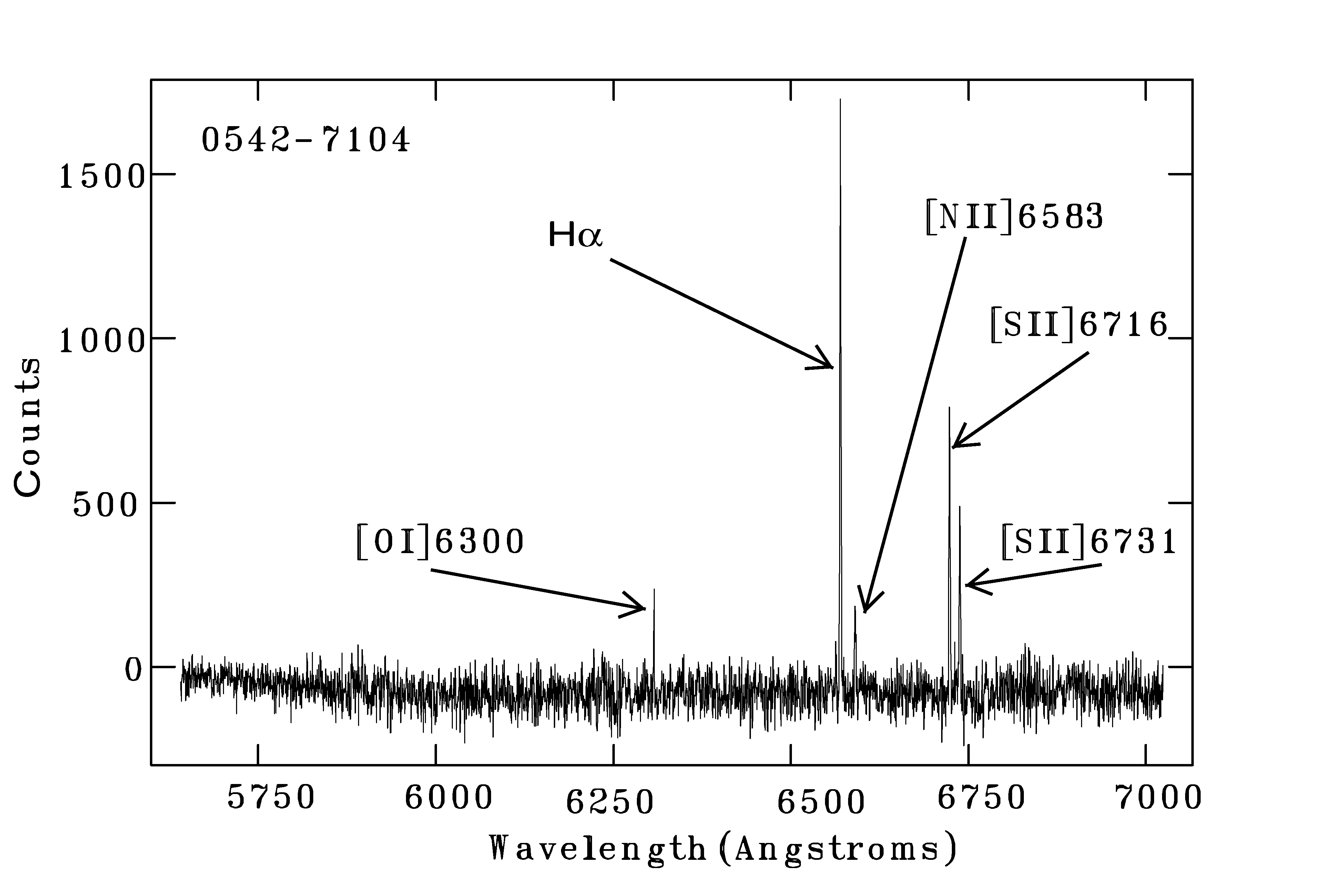}}
  \resizebox{1\linewidth}{!}{\includegraphics[trim = 0 0 0 0]{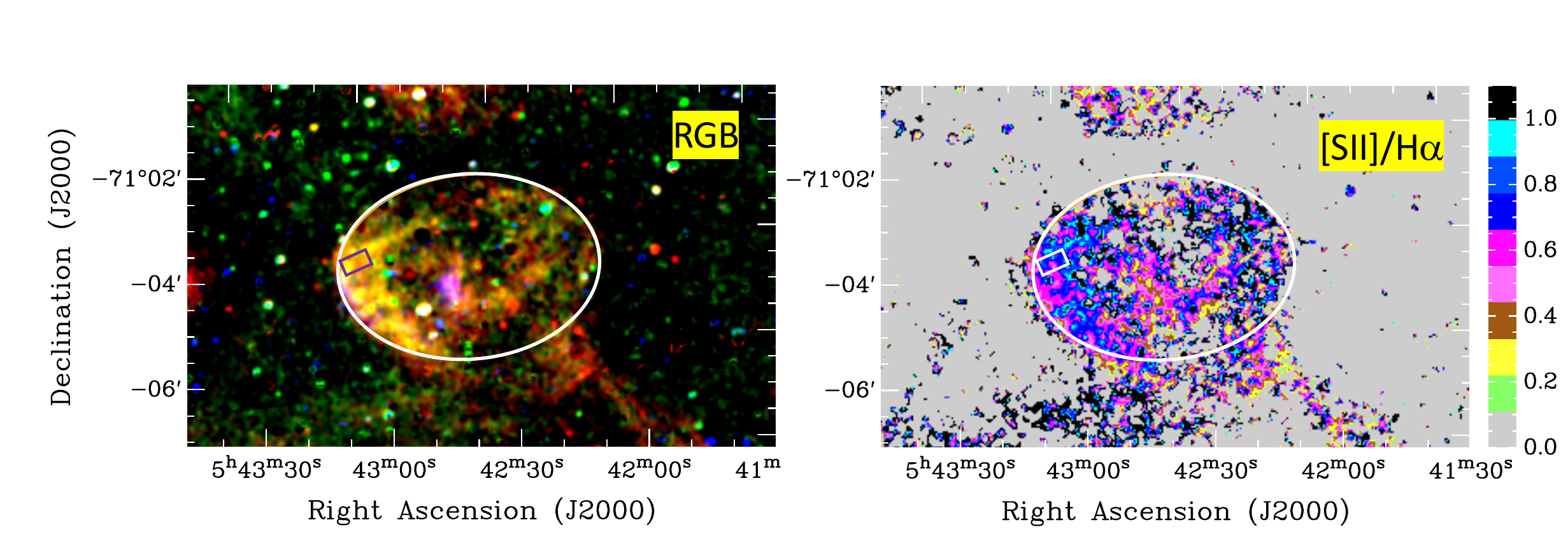}}
  \caption{MCSNR~J0542--7104: (Top) showing the spectra from both arms (left; blue, right; red) of the spectrograph; (Bottom) colour images produced from \ac{MCELS} data, where RGB corresponds to \Ha, \SII\ and \OIII\ while the ratio map is between \SII\ and \Ha. The rectangular box (magenta/white) represents an approximate position of the WiFeS slicer. The white ellipse indicates the location of the optical shell.}
  \label{fig:A18}
 \end{center}
\end{figure*}

\subsection{MCSNR~J0542--7104 (Figure~\ref{fig:A18})}
 \label{sec:0542-7104}
This is a large, elliptical \ac{SNR} candidate with a filled-in centre shell morphology (Figure~\ref{fig:A18}) that extends to 73$\times$51~pc. The \SII/\Ha\ ratio estimated from our spectroscopic observation is $\sim$0.82 which is typical of \ac{SNRs}. As for a number of other objects from our sample, this \ac{SNR} candidate does not show any detectable radio-continuum emission in our surveys. However, we note the detection of a source at this position in the {\it ROSAT} survey as HP[99]\,1235. Further {\it XMM-Newton} follow up observations of this source will be presented in Kavanagh~at~al.~(2020; in prep.) and we can consider this object as a bona-fide \ac{SNR}. We also note that only one but very distant OB star is found at some 90~pc distance from the centre of this object.

\begin{figure*}
 \begin{center}
 \resizebox{0.48\textwidth}{!}{\includegraphics[trim = 0 0 0 0]{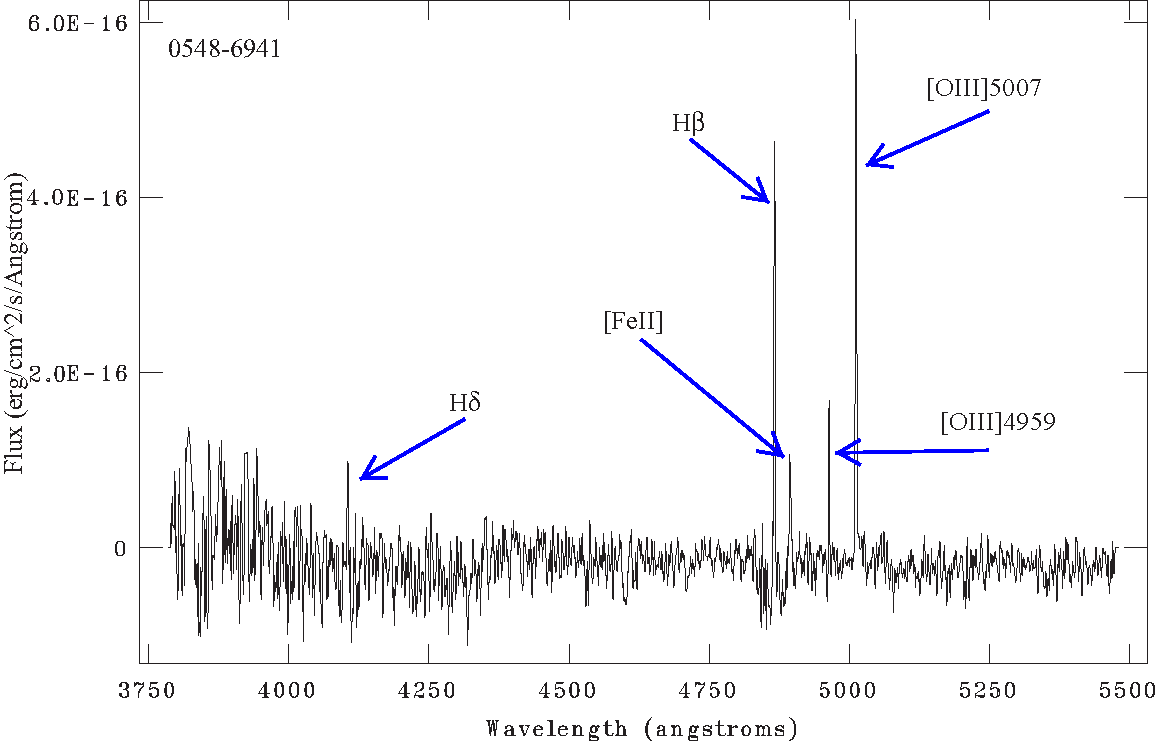}}
 \resizebox{0.48\textwidth}{!}{\includegraphics[trim = 173 0 0 0,clip]{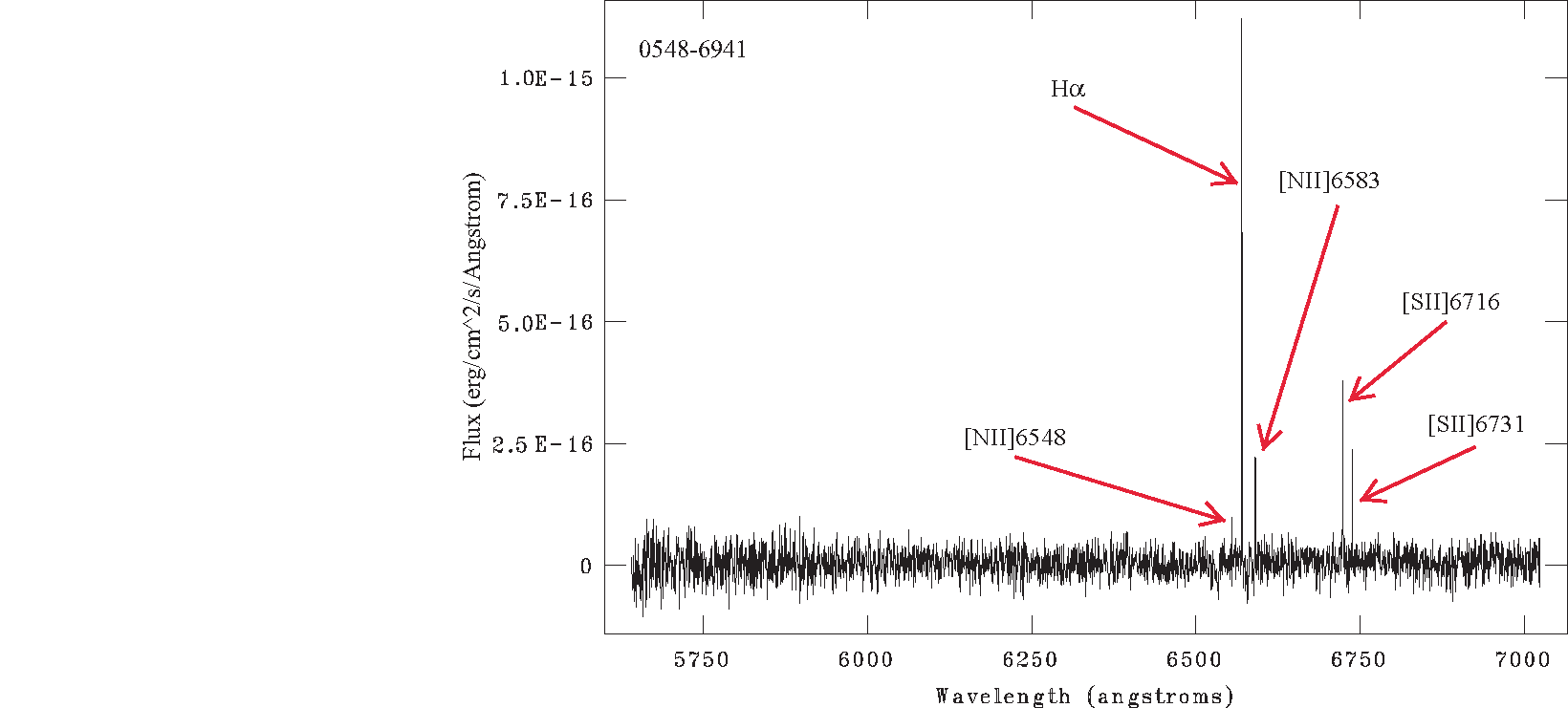}}
 \resizebox{1\linewidth}{!}{\includegraphics[trim = 0 0 0 0]{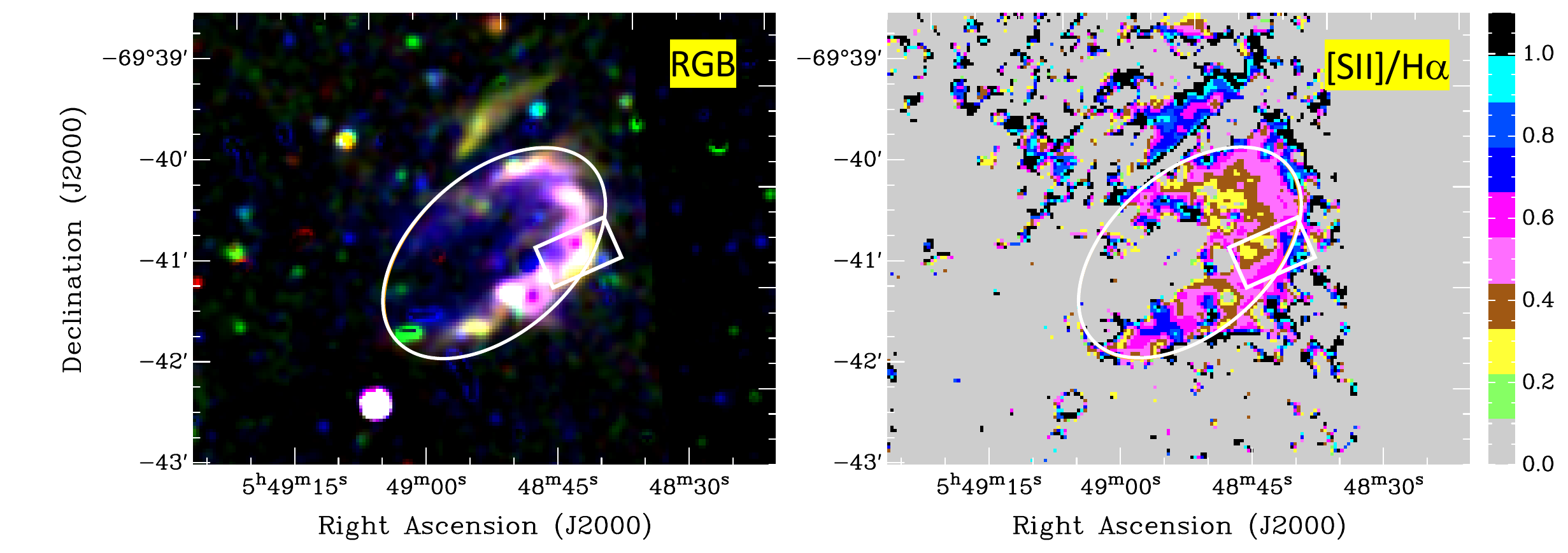}}
  \caption{J0548--6941: (Top) showing the spectra from both arms (left; blue, right; red) of the spectrograph; (Bottom) colour images produced from \ac{MCELS} data, where RGB corresponds to \Ha, \SII\ and \OIII\ while the ratio map is between \SII\ and \Ha. The rectangular box (white) represents an approximate position of the WiFeS slicer. The white ellipse indicates the location of the optical shell.}
  \label{fig:A19}
 \end{center}
\end{figure*}

\subsection{J0548--6941 (Figure~\ref{fig:A19})}
 \label{sec:J0548--6941}
 
This candidate has an obvious half shell elongated morphology to its south-west (Figure~\ref{fig:A19}) and a somewhat smaller size of 38$\times$23~pc. It has a \SII/\Ha\ ratio of $\sim$0.6 (from our WiFeS spectroscopic observations) and is the only \ac{SNR} candidate in the sample with a \FeII\ line. We find 4 distant OB stars in the area of \ac{SNR} candidate J0548--6941. This region is covered by {\it XMM-Newton} and reveals no X-ray emission.

\begin{figure*}
 \begin{center}
 \resizebox{1\linewidth}{!}{\includegraphics[trim = 0 0 0 0]{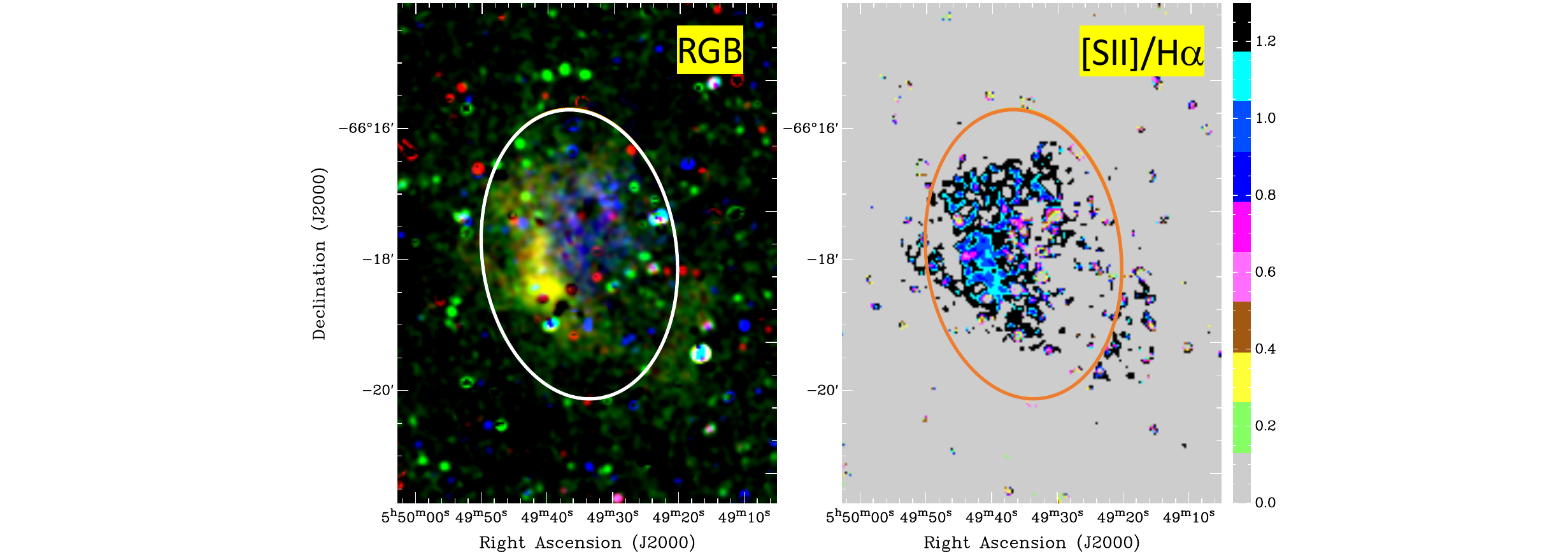}}
  \caption{J0549--6618: colour images produced from \ac{MCELS} data, where RGB corresponds to \Ha, \SII\ and \OIII\ while the ratio map is between \SII\ and \Ha. The white/orange ellipse indicates the location of the optical shell.}
  \label{fig:A20}
 \end{center}
\end{figure*}

\subsection{J0549--6618 (Figure~\ref{fig:A20})}
 \label{sec:J0549--6618}

J0549--6618 has no obvious shell morphology like many other objects in our sample but has a filled-in centre (Figure~\ref{fig:A20}). There are no spectra available for this candidate but our \SII/\Ha\ ratio estimates based on the \ac{MCELS} images is $\sim$1.0. The stellar environment is such that we can't see any nearby massive star.

\begin{figure*}
 \begin{center}
 \resizebox{0.34\textwidth}{!}{\includegraphics[trim = 0 0 0 0,clip]{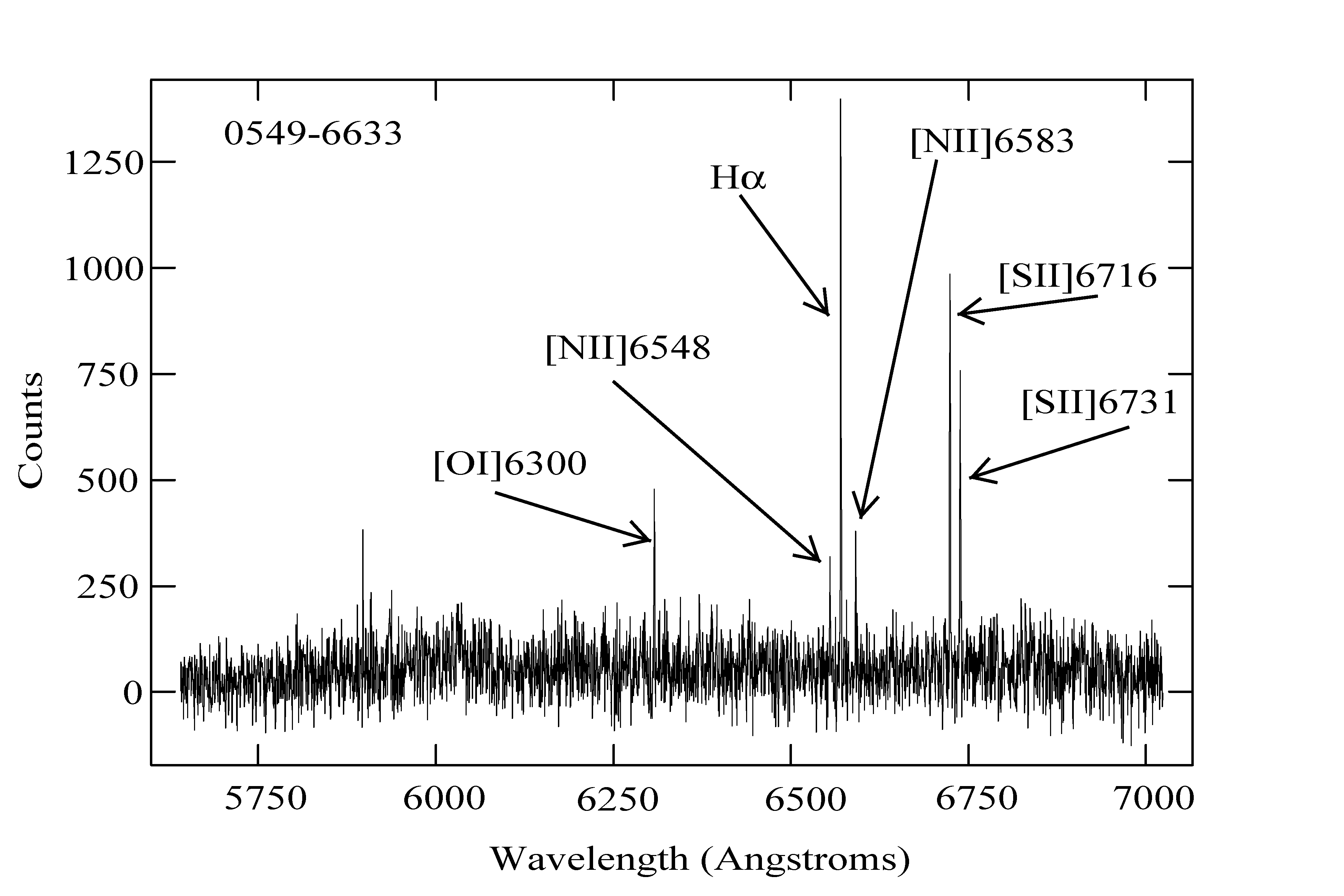}}
 \resizebox{0.65\linewidth}{!}{\includegraphics[trim = 0 0 0 0,clip]{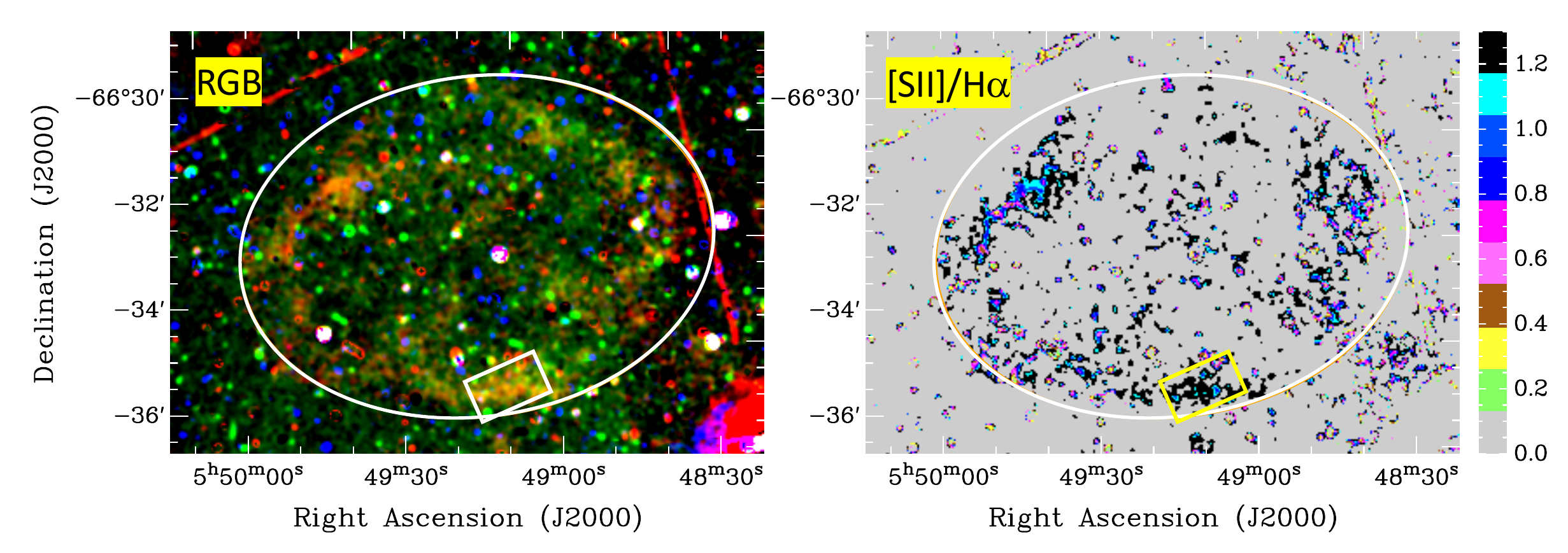}}
  \caption{0549--6633: (Left) showing the spectra from one arm (red) of the spectrograph; (Middle and right) colour images produced from \ac{MCELS} data, where RGB corresponds to \Ha, \SII\ and \OIII\ while the ratio map is between \SII\ and \Ha. The rectangular box (white/yellow) represents an approximate position of the WiFeS slicer. The white ellipse indicates the location of the optical shell.}
  \label{fig:A21}
 \end{center}
\end{figure*}

\subsection{J0549--6633 (Figure~\ref{fig:A21})}
\label{sec:0549-6633}

This \ac{SNR} candidate is chosen based on its very high \ac{MCELS} \SII/\Ha\ ratio of 1.4 (Figure~\ref{fig:A21}; Table~\ref{tab:snrflux}). While elongated (D=131$\times$95~pc), it also has a centre-filled-in shell morphology but without a distinctive peak at the centre. It is the largest among all of the objects studied here and it has the highest spectroscopic \SII/\Ha\ ratio at $\sim$1.38. The remnant is not only strong in \SII\ but is also detected in \OI. As shown in Figure~\ref{fig:2new}, this source is just outside of the established spectroscopic boundaries for being a \ac{SNR} and as such would qualify for a (super)bubble. However, as for J0549--6618, the stellar environment shows no nearby massive stars which is a strong argument against a (super)bubble classification.

\bsp	
\label{lastpage}
\end{document}